\pdfoutput=1

\documentclass{article}
\usepackage{iclr2027_conference,times}

\usepackage{amsmath,amsfonts,bm}

\def\eqref#1{equation~\ref{#1}}

\def\1{\bm{1}}

\DeclareMathAlphabet{\mathsfit}{\encodingdefault}{\sfdefault}{m}{sl}
\SetMathAlphabet{\mathsfit}{bold}{\encodingdefault}{\sfdefault}{bx}{n}

\DeclareMathOperator*{\argmax}{arg\,max}
\DeclareMathOperator*{\argmin}{arg\,min}

\usepackage{hyperref}
\usepackage{url}

\usepackage{amsmath}
\usepackage{amssymb}
\usepackage{amsthm}
\makeatletter
\renewenvironment{proof}[1][\proofname]{\par
  \pushQED{\qed}%
  \normalfont \topsep6\p@\@plus6\p@\relax
  \trivlist
  \item[\hskip\labelsep
        \itshape
    #1\@addpunct{.}]\ignorespaces
}{
  \popQED\endtrivlist\@endpefalse
}
\makeatother
\usepackage{multirow}
\usepackage{graphicx}
\usepackage{tabularx,booktabs}
\usepackage{newtxmath}
\usepackage{xcolor}
\usepackage{tcolorbox}
\newtcolorbox{promptbox}[1]{
    title=#1,
    fonttitle=\bfseries,
    colback=gray!7,
    colframe=black!100,
    colbacktitle=blue!50!black!50,
}
\usepackage{float}
\usepackage{makecell}
\usepackage{booktabs}
\usepackage{wrapfig}
\usepackage[table]{xcolor}
\definecolor{rosecolor}{HTML}{CC6677}
\definecolor{lightindigocolor}{HTML}{8D7CDE}
\usepackage{tikzsymbols}
\usepackage{wrapfig}
\usepackage{algpseudocode}
\usepackage{xcolor}
\usepackage{tcolorbox}
\usepackage{newtxmath}
\usepackage{algorithm}
\usepackage{setspace}
\usepackage[T1]{fontenc}

\usepackage{appendix}
\usepackage{titletoc}

\usepackage{appendix}
\usepackage{etoc}

\newtheorem{theorem}{Theorem}
\newtheorem{lemma}{Lemma}
\newtheorem{corollary}{Corollary}

\newtheorem{proposition}{Proposition}

\title{\textsc{SENECA}: \underline{S}mall-Sample Discrete \underline{En}tropy \underline{E}stimation via Self-\underline{C}onsistent Missing M\underline{a}ss}

\author{Lucas H. McCabe and H. Howie Huang \\
Department of Computer Science\\
George Washington University\\
Washington, DC 20052, USA \\
\texttt{\{lucasmccabe,howie\}@gwu.edu}
}

\iclrfinalcopy

\begin{document}

\maketitle

\begin{abstract}
Discrete entropy estimation is a classic information theory problem, wherein the average information content of a discrete random variable is estimated from samples alone.
Naive approaches, such as the plugin method, fail to account for the probability mass associated with members of the random variable's support that are unobserved in a given sample, known as the ``missing mass.''
The resulting systemic underestimation is particularly problematic when data is time-consuming or costly to gather.
We propose \textsc{SENECA}, an entropy estimation scheme based on a novel ``self-consistent'' missing mass calculation.
Extensive numerical experiments indicate that our approach outperforms many state-of-the-art alternatives overall in the small-sample setting.
We then apply \textsc{SENECA} to two practical use cases, namely biodiversity estimation and the detection of incorrect large language model responses, where our method is competitive with domain-specific approaches.
Our work advances \textsc{SENECA} as an effective drop-in replacement for small-sample entropy estimation, with broad utility across several domains.
\end{abstract}

\section{Introduction} \label{sec:introduction}

How surprising is a random variable? Can the expected amount of surprise of a random variable, known as information entropy, be reliably estimated from samples alone? 
Such are fundamental questions in information theory and uncertainty quantification, with consequences in machine learning and other data-driven disciplines.
A classic application is population ecology, where the entropy of species prevalence has served as an index of biodiversity for several decades \citep{sherwin2019introduction}.
Entropy estimators have also been employed to calculate information gain for decision trees \citep{nowozin2012improved} and mutual information for gene association network formulation \citep{hausser2009entropy}.
Extensive applications appear in natural language processing (NLP), as well, where samples may correspond to words of a vocabulary \citep{grignetti1964note, arora2022estimating}, meaning-clusters produced by a word sense induction system \citep{li-etal-2014-improved}, or class labels of text produced by a generative model \citep{kuhn2023semantic}.
An example of the latter is \textit{semantic entropy}, a popular method for predictive uncertainty quantification in large language models (LLMs), often used for hallucination detection in question-answering (QA) tasks \citep{Farquhar2024}.

As we discuss in Section \ref{sec:background}, discrete entropy estimators are biased \citep{paninski2003estimation}, and many have high error when sample size is small.
In practical scenarios, samples may be difficult to obtain, so methods well-suited for small sample sizes are crucial \citep{chao2012diversity}.
In population ecology, for instance, sample collection can be labor-intensive, involving manual collection and species identification \citep{janzen1973sweep1}.
In natural language processing applications, semantic entropy and other variability-based LLM uncertainty quantification (UQ) methods typically rely on only $10$ or so samples \citep{mccabe2026estimating}, for reasons that include the computational complexity of repeated text generation and response classification \citep{kuhn2023semantic, Farquhar2024}.
In this small-sample setting, \citet{mccabe2026estimating} demonstrated that the plugin method employed by \citet{Farquhar2024} in calculating ``black-box'' semantic entropy underestimates ``true'' semantic entropy.

\paragraph{Our contributions.} To address the above limitations, this work advances the following:
\begin{enumerate}
    \item We propose \textbf{\textsc{SENECA}} (\underline{s}mall-sample discrete \underline{en}tropy \underline{e}stimation via self-\underline{c}onsistent missing m\underline{a}ss), an entropy estimator based on shrinkage and fixed-point iteration (Section \ref{sec:sc}).
    \item We assess \textsc{SENECA} for entropy estimation, demonstrating \textbf{improved accuracy over existing estimators in the small-sample setting} in numerical experiments on $72$ unique distributions (Section \ref{sec:numerical}).
    Our method performs especially favorably in the under-sampled regime, where observation of all members of the support is impossible.
    \item We demonstrate broad interdisciplinary applications by extending our evaluation to \textbf{biodiversity estimation} and \textbf{LLM uncertainty quantification} (Section \ref{sec:applications}).
    Toward the former, \textbf{\textsc{SENECA} consistently outranks other methods} on $58$ real species-prevalence datasets.
    Regarding the latter, a specialized variant \textsc{SENECA-U} attains the \textbf{third-highest aggregated AUROC} among several competitive baselines, \textbf{outranking all other entropy-based methods but one}.
\end{enumerate}

\section{Background} \label{sec:background}

\paragraph{Entropy estimation.} Let $A^n = \{\{a_1, a_2, \dots, a_n\}\}$ be a multiset of $n$ independent and identically distributed (IID) samples drawn with replacement from a population of items.
A classification function $h: \mathcal{A} \rightarrow \mathcal{X}$ maps items to labels, inducing the multiset of label occurrences $X^n = \{\{x_1, x_2, \dots, x_n\}\} = \{\{h(a) | a \in A^n\}\}$.
Without knowing the multinomial label-prevalence distribution $P = \langle p_u | u \in \mathcal{X} \rangle$ or support $|\mathcal{X}|$, our goal is to estimate the entropy \citep{shannon1948mathematical}:
\begin{equation*}
\mathbb{H}(P) = - \sum_{u \in \mathcal{X}} p_u \log p_u.
\end{equation*}

\paragraph{Plugin entropy.}
Let the label-counting function $N_{u, h}(A^n) = \sum_{a \in A^n} \vmathbb{1}_{h(a)=u}$ denote the number of times items belonging to the label $u$ appear in $A^n$ (for brevity, we will use $N_{u, h}(A^n)$ and $N_{u}(X^n)$ interchangeably).
The so-called ``plugin'' entropy estimator relies on maximum-likelihood prevalence estimates $\hat{p}_u^{ML} = \frac{N_{u}(X^n)}{n}$ (i.e., the empirical frequencies of observed labels):
\begin{equation}\label{eq:plugin_entropy}
\mathbb{\hat{H}}^{plugin}(X^n) = - \sum_{u \in X^n_{\neq}} \hat{p}_u^{ML} \log \hat{p}_u^{ML},
\end{equation}
where $X^n_{\neq}$ denotes the set of unique items of $X^n$.
The plugin method is consistent \citep{antos2001convergence}, but negatively biased \citep{basharin1959statistical}, failing to account for unobserved labels (Figure \ref{fig:summary}A).
In the \textit{under-sampled regime}, where $|\mathcal{X}|>n$, unobserved labels are guaranteed.

\paragraph{Missing mass estimation.}
The total probability mass associated with unobserved labels, known as the \textit{missing mass}, corresponds with the probability that a subsequent observation $a_{n+1}$ will belong to a new, unobserved label (i.e., $h(a_{n+1}) \in U$, where $U = \mathcal{X} \setminus X^n_\neq$):
\begin{equation*}
M(X^n) = \sum_{u \in \mathcal{X}} p_u \vmathbb{1}_{N_{u}(n)=0}.
\end{equation*}
In our notation, $M(X^n)$ is the \textit{particular} missing mass associated with $X^n$, and $M$ is a random variable.
The classic Good-Turing missing mass estimator is given by $\hat{M}^{GT}(X^n) = \frac{\Phi_{1}(X^n)}{n}$, where $\Phi_{i}(X^n) = \sum_{u \in \mathcal{X}} \vmathbb{1}_{N_{u}(X^n) = i}$ \citep{good1953population}.
The multiset $\{\{ \Phi_i | i \in \mathbb{Z}_{[0, n]}\}\}$ denotes the frequency of frequencies (FoFs) or ``fingerprint'' of the sample \citep{valiant2017estimating, wu2019chebyshev}.
$\hat{M}^{GT}$ may be viewed as a leave-one-out method \citep{mcallester2001learning} that is unbiased with respect to $M(X^{n-1})$ \citep{juang2002bias}, but not $M(X^n)$ \citep{mcallester2000convergence, antos2001convergence}. 
Indeed, $\hat{M}^{GT}$ has bias at most $\frac{1}{n}$ \citep{mcallester2000convergence}, but its worst-case risk mean-squared error is at least $\frac{0.608}{n} + o(\frac{1}{n})$ \citep{rajaraman2017minimax}.

\paragraph{Coverage-adjustment.}
The complement of the missing mass is known as the sample coverage \citep{roswell2021conceptual}.
The coverage-adjusted entropy estimator of \citet{chao2003nonparametric} scales observed labels prevalence by Good-Turing-estimated sample coverage (i.e., $\hat{p}_u^{CS} = (1-\hat{M}^{GT}) \hat{p}_u^{ML}$) and applies a Horvitz–Thompson scheme \citep{horvitz1952generalization} to estimate the total entropy:
\begin{equation}
    \hat{\mathbb{H}}^{CS} = -\sum_{u \in X^n_{\neq}} \frac{\hat{p}_u^{CS}  \log \hat{p}_u^{CS} }{1-(1-\hat{p}_u^{CS})^n}. \label{eq:chao-shen}
\end{equation}
By way of $\hat{M}^{GT}$, $\hat{\mathbb{H}}^{CS}$ uses only one FoF: the number of ``singletons'' $\Phi_1$.
While $\hat{\mathbb{H}}^{CS}$ can excel for small sample sizes \citep{pinchas2024comparative}, its dependence on the singleton rate $\Phi_{1}$ can be problematic; if $\Phi_{1}(X^n) = n$, for example, the summand of Equation \ref{eq:chao-shen} is $\frac{0 \log 0}{0}$, and $\hat{p}_u^{CS} = \hat{p}_u^{ML}$ when $\Phi_{1}(X^n) = 0$.
Subsequent work by \citet{chao2013entropy} employs two FoFs: $\Phi_1$ and $\Phi_2$.
Such methods successfully reduce the bias associated with entropy estimation, but no unbiased estimator exists in general \citep{paninski2003estimation}, and low-bias methods may exhibit high variance \citep{hastie2009elements}.

\section{Entropy estimation via ``self-consistent'' missing mass}\label{sec:sc}

\begin{figure*}[t!]
    \centering
    \includegraphics[width=1.0\textwidth]{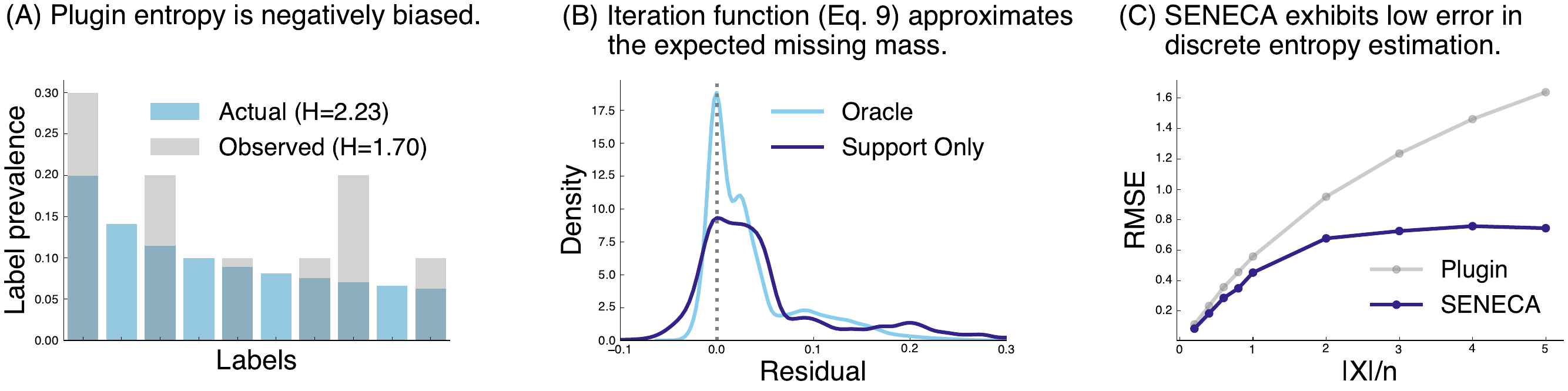}
    \caption{
    \textbf{Visual overview.}
    \textbf{(A) The plugin method can underestimate entropy when elements of the support go unobserved.}
    Shown: a Zipf distribution with shape parameter $0.5$ and support size of $10$ (cyan).
    A sample of size $10$ is drawn, and the empirical frequencies of observed labels are depicted (translucent grey overlay).
    Labels without grey overlay are unobserved; their cumulative probability mass is the ``missing mass.''
    \textbf{(B) When the support size and missing mass are known, Eq. \ref{eq:mu-fp} approximates the \textit{expected} missing mass.} 
    Shown: kernel density plots of residuals of two methods of estimating $\mathbb{E}_{X^n \sim P}[M]$ - $\mu(M(X^n), |U|, X^n)$ (Oracle) and the fixed-point missing mass calculated by \textsc{SENECA} with \textit{known} support size (Support Only) - under the simulation settings used in our numerical experiments (i.e., $n=10$, $72$ unique distributions, each over $1000$ trials; see Figure \ref{fig:entropy-rmse-n10}). 
    The grey dotted line highlights a residual of zero.
    \textbf{(C) \textsc{SENECA} uses fixed-point iteration and a support size \textit{estimate} to achieve low error.}
    Shown: simulation results for the Zipf-0.5 distribution only (for symmetry with panel A) over $9$ support sizes, each over $1000$ trials.
    }
    \label{fig:summary}
\end{figure*}

Herein, we motivate and introduce \textsc{SENECA}, which requires a sample size of at least $2$.
We begin by decomposing the expected missing mass of a multinomial label-prevalence distribution $P$ over support $\mathcal{X}$ for samples of size $n$ into ``observed'' and ``unobserved'' components, for a realized $X^n$:
\begin{equation}
    \mathbb{E}_{X^n \sim P}[M] =  \sum_{u \in \mathcal{X}} p_u  \Big(1-p_u\Big)^{n} = \underbrace{\Big( \sum_{u \in X^n_{\neq}} p_u  \Big(1-p_u\Big)^{n} \Big)}_{\text{``observed'' component}} + \underbrace{\Big( \sum_{u \in U} p_u  \Big(1-p_u\Big)^{n} \Big)}_{\text{``unobserved'' component}}. \label{eq:expected_m}
\end{equation}

We are left to estimate the label-prevalences within the observed and unobserved components of Eq. \ref{eq:expected_m}.
For the former, we notice that $\hat{p}_u^{ML}$ is conditionally positively biased on non-zero missing mass.
For this reason, we employ \textit{isotropic shrinkage} of the maximum likelihood prevalence estimator, scaling empirical frequencies of all observed labels by a single scalar (i.e., $\hat{p}_u^{IS} = \alpha \hat{p}_u^{ML} \forall u \in X^n_{\neq}$ for some $\alpha \in (0, 1)$).
If, hypothetically, $M(X^n)$ is known, then $\alpha = 1-M(X^n)$ preserves the missing mass, since $M(X^n)=1-\sum_{u \in X^n_{\neq}} \hat{p}_u^{IS} = 1-\alpha \sum_{u \in X^n_\neq}\hat{p}_u^{ML}=1-\alpha$.\footnote{Exchanging $\hat{M}^{GT}$ for $M(X^n)$ leads to the observed species prevalence estimates of \citet{chao2003nonparametric}.
}
Under \textit{natural estimation}, all labels appearing the same number of times in $X^n$ are assigned the same prevalence estimate \citep{orlitsky2015competitive}, an assumption made by the overwhelming majority of distribution estimators \citep{balasundaram2025estimating}.
Consequently, unobserved labels are assigned prevalence $\frac{M(X^n)}{|U|}$ if $|U|$ is known and at least one.
With the above in mind, we construct the following function of $m \in \mathbb{R}_{[0, 1]}$ and $\upsilon \in \mathbb{R}_{\geq0}$ and sample labels $X^n$, based on Eq. \ref{eq:expected_m}:
\begin{equation}
    \mu(m, \upsilon, X^n) = \underbrace{(1-m) \sum_{u \in X^n_{\neq}} \hat{p}_u^{ML}  \Big(1-(1-m)\hat{p}_u^{ML}\Big)^{n}}_{\text{``observed'' component}} + \underbrace{m \Big(1-\frac{m}{\max(\upsilon, 1)}\Big)^{n}}_{\text{``unobserved'' component}}, \label{eq:mu-fp}
\end{equation}
where the $\max(\upsilon, 1)$ ensures that the unobserved component is defined.
The sub-Gaussianity of missing mass \citep{skorski2023sub, li2024sub} implies large deviations are rare: realizations of $M$ are unlikely to substantially stray from expectation.
It stands to reason that $\mu(M(X^n), |U|, X^n) \approx \mathbb{E}_{X^n \sim P}[M]$, which we observe in practice; the cyan curve in Figure \ref{fig:summary}B (``Oracle'') depicts concentration near zero of residuals between $\mathbb{E}_{X^n \sim P}[M]$ and $\mu(M(X^n), |U|, X^n)$ over many settings.

Of course, the true values of $M(X^n)$ and $|U|$ are generally unknown.
Instead, given an estimate of the number of unobserved labels $\upsilon$, we propose a \textit{self-consistent} missing mass estimate:
\begin{equation}  \label{eq:mass-fp-def}
\hat{M}_{\upsilon}^{SC}(X^n) \triangleq m^*: \mu(m^*, \upsilon, X^n)=m^*.
\end{equation}
In the following theorems, we establish two complementary scenarios wherein such a fixed point exists uniquely (auxiliary results postponed to Appendix \ref{sec:proofs}):
\begin{theorem}\label{thm:existence-and-uniqueness-1}
    If $n \in \mathbb{Z}_{\geq 2}$ and $1 \leq \upsilon \leq \frac{n}{2}$, then Eq. \ref{eq:mass-fp-def} has a unique solution on $[0, 1]$.
\end{theorem}
\begin{proof}
    By Lemma~\ref{lem:fixed-point-existence}, such a fixed point exists. 
    Lemma~\ref{lem:fixed-point-uniqueness-all-labels-same} establishes uniqueness for $|X_\neq^n|=1$, so, for the remainder of this proof, we assume $|X_\neq^n|>1$.
    If no fixed point exists in $[0, \frac{\upsilon}{n+1})$, there must be exactly one in $[\frac{\upsilon}{n+1}, 1]$, by Lemma~\ref{lem:fixed-point-uniqueness-right-half}.
    If, on the other hand, a fixed point exists in $[0, \frac{\upsilon}{n+1})$, then it is the unique one on the entire $[0, 1]$ interval, by Lemma~\ref{lem:fixed-point-uniqueness-left-half}, and the result follows.
\end{proof}
\begin{theorem}\label{thm:existence-and-uniqueness-2}
    Define $B(n) \triangleq \min\Big(\frac{n(n-2)^2 + n - 2}{2n}, \Big((\frac{n+1}{n-1})^{n-1} - e + 1\Big)n \Big)$.
    If $n \in \mathbb{Z}_{\geq 4}$ and $\frac{n}{2} \leq \upsilon \leq B(n)$, then Eq. \ref{eq:mass-fp-def} has a unique solution on $[0, 1]$.
\end{theorem}
\begin{proof}
    Similar to Theorem~\ref{thm:existence-and-uniqueness-1}, existence follows from Lemma~\ref{lem:fixed-point-existence}, and Lemma~\ref{lem:fixed-point-uniqueness-all-labels-same-2} shows uniqueness for $|X_\neq^n|=1$.
    So, it suffices to show that at most one solution exists on $[0, 1]$ when $|X_\neq^n|>1$.
    In such cases, Proposition~\ref{prop:multiple-labels-no-zero} and Lemma~\ref{lem:sc-open-interval} establish that $m=0$ and $m=1$ are not fixed points, respectively.
    $\frac{\mu(m)}{m}$ is strictly decreasing on the remaining interval, per Lemma~\ref{lem:v-upper-bound-lemma}, so there is at most one $m^* \in (0, 1)$ such that $\frac{\mu(m^*)}{m^*} = 1$ (equivalently, $\mu(m^*) = m^*$, since $m^* \neq 0$).
\end{proof}
The indigo curve in Figure \ref{fig:summary}B (``Support Only'') illustrates that using the \textit{true} number of unobserved labels and performing fixed-point iteration on Eq. \ref{eq:mass-fp-def} induces similarly small, albeit slightly larger, residuals as the oracle setting.
In the absence of the true number unobserved labels, we propose the following recipe satisfying the constraints of Theorems~\ref{thm:existence-and-uniqueness-1} and \ref{thm:existence-and-uniqueness-2}: employing an estimate of the support size $\widehat{|\mathcal{X}|}$, set $\upsilon$ according to:\footnote{We further discuss this ``clipping'' strategy in Appendix~\ref{sec:v-clipping}, wherein Figure~\ref{fig:fingerprints-clipped} illustrates the acceptable region for $\upsilon$. Appendix~\ref{sec:b-subquadratic} confirms that the upper bound satisfies the $o(n^2)$ requirement of Theorem~\ref{thm:weak-consistency}.}
\begin{equation}
    \upsilon = \begin{cases}
            \max \Big( \min(\widehat{|\mathcal{X}|}-|X^n_{\neq}|, \frac{n}{2} ), 1 \Big) & n \in [2, 3] \\
            \max \Big( \min(\widehat{|\mathcal{X}|}-|X^n_{\neq}|, B(n) ), 1 \Big) & \text{otherwise,} \label{eq:def-v-clipping}
        \end{cases}
\end{equation}
and adjust empirical label-prevalences by $(1-\hat{M}_{\upsilon}^{SC}(X^n))$ in a Chao-Shen-like entropy estimator:
\begin{equation}
    \textsc{SENECA}(X^n, \upsilon) = -\Big(1-\hat{M}_{\upsilon}^{SC}(X^n)\Big)\sum_{u \in X^n_{\neq}} \frac{\hat{p}_u^{ML}  \log \Big((1-\hat{M}_{\upsilon}^{SC}(X^n)) \hat{p}_u^{ML}\Big)}{1-\Bigg(1-\Big((1-\hat{M}_{\upsilon}^{SC}(X^n)) \hat{p}_u^{ML}\Big)\Bigg)^n}. \label{eq:entropy-sc}
\end{equation}

Algorithm \ref{alg:sc} (Appendix \ref{sec:entropy-details}) summarizes our approach.
Except where stated otherwise, our implementation calculates $\widehat{|\mathcal{X}|}$ via a regularized weighted Chebyshev (RWC) estimator suitable for small-sample settings \citep{chien2019regularized, rana2022small}.
We provide further details in Appendix \ref{sec:support-size-details}, including an ablation confirming 
that our headline results are not particularly sensitive to our choice of support size estimator.
Although our priority is entropy estimation, we also evaluate the proposed missing mass estimator in its own right (Appendix \ref{sec:more-missing-mass-results}), finding that it outperforms several alternatives in our setting of interest.
\textsc{SENECA} aims to address the small-sample setting, which is the focus of the empirical results to follow.
For completeness,  we also provide a consistency result characterizing \textsc{SENECA}'s large-sample behavior:

\begin{theorem}\label{thm:weak-consistency}
    Let $X^n \sim P$ be a multiset of $n$ IID samples drawn from a multinomial distribution $P$ over finite support $\mathcal{X}$, with true entropy $\mathbb{H}(P)$. Additionally, let $1 \leq \upsilon \leq f(n)$, for some sub-quadratic function $f(n)=o(n^2)$.
    Then, as $n \rightarrow \infty$, \textsc{SENECA}$(X^n, \upsilon) \xrightarrow{p} \mathbb{H}(P)$.
\end{theorem}

\begin{proof}
Lemmas \ref{lem:sc-open-interval} and \ref{lem:msc-to-zero} establish that $\hat{M}^{SC}_{\upsilon}(X^n) \in [0, 1)$ and $\hat{M}_{\upsilon}^{SC} \rightarrow 0$ as $n \rightarrow \infty$, respectively.
By Lemma \ref{lem:generalized-vu}, any Horvitz-Thompson-style entropy estimator relying on a missing mass estimator with the two aforementioned properties is weakly consistent for distributions with finite entropy variance.
Since distributions over finite support have finite entropy variance, the result follows.
\end{proof}

\section{Assessing entropy estimators via simulation} \label{sec:numerical}

\begin{figure*}[t!]
    \centering
    \includegraphics[width=1.0\textwidth]{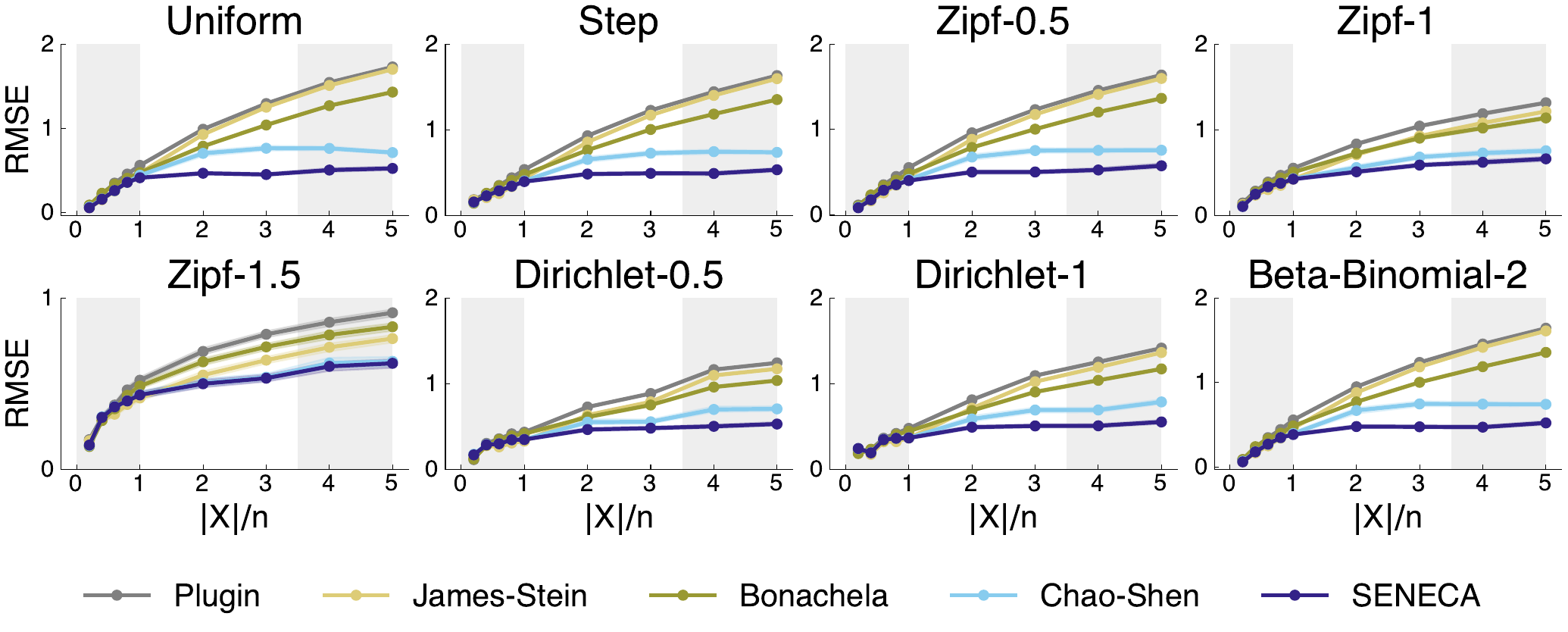}
    \caption{
    \textbf{Entropy estimation in small-sample item capture.} 
    For each of $8$ families of label-prevalence distribution, each item's capture probability is determined uniformly over its label's prevalence probability mass.
    The number of labels $|\mathcal{X}|$ varies from $2$ to $50$ across trials.
    For each trial, $n=10$ items are captured, and the entropy of the label-prevalence distribution is estimated. 
    Results are averaged over $1000$ trials per label count.
    Colored regions about each line plot depict $95\%$ bias-corrected and accelerated confidence intervals, based on $1000$ bootstrap replications.
    In each plot, the leftmost grey region corresponds to the well-sampled regime (i.e., $|\mathcal{X}|= \frac{n}{5}, \frac{2n}{5}, \frac{3n}{5}, \frac{4n}{5}, n$), where items to the right correspond to the under-sampled regime (i.e., $|\mathcal{X}|= 2n, 3n, 4n, 5n$).
    The rightmost grey region corresponds to the support-risky regime, where no consistent estimator for support size exists.
    Full results with all $8$ estimators may be found in Figure \ref{fig:entropy-rmse-n10-full} (Appendix \ref{sec:more-entropy-results}).
    }
    \label{fig:entropy-rmse-n10}
\end{figure*}

Our first batch of experiments assesses \textsc{SENECA} as a small-sample entropy estimator, alongside an array of prior methods found in information theory, ecological statistics, and physics literature.

\subsection{Methods} \label{sec:entropy-experiments-desc}

\paragraph{Numerical experiments using $72$ unique distributions.} 
We validate the performance of our entropy estimator across $9$ support sizes for each of $8$ families of prevalence distribution on a simulated item-capture task. 
For each distribution, we construct a synthetic population of $100$ items divided into a variable number of mutually exclusive labels (i.e., $|\mathcal{X}|=2, 4, 6, 8, 10, 20, 30, 40, 50$ in Figure \ref{fig:entropy-rmse-n10} and Table \ref{tab:entropy-rmse-n10}, $|\mathcal{X}|=4, 8, 12, 16, 20, 40, 60, 80, 100$ in Table \ref{tab:entropy-rmse-n20}).
Overall label-prevalences (i.e., $P = \{\{p_u | u \in \mathcal{X}\}\}$ follow distributions found in prior works. 
In particular, the first six are used repeatedly in related literature \citep{orlitsky2016optimal, hao2020optimal, leeHowMuchUnseen2025} and the following two are additional inclusions from other works \citep{orlitsky2015competitive, painsky2023generalized}: 
a uniform distribution (i.e., $p_u = \frac{1}{|\mathcal{X}|}$); 
a so-called ``step'' distribution, where half the labels have $p_u = \frac{1}{2 |\mathcal{X}|}$ and half have $p_u = \frac{3}{2 |\mathcal{X}|}$; 
three Zipf distributions (i.e., $p_u \propto u^{-\alpha}$), with shape parameters $\alpha = 0.5, 1, \text{ and } 1.5$; 
two distributions generated from Dirichlet priors (Dirichlet-$0.5$ and Dirichlet-$1$);
and a beta-binomial distribution whose parameters are both $2$ (i.e., $p_u = \binom{|\mathcal{X}|}{u} \frac{B(u+\alpha, |\mathcal{X}|-u+\beta)}{B(\alpha, \beta)}$), where $B$ denotes the beta function.
Item-level capture probabilities are uniform within each class.

\paragraph{Estimators and evaluation metrics.}
We compare against six alternative entropy estimators: the plugin estimator (Eq. \ref{eq:plugin_entropy}), the Grassberger estimator \citep{grassberger1988finite, grassberger2003entropy}, the James-Stein estimator \citep{hausser2009entropy}, the  Bonachela estimator \citep{bonachela2008entropy}, the Chao-Shen estimator (Eq. \ref{eq:chao-shen}) \citep{chao2003nonparametric}, the two-FoF Chao-Wang-Jost estimator \citep{chao2013entropy}, and the linear programming-based estimator of \citet{valiant2017estimating}.
The James-Stein and Bonachela estimators are explicitly specialized for small sample sizes, and the Chao-Shen and Chao-Wang-Jost methods are acclaimed for fast convergence to the ground truth \citep{la2025bee}.
Our primary metric for evaluating estimators is the root-mean-squared error (RMSE), which we later decompose into empirical estimates of bias and variance (Appendix \ref{sec:more-entropy-results}).
Because RMSE's sampling distribution can be biased, naive percentile bootstrap intervals may deviate from observed RMSEs.
For this reason, we calculate uncertainty about these evaluation metrics via Bias-Corrected and Accelerated (BCa) bootstrap confidence intervals \citep{efron1987better}.
BCa is not necessarily suitable for our aggregation of aggregate RMSEs across regimes, so we perform $1000$ bootstrap replications and calculate pivot confidence intervals, instead.
In Table \ref{tab:entropy-rmse-n10}, we report the larger of the two sides of each interval, making the corresponding confidence level slightly more conservative than $95\%$.\footnote{Due to the large number of trials, we anticipate these confidence intervals to be quite small \citep{faber1999estimating}.}

\subsection{Results} \label{sec:simulation-results}

\paragraph{\textsc{SENECA} achieves consistently low error across label distributions and sampling regimes.}
For visual clarity, Figure \ref{fig:entropy-rmse-n10} shows entropy estimation results for a subset of the considered baselines: the naive plugin estimator, the small-sample specialist James-Stein and Bonachela estimators, and the classic Chao-Shen estimator.
Full results with all considered estimators are shown in Figure \ref{fig:entropy-rmse-n10-full} (Appendix \ref{sec:more-entropy-results}).
In both figures, \textsc{SENECA} achieves consistently low error in the under-sampled regime.
We further examine these results in Table \ref{tab:entropy-rmse-n10}, averaging errors across domain sizes within the well-sampled ($|\mathcal{X}| \leq n$) and under-sampled ($|\mathcal{X}| > n$) regimes.
The James-Stein estimator, which excels in the well-sampled settings, is less reliable in the under-sampled regime.
\textsc{SENECA} performs well in both regimes, attaining the lowest average error in all under-sampled settings and lowest or second-lowest error in all but one well-sampled settings.
Practically, the regime is rarely known in advance, underscoring the importance of balanced performance across regimes.

\begin{table*}[t!]
    \caption{
        \textbf{Regime-averaged error for entropy estimation in small-sample item capture.}
        Using the data from Figure \ref{fig:entropy-rmse-n10}, we calculate the root-mean-squared error (RMSE $\downarrow$) for each estimator and average results within the well-sampled (Well, i.e., $|\mathcal{X}| \leq n$) and under-sampled (Under, i.e., $|\mathcal{X}| > n$) regimes (Reg.).
        The estimators are Plugin (ML), Grassberger (G), James-Stein (JS), Bonachela (B), Chao-Shen (CS), Chao-Wang-Jost (CWJ), Valiant-Valiant (VV), and \textsc{SENECA} (highlighted in blue).
        The distributions are Uniform, Step, Zipf-0.5, Zipf-1, Zipf-1.5, Dirichlet-0.5 (Diri-0.5), Dirichlet-1 (Diri-1), and Beta-Binomial-2 (BB-2).
        The lowest and second-lowest values in each row, prior to rounding, are bold and underlined, respectively.
        Bootstrapped confidence intervals (grey text) are slightly more conservative than $95\%$, as described in Section \ref{sec:entropy-experiments-desc}.
        In parentheses, we show the relative difference between \textsc{SENECA}'s averaged RMSE and the lowest value in each row.
    }
    \label{tab:entropy-rmse-n10}
    \centering
    \small
    \setlength{\tabcolsep}{4.25pt}
    \begin{tabular}{lcccccccc>{\columncolor{lightindigocolor!10}}c}
\toprule
Distr. & Reg. & ML & G & JS & B & CS & CWJ & VV & SENECA \\
\midrule
\multirow{2}{*}{Uniform} & Well & 0.33 \tiny{\textcolor{gray}{$\pm$.00}} & 0.25 \tiny{\textcolor{gray}{$\pm$.00}} & 0.26 \tiny{\textcolor{gray}{$\pm$.00}} & 0.30 \tiny{\textcolor{gray}{$\pm$.00}} & 0.26 \tiny{\textcolor{gray}{$\pm$.01}} & \textbf{0.25} \tiny{\textcolor{gray}{$\pm$.01}} & 0.36 \tiny{\textcolor{gray}{$\pm$.01}} & \underline{0.25} \tiny{\textcolor{gray}{$\pm$.01}} (0.7\%) \\
 & Under & 1.39 \tiny{\textcolor{gray}{$\pm$.00}} & 0.66 \tiny{\textcolor{gray}{$\pm$.01}} & 1.35 \tiny{\textcolor{gray}{$\pm$.00}} & 1.13 \tiny{\textcolor{gray}{$\pm$.01}} & 0.73 \tiny{\textcolor{gray}{$\pm$.01}} & \underline{0.56} \tiny{\textcolor{gray}{$\pm$.01}} & 0.62 \tiny{\textcolor{gray}{$\pm$.01}} & \textbf{0.49} \tiny{\textcolor{gray}{$\pm$.01}} (0.0\%) \\
\midrule
\multirow{2}{*}{Step} & Well & 0.35 \tiny{\textcolor{gray}{$\pm$.01}} & 0.29 \tiny{\textcolor{gray}{$\pm$.01}} & \underline{0.28} \tiny{\textcolor{gray}{$\pm$.01}} & 0.33 \tiny{\textcolor{gray}{$\pm$.01}} & 0.29 \tiny{\textcolor{gray}{$\pm$.01}} & 0.29 \tiny{\textcolor{gray}{$\pm$.01}} & 0.38 \tiny{\textcolor{gray}{$\pm$.01}} & \textbf{0.28} \tiny{\textcolor{gray}{$\pm$.01}} (0.0\%) \\
 & Under & 1.31 \tiny{\textcolor{gray}{$\pm$.01}} & 0.64 \tiny{\textcolor{gray}{$\pm$.01}} & 1.26 \tiny{\textcolor{gray}{$\pm$.00}} & 1.08 \tiny{\textcolor{gray}{$\pm$.01}} & 0.72 \tiny{\textcolor{gray}{$\pm$.02}} & \underline{0.58} \tiny{\textcolor{gray}{$\pm$.01}} & 0.64 \tiny{\textcolor{gray}{$\pm$.01}} & \textbf{0.50} \tiny{\textcolor{gray}{$\pm$.01}} (0.0\%) \\
\midrule
\multirow{2}{*}{Zipf-0.5} & Well & 0.34 \tiny{\textcolor{gray}{$\pm$.01}} & 0.26 \tiny{\textcolor{gray}{$\pm$.01}} & 0.26 \tiny{\textcolor{gray}{$\pm$.01}} & 0.31 \tiny{\textcolor{gray}{$\pm$.01}} & 0.26 \tiny{\textcolor{gray}{$\pm$.01}} & \underline{0.26} \tiny{\textcolor{gray}{$\pm$.01}} & 0.36 \tiny{\textcolor{gray}{$\pm$.01}} & \textbf{0.26} \tiny{\textcolor{gray}{$\pm$.01}} (0.0\%) \\
 & Under & 1.32 \tiny{\textcolor{gray}{$\pm$.01}} & 0.66 \tiny{\textcolor{gray}{$\pm$.01}} & 1.27 \tiny{\textcolor{gray}{$\pm$.00}} & 1.09 \tiny{\textcolor{gray}{$\pm$.01}} & 0.74 \tiny{\textcolor{gray}{$\pm$.02}} & \underline{0.59} \tiny{\textcolor{gray}{$\pm$.01}} & 0.66 \tiny{\textcolor{gray}{$\pm$.01}} & \textbf{0.53} \tiny{\textcolor{gray}{$\pm$.01}} (0.0\%) \\
\midrule
\multirow{2}{*}{Zipf-1} & Well & 0.37 \tiny{\textcolor{gray}{$\pm$.01}} & 0.32 \tiny{\textcolor{gray}{$\pm$.01}} & \textbf{0.29} \tiny{\textcolor{gray}{$\pm$.01}} & 0.34 \tiny{\textcolor{gray}{$\pm$.01}} & 0.30 \tiny{\textcolor{gray}{$\pm$.01}} & 0.32 \tiny{\textcolor{gray}{$\pm$.01}} & 0.38 \tiny{\textcolor{gray}{$\pm$.01}} & \underline{0.30} \tiny{\textcolor{gray}{$\pm$.01}} (3.1\%) \\
 & Under & 1.10 \tiny{\textcolor{gray}{$\pm$.01}} & \underline{0.62} \tiny{\textcolor{gray}{$\pm$.01}} & 0.98 \tiny{\textcolor{gray}{$\pm$.01}} & 0.95 \tiny{\textcolor{gray}{$\pm$.01}} & 0.68 \tiny{\textcolor{gray}{$\pm$.01}} & 0.63 \tiny{\textcolor{gray}{$\pm$.01}} & 0.69 \tiny{\textcolor{gray}{$\pm$.01}} & \textbf{0.60} \tiny{\textcolor{gray}{$\pm$.01}} (0.0\%) \\
\midrule
\multirow{2}{*}{Zipf-1.5} & Well & 0.37 \tiny{\textcolor{gray}{$\pm$.01}} & 0.35 \tiny{\textcolor{gray}{$\pm$.01}} & \textbf{0.32} \tiny{\textcolor{gray}{$\pm$.01}} & 0.34 \tiny{\textcolor{gray}{$\pm$.01}} & 0.33 \tiny{\textcolor{gray}{$\pm$.01}} & 0.35 \tiny{\textcolor{gray}{$\pm$.01}} & 0.37 \tiny{\textcolor{gray}{$\pm$.01}} & \underline{0.33} \tiny{\textcolor{gray}{$\pm$.01}} (4.7\%) \\
 & Under & 0.81 \tiny{\textcolor{gray}{$\pm$.01}} & \underline{0.56} \tiny{\textcolor{gray}{$\pm$.01}} & 0.67 \tiny{\textcolor{gray}{$\pm$.01}} & 0.74 \tiny{\textcolor{gray}{$\pm$.01}} & 0.57 \tiny{\textcolor{gray}{$\pm$.01}} & 0.60 \tiny{\textcolor{gray}{$\pm$.01}} & 0.66 \tiny{\textcolor{gray}{$\pm$.01}} & \textbf{0.56} \tiny{\textcolor{gray}{$\pm$.01}} (0.0\%) \\
\midrule
\multirow{2}{*}{Diri-0.5} & Well & 0.33 \tiny{\textcolor{gray}{$\pm$.01}} & 0.30 \tiny{\textcolor{gray}{$\pm$.01}} & \textbf{0.26} \tiny{\textcolor{gray}{$\pm$.01}} & 0.31 \tiny{\textcolor{gray}{$\pm$.01}} & 0.29 \tiny{\textcolor{gray}{$\pm$.01}} & \underline{0.29} \tiny{\textcolor{gray}{$\pm$.01}} & 0.35 \tiny{\textcolor{gray}{$\pm$.01}} & 0.29 \tiny{\textcolor{gray}{$\pm$.01}} (10.8\%) \\
 & Under & 1.01 \tiny{\textcolor{gray}{$\pm$.01}} & \underline{0.52} \tiny{\textcolor{gray}{$\pm$.01}} & 0.92 \tiny{\textcolor{gray}{$\pm$.01}} & 0.84 \tiny{\textcolor{gray}{$\pm$.01}} & 0.63 \tiny{\textcolor{gray}{$\pm$.02}} & 0.55 \tiny{\textcolor{gray}{$\pm$.01}} & 0.62 \tiny{\textcolor{gray}{$\pm$.01}} & \textbf{0.50} \tiny{\textcolor{gray}{$\pm$.01}} (0.0\%) \\
\midrule
\multirow{2}{*}{Diri-1} & Well & 0.34 \tiny{\textcolor{gray}{$\pm$.01}} & 0.31 \tiny{\textcolor{gray}{$\pm$.01}} & \textbf{0.29} \tiny{\textcolor{gray}{$\pm$.01}} & 0.32 \tiny{\textcolor{gray}{$\pm$.01}} & 0.31 \tiny{\textcolor{gray}{$\pm$.01}} & 0.31 \tiny{\textcolor{gray}{$\pm$.01}} & 0.37 \tiny{\textcolor{gray}{$\pm$.01}} & \underline{0.30} \tiny{\textcolor{gray}{$\pm$.01}} (6.4\%) \\
 & Under & 1.15 \tiny{\textcolor{gray}{$\pm$.01}} & \underline{0.58} \tiny{\textcolor{gray}{$\pm$.01}} & 1.07 \tiny{\textcolor{gray}{$\pm$.01}} & 0.95 \tiny{\textcolor{gray}{$\pm$.01}} & 0.69 \tiny{\textcolor{gray}{$\pm$.02}} & 0.59 \tiny{\textcolor{gray}{$\pm$.01}} & 0.65 \tiny{\textcolor{gray}{$\pm$.01}} & \textbf{0.52} \tiny{\textcolor{gray}{$\pm$.01}} (0.0\%) \\
\midrule
\multirow{2}{*}{BB-2} & Well & 0.33 \tiny{\textcolor{gray}{$\pm$.01}} & 0.26 \tiny{\textcolor{gray}{$\pm$.00}} & 0.26 \tiny{\textcolor{gray}{$\pm$.01}} & 0.31 \tiny{\textcolor{gray}{$\pm$.01}} & 0.25 \tiny{\textcolor{gray}{$\pm$.01}} & \textbf{0.25} \tiny{\textcolor{gray}{$\pm$.01}} & 0.36 \tiny{\textcolor{gray}{$\pm$.01}} & \underline{0.25} \tiny{\textcolor{gray}{$\pm$.01}} (0.3\%) \\
 & Under & 1.32 \tiny{\textcolor{gray}{$\pm$.00}} & 0.64 \tiny{\textcolor{gray}{$\pm$.01}} & 1.27 \tiny{\textcolor{gray}{$\pm$.00}} & 1.08 \tiny{\textcolor{gray}{$\pm$.01}} & 0.73 \tiny{\textcolor{gray}{$\pm$.01}} & \underline{0.57} \tiny{\textcolor{gray}{$\pm$.01}} & 0.63 \tiny{\textcolor{gray}{$\pm$.01}} & \textbf{0.49} \tiny{\textcolor{gray}{$\pm$.01}} (0.0\%) \\
\bottomrule
\end{tabular}
\end{table*}

\paragraph{\textsc{SENECA} excels even where support estimation is risky.}
A result of \citet{wu2019chebyshev} indicates that consistent estimation of $|\mathcal{X}|$ is only possible with at least $\frac{|\mathcal{X}|}{\log |\mathcal{X}|}$ samples.
We therefore denote the \textit{support-risky} regime as that for which $|\mathcal{X}| > \gamma$, where $\gamma$ is the largest integer for which the sample size $n > \frac{\gamma}{\log \gamma}$ (e.g., $\gamma=35$ for $n=10$).
We illustrate this support-risky regime in the rightmost grey regions of each subplot in Figure \ref{fig:entropy-rmse-n10}.
Despite reliance on a support size estimate, our method maintains its top-end performance, with low RMSE well into the support-risky regime.

\paragraph{\textsc{SENECA} balances bias and variance to achieve low error.}
The well-known bias-variance tradeoff \citep{hastie2009elements} is no less applicable to entropy estimation: minimizing error warrants management of both components.
An empirical bias-variance decomposition of the results in Figure \ref{fig:entropy-rmse-n10-full} is provided in Figure \ref{fig:entropy-bias-variance-n10} (Appendix \ref{sec:more-entropy-results}). 
In general, the estimators with the lowest bias in the under-sampled regime tend to exhibit the highest variance (e.g., Chao-Shen), and vice-versa (e.g., Plugin) in the under-sampled regime. 
On the other hand, \textsc{SENECA} maintains comparably low or moderate bias and variance across settings, resulting in consistently low RMSE overall.

\section{Applications} \label{sec:applications}

We extend our comparison of entropy estimators to biodiversity quantification, where the entropy of species-prevalence distributions is a standard metric \citep{sherwin2019introduction}.
Here, ``items'' in our problem statement correspond to organisms, and ``labels'' are their species.
We then conclude our experiments by applying \textsc{SENECA} to incorrect LLM response detection.
Variability-based uncertainty quantification techniques operate under the assumption that LLMs may be more uncertain about a query when they express greater semantic variability under repeated sampling.

\subsection{Methods} \label{sec:applications-methods}

\subsubsection{Biodiversity estimation} \label{sec:biodiversity-experiments-desc}

\paragraph{Datasets.}
\citet{janzen1973sweep2} meticulously reports insect species frequencies observed in sweep samples across 25 tropical sites under various times of day (i.e., day, night) and conditions (i.e., wet, dry); the dataset is described in further detail in \citet{janzen1973sweep1}.
We treat each of the $58$ collections of true bugs (order Hemiptera) as a complete population, similar to \citet{chao2003nonparametric}.
The corresponding species-prevalence distributions vary substantially in population size ($|A| \leq 1407$) and support size ($|\mathcal{X}| \leq 67$).
The distributions' coefficients of variation range from $0$ to approximately $3.5$, indicating that the considered populations consist of both heavy-tailed and light-tailed distributions.
We draw samples of various sizes ($n=10, 20, 30, 40, 50$) and estimate the entropy of the complete species distributions accordingly, employing the following methods.

\paragraph{Estimators and evaluation metrics.}
We use the same entropy estimators itemized in Section \ref{sec:entropy-experiments-desc}, comparing the entropy estimated from samples to the entropy of the ``complete'' population.
To select an estimator that performs best overall, we seek reliability across numerous distributions and sample sizes, where RMSE magnitudes vary substantially.
Therefore, we treat the RMSE-based ranking of estimators for each collection and sample size as a ``ballot'' and report the Borda count, which allocates points to candidates (estimators) according to the number of other candidates they outrank in each ballot.
Consequently, the estimator with the highest Borda count consistently achieves lower RMSE than others, across the populations and sample sizes considered.
The notable source of uncertainty in this aggregated count arises from the underlying species prevalence collections, so we calculate $95\%$ pivot confidence intervals by bootstrapping over collections.

\subsection{LLM predictive uncertainty quantification} \label{sec:llm-experiments-desc}

\paragraph{Models.} 
We select five LLMs, ranging from $3.8$ to $32$ billion total parameters: the 3.8B-parameter Phi-4-mini-instruct \citep{abouelenin2025phi},
Mistral-7B-Instruct-v0.3,
the 14B-parameter Phi-4 \citep{abdin2024phi},
Granite-3.0-8B-Instruct \citep{granite2024granite},
and the 32B-parameter Granite-4.0-H-Small \citep{granite2025}.
The first four are dense models, and the latter is a ``hybrid'' mixture-of-experts model with 9B active parameters.
Additional details are found in Appendix \ref{sec:text-generation-details}.

\paragraph{Datasets.}
Our experiments use six permissively licensed datasets concerning biomedical subjects, mathematics word problems, common misconceptions, and general question-answering, namely 
BioASQ Task 11b \citep{krithara2023bioasq},
SQuAD 2.0 \citep{rajpurkar-etal-2016-squad},
SVAMP \citep{patel-etal-2021-nlp},
HotpotQA \citep{yang2018hotpotqa},
TruthfulQA \citep{lin-etal-2022-truthfulqa}, and 
SimpleQA-Verified \citep{haas2025simpleqa}, using the first $1000$ exemplars from each dataset.
TruthfulQA has fewer than $1000$ total items, so we use the full dataset, with the exception of the several prompts for which the reference answer is listed as ``I have no comment.''
Following \citet{Farquhar2024}, we make the QA task more challenging by omitting context passages in those datasets that offer them.

\paragraph{Baselines.}
We consider several black and grey-box methods (i.e., not using and using response log-probabilities, respectively) for LLM UQ.
The established black-box methods are discrete semantic entropy (DSE, \citep{Farquhar2024}), $U_{EigV}$ \citep{lin2023generating}, kernel language entropy (KLE, \citep{nikitin2024kernel}), hybrid entropy (HE, \citet{mccabe2026estimating}), and \textsc{SENECA}.
The grey-box methods are predictive entropy (PE, \citep{kadavath2022language}), semantic entropy (SE, \citep{kuhn2023semantic}), G-NLL \citep{aichberger2026rethinking}, and surprise (which is similar to PE, \citep{ismayilzada-etal-2025-creative}).
PE, DSE, SE, KLE, and the SENECA methods are entropy estimators of various forms.
DSE is equivalent to the established Plugin method, carried out over semantic equivalence classes.
SE is similar to DSE, but class prevalences are based on sequence log-probabilities, rather than empirical frequencies.
$U_{EigV}$ is a support size estimator that incorporates semantic information from the responses.
For this reason, we also consider a simple \textsc{SENECA} variant (namely, \textsc{SENECA-U}) that uses $U_{EigV}$ in place of RWC.
Finally, we include all entropy estimators discussed in the prior sections, which are also black-box methods.
Baselines are detailed further in Appendix \ref{sec:llm-uq-details}.

\paragraph{Semantic clustering.}
SE, DSE, and \textsc{SENECA} calculate entropy over semantic equivalence classes, which are obtained via semantic clustering of responses.
For all three, we follow the bidirectional entailment classification method of \citet{kuhn2023semantic} (Appendix \ref{sec:semantic-clustering-details}).
KLE and $U_{EigV}$ do not explicitly perform semantic clustering, but they do rely on natural language inference (NLI)-based semantic graphs.
For consistency, we use the same NLI model for all five above methods.

\paragraph{LLM-as-judge.}
To assess LLM correctness in response to a query, we draw a low-temperature ($\tau=0.1$) ``best guess'' sample, as seen in prior works \citep{Farquhar2024, nikitin2024kernel, nguyen-etal-2025-beyond}, and prompt an LLM judge to gauge its consistency with a reference answer \citep{lin2023generating}; further details are provided in Appendix \ref{sec:llm-judge-details}.
In Appendix \ref{sec:more-llm-results}, we also consider a binary (i.e., yes/no) LLM judge of correctness, with qualitatively indistinguishable results.

\paragraph{Metrics.}
We measure success in classifying incorrect responses by the area under the receiver operating characteristic curve (AUROC), as seen in prior works \citep{kuhn2023semantic, lin2023generating, Farquhar2024, aichberger2026rethinking}.
Empirical AUROC is a parameter estimate - in this case, that of the probability that an UQ method assigns a higher uncertainty to a randomly-selected incorrect response than a random correct one.
As such, we represent the uncertainty of each AUROC point estimate by 95\% confidence intervals (CIs), calculated using the DeLong method \citep{delong1988comparing, sun2014fast}.
Because each model-dataset pair represents a distinct setting, we aggregate the AUROC estimates and CIs for methods (i.e., for Figure \ref{fig:applications-summary-uq-consistency}) using DerSimonian-Laird random effects meta-analysis \citep{dersimonian1986meta} with Hartung-Knapp-Sidik-Jonkman error adjustment \citep{hartung2001tests, sidik2002simple, inthout2014hartung}.

\subsection{Results} \label{sec:applications-results}

\begin{wrapfigure}[35]{r}{0.49\textwidth}
    \centering
    \includegraphics[width=0.49\textwidth]{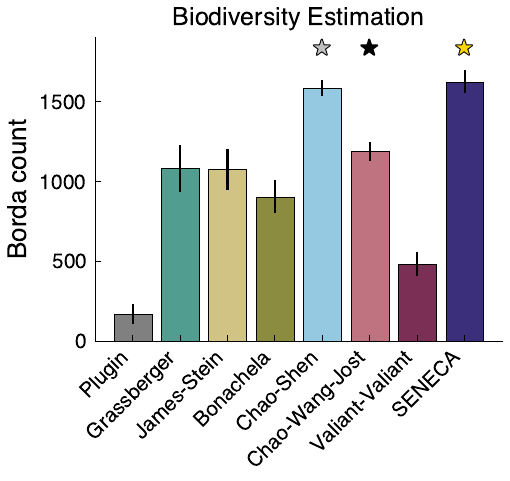}
    \vspace{-0.35cm}
    \caption{
    \textbf{Summary of entropy methods for biodiversity estimation across $58$ tropical insect collections.}
    Treating each collection of true bugs reported by \citet{janzen1973sweep2} as a complete population, we draw samples ($n=10, 20, 30, 40, 50$) and estimate the entropy of the population's species-prevalence distribution.
    RMSE results (Figure \ref{fig:biodiversity-rmse}) are averaged over $1000$ trials per population and sample size.
    Results are summarized by Borda count.
    Error bars reflect $95\%$ pivot confidence intervals, obtained via $1000$ bootstrap replicates over the collections.
    Gold, silver, and black stars indicate the methods with highest, second-highest, and third-highest aggregated parameter estimates, respectively.
    }
    \label{fig:applications-summary-biodiversity}
\end{wrapfigure}

\paragraph{\textsc{SENECA} consistently outranks other methods for biodiversity quantification across $58$ ecological collections.}
We apply the methods described herein to estimate tropical true bug biodiversity, in terms of the entropy their corresponding species-prevalence distributions.
We summarize the results in Figure \ref{fig:applications-summary-biodiversity}, where our method most consistently achieves low RMSE, as judged by Borda count.
The next highest-scoring estimators are Chao-Shen and Chao-Wang-Jost, consistent with the under-sampling results of our simulation study, since many of these species-prevalence distributions are heavy-tailed.
Chao-Shen and SENECA both take a Horvitz-Thompson-style approach, but SENECA's self-consistent missing mass estimator leads to a Borda count several dozen larger than the former's.
Full results for each population are shown in Figure \ref{fig:biodiversity-rmse} (Appendix \ref{sec:more-entropy-results}).

\begin{wrapfigure}[28]{r}{0.49\textwidth}
    \centering
    \includegraphics[width=0.49\textwidth]{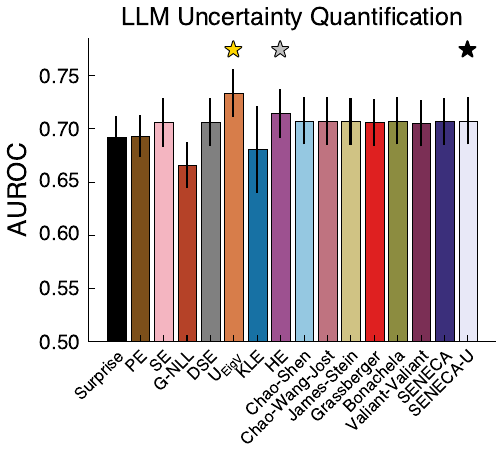}
    \vspace{-0.575cm}
    \caption{
    \textbf{Summary of LLM UQ results using consistency-based LLM judge.}
    AUROC values and parameter uncertainties are aggregated across models and datasets, as described in Section \ref{sec:llm-experiments-desc}, using the consistency-based LLM-as-judge (Appendix \ref{sec:llm-judge-details}).
    Gold, silver, and black stars indicate the methods with highest, second-highest, and third-highest aggregated parameter estimates, respectively.
    }
    \label{fig:applications-summary-uq-consistency}
\end{wrapfigure}

\paragraph{On the LLM uncertainty quantification task, most alternative entropy estimators from Section~\ref{sec:entropy-experiments-desc}, including \textsc{SENECA}, outrank SE and DSE,}
as measured by aggregated AUROC point estimate (Figure \ref{fig:applications-summary-uq-consistency}).
Full results are shown in Figure \ref{fig:auroc}.
SE and DSE are canonical notions of the aleatoric component of semantic uncertainty \citep{park2026efficient}; in this setting, these alternative estimators approximate the same estimand.
The aggregated AUROC of the Chao-Shen, Chao-Wang-Jost, James-Stein, Bonachela, and \textsc{SENECA} estimators (differing mutually by at most $\approx 0.00055$) all exceed those of SE and DSE.
We conclude that aleatoric uncertainty estimation in LLMs is enhanced by these alternative entropy estimation strategies.
Total predictive uncertainty also includes an epistemic term, which recent works have elicited via methods that include model weight perturbation \citep{liu2026enhancing}, diverse sampling \citep{park2026efficient}, or semantic similarity of responses from an assortment of other models \citep{hamidieh2026complementing}.
Generally, these approaches are computationally involved or call for white-box model access.
Nonetheless, \textsc{SENECA} is complementary to such techniques by improving the estimation of the aleatoric term.

\paragraph{\textsc{SENECA-U} is competitive with purpose-built LLM hallucination detection methods,} such as $U_{EigV}$, KLE, G-NLL, and HE.
On the incorrectness detection task, \textsc{SENECA-U} attains the third-highest aggregated AUROC point estimate (Figure \ref{fig:applications-summary-uq-consistency}), outranked only by support size estimator $U_{EigV}$ and ``hybrid'' entropy estimator HE.
Among entropy-based methods, only HE outranks \textsc{SENECA-U}, highlighting the flexibility and broad applicability of the methodology presented herein.

\section{Discussion} \label{sec:discussion}

\subsection{Conclusion} \label{sec:conclusion}
We present \textsc{SENECA}, an entropy estimation scheme that excels in the small-sample setting, especially in the presence of under-sampling.
In numerical experiments, we demonstrate consistently low RMSE across $72$ unique label-prevalence distributions, compared to seven baselines (Section \ref{sec:simulation-results}).
We extend this evaluation to $58$ ecological collections transcribed from \citet{janzen1973sweep2}, where \textsc{SENECA} also excels.
In application to the classification of incorrect LLM responses, the point estimate for \textsc{SENECA-U}'s overall AUROC is third-highest among several purpose-built baselines (Section \ref{sec:applications-results}).
\textsc{SENECA}'s underlying missing mass estimator may be of independent interest, as well (Appendix \ref{sec:more-missing-mass-results}); it estimates the probability that a subsequently sampled item deserves a label that has not yet been observed.

Finally, we observe that any advantage \textsc{SENECA} demonstrates in Section~\ref{sec:applications} appears more modest than that of Section~\ref{sec:numerical}.
We attribute this to two factors.
First, \textsc{SENECA} targets a particular setting: the small-sample and under-sampled regime, but many of the species distributions in the biodiversity study are well-sampled ($\approx 29\%$ for $n=10$).
With support sizes as low as $2$, the degenerate case of $|X_\neq^n|=1$ is possible, where \textsc{SENECA} is equivalent to the plugin estimator (Corollary~\ref{cor:fixed-point-uniqueness-all-labels-same}).
Second, the LLM UQ study is a bipartite ranking task that does not strictly assess entropy estimation ability (note the highest-ranking method is a support size estimator).
In general, there exists no discrete entropy estimator that is optimal in every setting \citep{la2025bee}.
Nonetheless, our results indicate that \textsc{SENECA} has broadly favorable performance and may be utile to many diverse technical disciplines.

\subsection{Limitations and future work} \label{sec:limitations}
In this work, our priority is the small-sample setting, where the effects of under-sampling and estimator variance are especially pertinent.
Although the method proposed herein exhibits low error in our experiments, it is not unbiased. 
Indeed, unbiased discrete entropy estimators do not exist in general \citep{paninski2003estimation}, and estimators classically may trade off bias and variance \citep{hastie2009elements}.
As a consequence, there are scenarios for which our estimator may not be best-suited. For instance, in the unusual event that the regime is known to be well-sampled before estimation, our Table \ref{tab:entropy-rmse-n10} indicates that the James-Stein estimator may be preferable.
Overall performance may also differ in the large-sample setting, since several entropy estimators are consistent (e.g., Plugin \citep{antos2001convergence, arora2022estimating} and Chao-Shen \citep{vu2007coverage}).

\subsection{Reproducibility statement} \label{sec:reproducibility}
Experimental settings, datasets, and baselines are outlined in Sections \ref{sec:entropy-experiments-desc} and \ref{sec:applications-methods}, with additional implementation details in Appendix \ref{sec:implementation-details}.
Computational resources for LLM experiments are described in Appendix \ref{sec:text-generation-details}.
Use of commercial serverless endpoints with multiple workers make exact compute consumption difficult to track precisely, but we estimate that the LLM experiments took between $30$ and $35$ GPU-hours.
Semantic clustering was performed on a CPU (24 GB RAM), running continuously for roughly $3$ days.
All remaining experiments and calculations occurred on the same CPU, requiring a few hours of total execution time.
The two Phi models are released under the MIT license, and Mistral, Granite, and DeBERTa models under Apache 2.0.
The BioASQ dataset is licensed under CC BY 2.5, SQuAD 2.0 and HotpotQA under CC BY SA 4.0, SVAMP and SimpleQA-Verified under MIT, and TruthfulQA under Apache $2.0$.
Section \ref{sec:biodiversity-experiments-desc} describes usage of species-prevalence records, similar to \citep{chao2003nonparametric, chao2013entropy}.
We do not re-release these records, but they are visible in their corresponding publication \citep{janzen1973sweep2}.
We will release a code repository upon publication.

\subsection{Acknowledgements}
LHM thanks Naoki Masuda and Rimon Melamed for helpful comments in the drafting of this work.
The authors report no funding to disclose.

\clearpage
\bibliography{references}

\begin{thebibliography}{83}
\providecommand{\natexlab}[1]{#1}
\providecommand{\url}[1]{\texttt{#1}}
\expandafter\ifx\csname urlstyle\endcsname\relax
  \providecommand{\doi}[1]{doi: #1}\else
  \providecommand{\doi}{doi: \begingroup \urlstyle{rm}\Url}\fi

\bibitem[Abdin et~al.(2024)Abdin, Aneja, Behl, Bubeck, Eldan, Gunasekar, Harrison, Hewett, Javaheripi, Kauffmann, et~al.]{abdin2024phi}
Marah Abdin, Jyoti Aneja, Harkirat Behl, S{\'e}bastien Bubeck, Ronen Eldan, Suriya Gunasekar, Michael Harrison, Russell~J Hewett, Mojan Javaheripi, Piero Kauffmann, et~al.
\newblock {Phi-4 Technical Report}.
\newblock \emph{arXiv preprint arXiv:2412.08905}, 2024.

\bibitem[Abouelenin et~al.(2025)Abouelenin, Ashfaq, Atkinson, Awadalla, Bach, Bao, Benhaim, Cai, Chaudhary, Chen, et~al.]{abouelenin2025phi}
Abdelrahman Abouelenin, Atabak Ashfaq, Adam Atkinson, Hany Awadalla, Nguyen Bach, Jianmin Bao, Alon Benhaim, Martin Cai, Vishrav Chaudhary, Congcong Chen, et~al.
\newblock {Phi-4-Mini Technical Report: Compact yet Powerful Multimodal Language Models via Mixture-of-LoRAs}.
\newblock \emph{arXiv preprint arXiv:2503.01743}, 2025.

\bibitem[Aichberger et~al.(2026)Aichberger, Schweighofer, and Hochreiter]{aichberger2026rethinking}
Lukas Aichberger, Kajetan Schweighofer, and Sepp Hochreiter.
\newblock {Rethinking Uncertainty Estimation in {LLM}s: A Principled Single-Sequence Measure}.
\newblock In \emph{The Fourteenth International Conference on Learning Representations}, 2026.
\newblock URL \url{https://openreview.net/forum?id=wdhruVcRx1}.

\bibitem[Antos \& Kontoyiannis(2001)Antos and Kontoyiannis]{antos2001convergence}
Andr{\'a}s Antos and Ioannis Kontoyiannis.
\newblock Convergence properties of functional estimates for discrete distributions.
\newblock \emph{Random Structures \& Algorithms}, 19\penalty0 (3-4):\penalty0 163--193, 2001.

\bibitem[Arora et~al.(2022)Arora, Meister, and Cotterell]{arora2022estimating}
Aryaman Arora, Clara Meister, and Ryan Cotterell.
\newblock {Estimating the Entropy of Linguistic Distributions}.
\newblock In \emph{Proceedings of the 60th Annual Meeting of the Association for Computational Linguistics (Volume 2: Short Papers)}, pp.\  175--195, 2022.

\bibitem[Balasundaram \& Thangaraj(2025)Balasundaram and Thangaraj]{balasundaram2025estimating}
Haricharan Balasundaram and Andrew Thangaraj.
\newblock {Estimating Error in Natural Distribution Estimation}.
\newblock In \emph{2025 61st Allerton Conference on Communication, Control, and Computing Proceedings}. Allerton Conference on Communication, Control, and Computing, 2025.

\bibitem[Basharin(1959)]{basharin1959statistical}
Georgij~P Basharin.
\newblock {On a Statistical Estimate for the Entropy of a Sequence of Independent Random Variables}.
\newblock \emph{Theory of Probability \& Its Applications}, 4\penalty0 (3):\penalty0 333--336, 1959.

\bibitem[Bonachela et~al.(2008)Bonachela, Hinrichsen, and Munoz]{bonachela2008entropy}
Juan~A Bonachela, Haye Hinrichsen, and Miguel~A Munoz.
\newblock Entropy estimates of small data sets.
\newblock \emph{Journal of Physics A: Mathematical and Theoretical}, 41\penalty0 (20):\penalty0 202001, 2008.

\bibitem[Burden \& Faires(2011)Burden and Faires]{burden2011numerical}
Richard~L Burden and J~Douglas Faires.
\newblock {Accelerating Convergence}.
\newblock In \emph{{Numerical Analysis}}, pp.\ ~89. Brooks/Cole, Cengage Learning, 9th edition, 2011.
\newblock ISBN 978-0538733519.

\bibitem[B{\"u}th et~al.(2025)B{\"u}th, Acharya, and Zanin]{buth2025infomeasure}
Carlson~Moses B{\"u}th, Kishor Acharya, and Massimiliano Zanin.
\newblock {infomeasure: a comprehensive Python package for information theory measures and estimators}.
\newblock \emph{Scientific Reports}, 15\penalty0 (1):\penalty0 29323, 2025.

\bibitem[Chao(1984)]{chao1984nonparametric}
Anne Chao.
\newblock {Nonparametric Estimation of the Number of Classes in a Population}.
\newblock \emph{Scandinavian Journal of Statistics}, pp.\  265--270, 1984.

\bibitem[Chao(2005)]{chao2005speciesrichness}
Anne Chao.
\newblock Species estimation and applications.
\newblock In Samuel Kotz, Campbell~B Read, Brani Vidakovic, and Norman~L. Johnson (eds.), \emph{Encyclopedia of Statistical Sciences}, volume~12, pp.\  7907--7916. Wiley, New York, 2 edition, 2005.

\bibitem[Chao \& Jost(2012)Chao and Jost]{chao2012diversity}
Anne Chao and Lou Jost.
\newblock Diversity measures.
\newblock \emph{Encyclopedia of Theoretical Ecology}, pp.\  203--207, 2012.

\bibitem[Chao \& Shen(2003)Chao and Shen]{chao2003nonparametric}
Anne Chao and Tsung-Jen Shen.
\newblock {Nonparametric estimation of Shannon’s index of diversity when there are unseen species in sample}.
\newblock \emph{{Environmental and Ecological Statistics}}, 10:\penalty0 429--443, 2003.

\bibitem[Chao et~al.(2013)Chao, Wang, and Jost]{chao2013entropy}
Anne Chao, Yun~Tao Wang, and Lou Jost.
\newblock {Entropy and the species accumulation curve: a novel entropy estimator via discovery rates of new species}.
\newblock \emph{Methods in Ecology and Evolution}, 4\penalty0 (11):\penalty0 1091--1100, 2013.

\bibitem[Chien \& Milenkovic(2019)Chien and Milenkovic]{chien2019regularized}
I~Chien and Olgica Milenkovic.
\newblock {Regularized Weighted Chebyshev Approximations for Support Estimation}.
\newblock \emph{arXiv preprint arXiv:1901.07506}, 2019.

\bibitem[DeLong et~al.(1988)DeLong, DeLong, and Clarke-Pearson]{delong1988comparing}
Elizabeth~R DeLong, David~M DeLong, and Daniel~L Clarke-Pearson.
\newblock {Comparing the Areas under Two or More Correlated Receiver Operating Characteristic Curves: A Nonparametric Approach}.
\newblock \emph{{Biometrics}}, pp.\  837--845, 1988.

\bibitem[Deng et~al.(2024)Deng, Umbach, and Neufeld]{deng2024nonparametric}
Yongcui Deng, Alexander~K Umbach, and Josh~D Neufeld.
\newblock {Nonparametric richness estimators Chao1 and ACE must not be used with amplicon sequence variant data}.
\newblock \emph{The ISME Journal}, 18\penalty0 (1):\penalty0 wrae106, 2024.

\bibitem[DerSimonian \& Laird(1986)DerSimonian and Laird]{dersimonian1986meta}
Rebecca DerSimonian and Nan Laird.
\newblock Meta-analysis in clinical trials.
\newblock \emph{{Controlled Clinical Trials}}, 7\penalty0 (3):\penalty0 177--188, 1986.

\bibitem[Efron(1987)]{efron1987better}
Bradley Efron.
\newblock {Beiter Bootstrap Confidence Intervals}.
\newblock \emph{Journal of the American statistical Association}, 82\penalty0 (397):\penalty0 171--185, 1987.

\bibitem[Elena~Schmitz \& Rahmann(2025)Elena~Schmitz and Rahmann]{elena2025comprehensive}
Johanna Elena~Schmitz and Sven Rahmann.
\newblock {A comprehensive review and evaluation of species richness estimation}.
\newblock \emph{Briefings in Bioinformatics}, 26\penalty0 (2):\penalty0 bbaf158, 2025.

\bibitem[Faber(1999)]{faber1999estimating}
Nicolaas (Klaas)~M Faber.
\newblock Estimating the uncertainty in estimates of root mean square error of prediction: application to determining the size of an adequate test set in multivariate calibration.
\newblock \emph{Chemometrics and Intelligent Laboratory Systems}, 49\penalty0 (1):\penalty0 79--89, 1999.

\bibitem[Farquhar et~al.(2024)Farquhar, Kossen, Kuhn, and Gal]{Farquhar2024}
Sebastian Farquhar, Joel Kossen, Lukas Kuhn, and Yarin Gal.
\newblock {Detecting hallucinations in large language models using semantic entropy}.
\newblock \emph{Nature}, 630:\penalty0 625--630, 2024.
\newblock \doi{10.1038/s41586-024-07421-0}.
\newblock URL \url{https://doi.org/10.1038/s41586-024-07421-0}.

\bibitem[Good(1953)]{good1953population}
Irving~J Good.
\newblock {The Population Frequencies of Species and the Estimation of Population Parameters}.
\newblock \emph{{Biometrika}}, 40\penalty0 (3-4):\penalty0 237--264, 1953.

\bibitem[Granite~Team(2024)]{granite2024granite}
IBM Granite~Team.
\newblock {Granite 3.0 Language Models}, October 2024.
\newblock URL \url{https://github.com/ibm-granite/granite-3.0-language-models/}.

\bibitem[Grassberger(1988)]{grassberger1988finite}
Peter Grassberger.
\newblock Finite sample corrections to entropy and dimension estimates.
\newblock \emph{Physics Letters A}, 128\penalty0 (6-7):\penalty0 369--373, 1988.

\bibitem[Grassberger(2003)]{grassberger2003entropy}
Peter Grassberger.
\newblock {Entropy Estimates from Insufficient Samplings}.
\newblock \emph{arXiv preprint physics/0307138}, 2003.

\bibitem[Grignetti(1964)]{grignetti1964note}
Mario~C Grignetti.
\newblock {A Note on the Entropy of Words in Printed English}.
\newblock \emph{Information and Control}, 7\penalty0 (3):\penalty0 304--306, 1964.

\bibitem[Haas et~al.(2025)Haas, Yona, D'Antonio, Goldshtein, and Das]{haas2025simpleqa}
Lukas Haas, Gal Yona, Giovanni D'Antonio, Sasha Goldshtein, and Dipanjan Das.
\newblock {SimpleQA Verified: A Reliable Factuality Benchmark to Measure Parametric Knowledge}.
\newblock \emph{arXiv preprint arXiv:2509.07968}, 2025.

\bibitem[Hamidieh et~al.(2026)Hamidieh, Thost, Gerych, Yurochkin, and Ghassemi]{hamidieh2026complementing}
Kimia Hamidieh, Veronika Thost, Walter Gerych, Mikhail Yurochkin, and Marzyeh Ghassemi.
\newblock {Complementing Self-Consistency with Cross-Model Disagreement for Uncertainty Quantification}.
\newblock In \emph{The Fourteenth International Conference on Learning Representations}, 2026.
\newblock URL \url{https://openreview.net/forum?id=lOoRJo8xWy}.

\bibitem[Hao \& Li(2020)Hao and Li]{hao2020optimal}
Yi~Hao and Ping Li.
\newblock {Optimal Prediction of the Number of Unseen Species with Multiplicity}.
\newblock \emph{{Advances in Neural Information Processing Systems}}, 33:\penalty0 8553--8564, 2020.

\bibitem[Hartung \& Knapp(2001)Hartung and Knapp]{hartung2001tests}
Joachim Hartung and Guido Knapp.
\newblock {On tests of the overall treatment effect in meta-analysis with normally distributed responses}.
\newblock \emph{Statistics in Medicine}, 20\penalty0 (12):\penalty0 1771--1782, 2001.

\bibitem[Hastie et~al.(2009)Hastie, Tibshirani, and Friedman]{hastie2009elements}
Trevor Hastie, Robert Tibshirani, and Jerome Friedman.
\newblock {Model Selection and the Bias-Variance Tradeoff}.
\newblock In \emph{The Elements of Statistical Learning: Data Mining, Inference, and Prediction}, pp.\  37--38. Springer, 2009.
\newblock ISBN 978-0387848570.

\bibitem[Hausser \& Strimmer(2009)Hausser and Strimmer]{hausser2009entropy}
Jean Hausser and Korbinian Strimmer.
\newblock {Entropy Inference and the James-Stein Estimator, with Application to Nonlinear Gene Association Networks}.
\newblock \emph{Journal of Machine Learning Research}, 10\penalty0 (7), 2009.

\bibitem[He et~al.(2021)He, Gao, and Chen]{he2021debertav3}
Pengcheng He, Jianfeng Gao, and Weizhu Chen.
\newblock {DeBERTaV3: Improving DeBERTa using ELECTRA-Style Pre-Training with Gradient-Disentangled Embedding Sharing}.
\newblock In \emph{{The Eleventh International Conference on Learning Representations}}, 2021.

\bibitem[Horvitz \& Thompson(1952)Horvitz and Thompson]{horvitz1952generalization}
Daniel~G Horvitz and Donovan~J Thompson.
\newblock {A Generalization of Sampling Without Replacement from a Finite Universe}.
\newblock \emph{{Journal of the American Statistical Association}}, 47\penalty0 (260):\penalty0 663--685, 1952.

\bibitem[{IBM Research}(2025)]{granite2025}
{IBM Research}.
\newblock Granite 4.0 language models.
\newblock \url{https://github.com/ibm-granite/granite-4.0-language-models}, 2025.

\bibitem[IntHout et~al.(2014)IntHout, Ioannidis, and Borm]{inthout2014hartung}
Joanna IntHout, John Ioannidis, and George~F Borm.
\newblock {The Hartung-Knapp-Sidik-Jonkman method for random effects meta-analysis is straightforward and considerably outperforms the standard DerSimonian-Laird method}.
\newblock \emph{BMC Medical Research Methodology}, 14\penalty0 (1):\penalty0 25, 2014.

\bibitem[Ismayilzada et~al.(2025)Ismayilzada, Laverghetta~Jr., Luchini, Patel, Bosselut, Plas, and Beaty]{ismayilzada-etal-2025-creative}
Mete Ismayilzada, Antonio Laverghetta~Jr., Simone~A. Luchini, Reet Patel, Antoine Bosselut, Lonneke Van~Der Plas, and Roger~E. Beaty.
\newblock {Creative Preference Optimization}.
\newblock In Christos Christodoulopoulos, Tanmoy Chakraborty, Carolyn Rose, and Violet Peng (eds.), \emph{Findings of the Association for Computational Linguistics: EMNLP 2025}, pp.\  9580--9609, Suzhou, China, November 2025. Association for Computational Linguistics.
\newblock ISBN 979-8-89176-335-7.
\newblock \doi{10.18653/v1/2025.findings-emnlp.509}.
\newblock URL \url{https://aclanthology.org/2025.findings-emnlp.509/}.

\bibitem[Janzen(1973{\natexlab{a}})]{janzen1973sweep1}
Daniel~H Janzen.
\newblock {Sweep Samples of Tropical Foliage Insects: Description of Study Sites, With Data on Species Abundances and Size Distributions}.
\newblock \emph{Ecology}, 54\penalty0 (3):\penalty0 659--686, 1973{\natexlab{a}}.

\bibitem[Janzen(1973{\natexlab{b}})]{janzen1973sweep2}
Daniel~H Janzen.
\newblock {Sweep Samples of Tropical Foliage Insects: Effects of Seasons, Vegetation Types, Elevation, Time of Day, and Insularity}.
\newblock \emph{Ecology}, 54\penalty0 (3):\penalty0 687--708, 1973{\natexlab{b}}.

\bibitem[Juang \& Lo(2002)Juang and Lo]{juang2002bias}
Biing-Hwang Juang and SH~Lo.
\newblock {On the bias of the Turing-Good estimate of probabilities}.
\newblock \emph{{IEEE Transactions on Signal Processing}}, 42\penalty0 (2):\penalty0 496--498, 2002.

\bibitem[Kadavath et~al.(2022)Kadavath, Conerly, Askell, Henighan, Drain, Perez, Schiefer, Hatfield-Dodds, DasSarma, Tran-Johnson, et~al.]{kadavath2022language}
Saurav Kadavath, Tom Conerly, Amanda Askell, Tom Henighan, Dawn Drain, Ethan Perez, Nicholas Schiefer, Zac Hatfield-Dodds, Nova DasSarma, Eli Tran-Johnson, et~al.
\newblock {Language Models (Mostly) Know What They Know}.
\newblock \emph{arXiv preprint arXiv:2207.05221}, 2022.

\bibitem[Krithara et~al.(2023)Krithara, Nentidis, Bougiatiotis, and Paliouras]{krithara2023bioasq}
Anastasia Krithara, Anastasios Nentidis, Konstantinos Bougiatiotis, and Georgios Paliouras.
\newblock {BioASQ-QA: A manually curated corpus for Biomedical Question Answering}.
\newblock \emph{Scientific Data}, 10\penalty0 (1):\penalty0 170, 2023.

\bibitem[Kuhn et~al.(2023)Kuhn, Gal, and Farquhar]{kuhn2023semantic}
Lorenz Kuhn, Yarin Gal, and Sebastian Farquhar.
\newblock {Semantic Uncertainty: Linguistic Invariances for Uncertainty Estimation in Natural Language Generation}.
\newblock In \emph{{The Eleventh International Conference on Learning Representations}}, 2023.

\bibitem[Kwon et~al.(2023)Kwon, Li, Zhuang, Sheng, Zheng, Yu, Gonzalez, Zhang, and Stoica]{kwon2023efficient}
Woosuk Kwon, Zhuohan Li, Siyuan Zhuang, Ying Sheng, Lianmin Zheng, Cody~Hao Yu, Joseph~E. Gonzalez, Hao Zhang, and Ion Stoica.
\newblock {Efficient Memory Management for Large Language Model Serving with PagedAttention}.
\newblock In \emph{Proceedings of the ACM SIGOPS 29th Symposium on Operating Systems Principles}, 2023.

\bibitem[La~Torre et~al.(2025)La~Torre, Kelly, Menendez, and Clark]{la2025bee}
Ilaria~Pia La~Torre, David~A Kelly, Hector~D Menendez, and David Clark.
\newblock {To BEE or Not to BEE: Estimating more than Entropy with Biased Entropy Estimators}.
\newblock \emph{arXiv preprint arXiv:2501.11395}, 2025.

\bibitem[Lee \& B{\"o}hme(2025)Lee and B{\"o}hme]{leeHowMuchUnseen2025}
Seongmin Lee and Marcel B{\"o}hme.
\newblock {How Much Is Unseen Depends Chiefly on Information About the Seen}.
\newblock In \emph{Proceedings of the 13th International Conference on Learning Representations}, May 2025.

\bibitem[Li et~al.(2014)Li, Titov, and Sporleder]{li-etal-2014-improved}
Linlin Li, Ivan Titov, and Caroline Sporleder.
\newblock {Improved Estimation of Entropy for Evaluation of Word Sense Induction}.
\newblock \emph{Computational Linguistics}, 40\penalty0 (3):\penalty0 671--685, September 2014.
\newblock \doi{10.1162/COLI_a_00196}.
\newblock URL \url{https://aclanthology.org/J14-3007/}.

\bibitem[Li \& Tian(2024)Li and Tian]{li2024sub}
Yanpeng Li and Boping Tian.
\newblock {On the sub-Gaussianity of the missing mass}.
\newblock \emph{Statistics \& Probability Letters}, 206:\penalty0 109991, 2024.

\bibitem[Lin et~al.(2022)Lin, Hilton, and Evans]{lin-etal-2022-truthfulqa}
Stephanie Lin, Jacob Hilton, and Owain Evans.
\newblock {TruthfulQA: Measuring How Models Mimic Human Falsehoods}.
\newblock In Smaranda Muresan, Preslav Nakov, and Aline Villavicencio (eds.), \emph{Proceedings of the 60th Annual Meeting of the Association for Computational Linguistics (Volume 1: Long Papers)}, pp.\  3214--3252, Dublin, Ireland, May 2022. Association for Computational Linguistics.
\newblock \doi{10.18653/v1/2022.acl-long.229}.
\newblock URL \url{https://aclanthology.org/2022.acl-long.229/}.

\bibitem[Lin et~al.(2024)Lin, Trivedi, and Sun]{lin2023generating}
Zhen Lin, Shubhendu Trivedi, and Jimeng Sun.
\newblock {Generating with Confidence: Uncertainty Quantification for Black-box Large Language Models}.
\newblock \emph{{Transactions on Machine Learning Research}}, 2024, 2024.
\newblock ISSN 2835-8856.

\bibitem[Liu et~al.(2026)Liu, Pourreza, Panchal, Bhattacharyya, Jian, Qin, and Memisevic]{liu2026enhancing}
Litian Liu, Reza Pourreza, Sunny Panchal, Apratim Bhattacharyya, Yubing Jian, Yao Qin, and Roland Memisevic.
\newblock {Enhancing Hallucination Detection through Noise Injection}.
\newblock In \emph{The Fourteenth International Conference on Learning Representations}, 2026.
\newblock URL \url{https://openreview.net/forum?id=WnM3sluiVn}.

\bibitem[McAllester \& Schapire(2001)McAllester and Schapire]{mcallester2001learning}
David McAllester and Robert~E Schapire.
\newblock {Learning theory and language modeling}.
\newblock In \emph{{Seventeenth International Joint Conference on Artificial Intelligence}}, 2001.

\bibitem[McAllester \& Schapire(2000)McAllester and Schapire]{mcallester2000convergence}
David~A McAllester and Robert~E Schapire.
\newblock {On the Convergence Rate of Good-Turing Estimators}.
\newblock In \emph{Conference on Learning Theory}, pp.\  1--6, 2000.

\bibitem[McCabe et~al.(2026)McCabe, Melamed, Hartvigsen, and Huang]{mccabe2026estimating}
Lucas~Hurley McCabe, Rimon Melamed, Thomas Hartvigsen, and H~Howie Huang.
\newblock {Estimating Semantic Alphabet Size for {LLM} Uncertainty Quantification}.
\newblock In \emph{The Fourteenth International Conference on Learning Representations}, 2026.
\newblock URL \url{https://openreview.net/forum?id=uYK6GPVg1O}.

\bibitem[Nemenman et~al.(2001)Nemenman, Shafee, and Bialek]{nemenman2001entropy}
Ilya Nemenman, Fariel Shafee, and William Bialek.
\newblock {Entropy and Inference, Revisited}.
\newblock \emph{Advances in Neural Information Processing Systems}, 14, 2001.

\bibitem[Nguyen et~al.(2025)Nguyen, Payani, and Mirzasoleiman]{nguyen-etal-2025-beyond}
Dang Nguyen, Ali Payani, and Baharan Mirzasoleiman.
\newblock {Beyond Semantic Entropy: Boosting {LLM} Uncertainty Quantification with Pairwise Semantic Similarity}.
\newblock In Wanxiang Che, Joyce Nabende, Ekaterina Shutova, and Mohammad~Taher Pilehvar (eds.), \emph{{Findings of the Association for Computational Linguistics: ACL 2025}}, pp.\  4530--4540, Vienna, Austria, July 2025. Association for Computational Linguistics.
\newblock ISBN 979-8-89176-256-5.

\bibitem[Nikitin et~al.(2024)Nikitin, Kossen, Gal, and Marttinen]{nikitin2024kernel}
Alexander Nikitin, Jannik Kossen, Yarin Gal, and Pekka Marttinen.
\newblock {Kernel Language Entropy: Fine-grained Uncertainty Quantification for LLMs from Semantic Similarities}.
\newblock In \emph{{Advances in Neural Information Processing Systems}}, volume~37, pp.\  8901--8929, 2024.

\bibitem[Nowozin(2012)]{nowozin2012improved}
Sebastian Nowozin.
\newblock {Improved Information Gain Estimates for Decision Tree Induction}.
\newblock In \emph{Proceedings of the 29th International Coference on International Conference on Machine Learning}, pp.\  571--578, 2012.

\bibitem[Orlitsky \& Suresh(2015)Orlitsky and Suresh]{orlitsky2015competitive}
Alon Orlitsky and Ananda~Theertha Suresh.
\newblock {Competitive Distribution Estimation: Why is Good-Turing Good}.
\newblock \emph{{Advances in Neural Information Processing Systems}}, 28, 2015.

\bibitem[Orlitsky et~al.(2016)Orlitsky, Suresh, and Wu]{orlitsky2016optimal}
Alon Orlitsky, Ananda~Theertha Suresh, and Yihong Wu.
\newblock {Optimal prediction of the number of unseen species}.
\newblock \emph{Proceedings of the National Academy of Sciences}, 113\penalty0 (47):\penalty0 13283--13288, 2016.

\bibitem[Painsky(2023)]{painsky2023generalized}
Amichai Painsky.
\newblock {Generalized Good-Turing Improves Missing Mass Estimation}.
\newblock \emph{{Journal of the American Statistical Association}}, 118\penalty0 (543):\penalty0 1890--1899, 2023.

\bibitem[Paninski(2003)]{paninski2003estimation}
Liam Paninski.
\newblock {Estimation of Entropy and Mutual Information}.
\newblock \emph{{Neural Computation}}, 15\penalty0 (6):\penalty0 1191--1253, 2003.

\bibitem[Park \& Cho(2026)Park and Cho]{park2026efficient}
Ji~Won Park and Kyunghyun Cho.
\newblock Efficient semantic uncertainty quantification in language models via diversity-steered sampling.
\newblock In \emph{The Thirty-ninth Annual Conference on Neural Information Processing Systems}, 2026.
\newblock URL \url{https://openreview.net/forum?id=IiEtQPGVyV}.

\bibitem[Patel et~al.(2021)Patel, Bhattamishra, and Goyal]{patel-etal-2021-nlp}
Arkil Patel, Satwik Bhattamishra, and Navin Goyal.
\newblock {Are {NLP} Models really able to Solve Simple Math Word Problems?}
\newblock In \emph{Proceedings of the 2021 Conference of the North American Chapter of the Association for Computational Linguistics: Human Language Technologies}, pp.\  2080--2094, Online, June 2021. Association for Computational Linguistics.
\newblock \doi{10.18653/v1/2021.naacl-main.168}.
\newblock URL \url{https://aclanthology.org/2021.naacl-main.168}.

\bibitem[Pinchas et~al.(2024)Pinchas, Ben-Gal, and Painsky]{pinchas2024comparative}
Assaf Pinchas, Irad Ben-Gal, and Amichai Painsky.
\newblock {A Comparative Analysis of Discrete Entropy Estimators for Large-Alphabet Problems}.
\newblock \emph{{Entropy}}, 26\penalty0 (5):\penalty0 369, 2024.

\bibitem[Rajaraman et~al.(2017)Rajaraman, Thangaraj, and Suresh]{rajaraman2017minimax}
Nikhilesh Rajaraman, Andrew Thangaraj, and Ananda~Theertha Suresh.
\newblock {Minimax Risk for Missing Mass Estimation}.
\newblock In \emph{2017 IEEE International Symposium on Information Theory (ISIT)}, pp.\  3025--3029. IEEE, 2017.

\bibitem[Rajpurkar et~al.(2016)Rajpurkar, Zhang, Lopyrev, and Liang]{rajpurkar-etal-2016-squad}
Pranav Rajpurkar, Jian Zhang, Konstantin Lopyrev, and Percy Liang.
\newblock {{SQ}u{AD}: 100,000+ Questions for Machine Comprehension of Text}.
\newblock In Jian Su, Kevin Duh, and Xavier Carreras (eds.), \emph{{Proceedings of the 2016 Conference on Empirical Methods in Natural Language Processing}}, pp.\  2383--2392, Austin, Texas, November 2016. Association for Computational Linguistics.
\newblock \doi{10.18653/v1/D16-1264}.
\newblock URL \url{https://aclanthology.org/D16-1264}.

\bibitem[Rana et~al.(2022)Rana, Chien, Peng, and Milenkovic]{rana2022small}
Vishal Rana, Eli Chien, Jianhao Peng, and Olgica Milenkovic.
\newblock {Small-Sample Estimation of the Mutational Support and Distribution of SARS-CoV-2}.
\newblock \emph{IEEE/ACM Transactions on Computational Biology and Bioinformatics}, 20\penalty0 (1):\penalty0 668--682, 2022.

\bibitem[Roswell et~al.(2021)Roswell, Dushoff, and Winfree]{roswell2021conceptual}
Michael Roswell, Jonathan Dushoff, and Rachael Winfree.
\newblock A conceptual guide to measuring species diversity.
\newblock \emph{Oikos}, 130\penalty0 (3):\penalty0 321--338, 2021.

\bibitem[Shannon(1948)]{shannon1948mathematical}
Claude~Elwood Shannon.
\newblock {A Mathematical Theory of Communication}.
\newblock \emph{The Bell System Technical Journal}, 27\penalty0 (3):\penalty0 379--423, 1948.

\bibitem[Sherwin \& Prat~i Fornells(2019)Sherwin and Prat~i Fornells]{sherwin2019introduction}
William~B Sherwin and Narcis Prat~i Fornells.
\newblock {The Introduction of Entropy and Information Methods to Ecology by Ramon Margalef}.
\newblock \emph{Entropy}, 21\penalty0 (8):\penalty0 794, 2019.

\bibitem[Sidik \& Jonkman(2002)Sidik and Jonkman]{sidik2002simple}
Kurex Sidik and Jeffrey~N Jonkman.
\newblock A simple confidence interval for meta-analysis.
\newblock \emph{Statistics in Medicine}, 21\penalty0 (21):\penalty0 3153--3159, 2002.

\bibitem[Skorski(2021)]{skorski2021mean}
Maciej Skorski.
\newblock {Mean-Squared Accuracy of Good-Turing Estimator}.
\newblock In \emph{IEEE International Symposium on Information Theory}, pp.\  2846--2851. IEEE, 2021.

\bibitem[Skorski(2023)]{skorski2023sub}
Maciej Skorski.
\newblock {On Sub-Gaussian Concentration of Missing Mass}.
\newblock \emph{Theory of Probability \& Its Applications}, 68\penalty0 (2):\penalty0 324--329, 2023.

\bibitem[Sun \& Xu(2014)Sun and Xu]{sun2014fast}
Xu~Sun and Weichao Xu.
\newblock {Fast Implementation of DeLong’s Algorithm for Comparing the Areas Under Correlated Receiver Operating Characteristic Curves}.
\newblock \emph{{IEEE Signal Processing Letters}}, 21\penalty0 (11):\penalty0 1389--1393, 2014.

\bibitem[Valiant \& Valiant(2017)Valiant and Valiant]{valiant2017estimating}
Gregory Valiant and Paul Valiant.
\newblock {Estimating the Unseen: Improved Estimators for Entropy and Other Properties}.
\newblock \emph{Journal of the ACM (JACM)}, 64\penalty0 (6):\penalty0 1--41, 2017.

\bibitem[Virtanen et~al.(2020)Virtanen, Gommers, Oliphant, Haberland, Reddy, Cournapeau, Burovski, Peterson, Weckesser, Bright, {van der Walt}, Brett, Wilson, Millman, Mayorov, Nelson, Jones, Kern, Larson, Carey, Polat, Feng, Moore, {VanderPlas}, Laxalde, Perktold, Cimrman, Henriksen, Quintero, Harris, Archibald, Ribeiro, Pedregosa, {van Mulbregt}, and {SciPy 1.0 Contributors}]{2020SciPy-NMeth}
Pauli Virtanen, Ralf Gommers, Travis~E. Oliphant, Matt Haberland, Tyler Reddy, David Cournapeau, Evgeni Burovski, Pearu Peterson, Warren Weckesser, Jonathan Bright, St{\'e}fan~J. {van der Walt}, Matthew Brett, Joshua Wilson, K.~Jarrod Millman, Nikolay Mayorov, Andrew R.~J. Nelson, Eric Jones, Robert Kern, Eric Larson, C~J Carey, {\.I}lhan Polat, Yu~Feng, Eric~W. Moore, Jake {VanderPlas}, Denis Laxalde, Josef Perktold, Robert Cimrman, Ian Henriksen, E.~A. Quintero, Charles~R. Harris, Anne~M. Archibald, Ant{\^o}nio~H. Ribeiro, Fabian Pedregosa, Paul {van Mulbregt}, and {SciPy 1.0 Contributors}.
\newblock {{SciPy} 1.0: Fundamental Algorithms for Scientific Computing in Python}.
\newblock \emph{Nature Methods}, 17:\penalty0 261--272, 2020.
\newblock \doi{10.1038/s41592-019-0686-2}.

\bibitem[Vu et~al.(2007)Vu, Yu, and Kass]{vu2007coverage}
Vincent~Q Vu, Bin Yu, and Robert~E Kass.
\newblock {Coverage-adjusted entropy estimation}.
\newblock \emph{{Statistics in Medicine}}, 26\penalty0 (21):\penalty0 4039--4060, 2007.

\bibitem[Wu \& Yang(2019)Wu and Yang]{wu2019chebyshev}
Yihong Wu and Pengkun Yang.
\newblock Chebyshev polynomials, moment matching, and optimal estimation of the unseen.
\newblock \emph{The Annals of Statistics}, 47\penalty0 (2):\penalty0 857--883, 2019.

\bibitem[Wyner \& Foster(2003)Wyner and Foster]{wyner2003lower}
Abraham~J Wyner and Dean Foster.
\newblock On the lower limits of entropy estimation.
\newblock \emph{IEEE Transactions on Information Theory}, 2003.

\bibitem[Yang et~al.(2018)Yang, Qi, Zhang, Bengio, Cohen, Salakhutdinov, and Manning]{yang2018hotpotqa}
Zhilin Yang, Peng Qi, Saizheng Zhang, Yoshua Bengio, William~W Cohen, Ruslan Salakhutdinov, and Christopher~D Manning.
\newblock {HotpotQA: A Dataset for Diverse, Explainable Multi-hop Question Answering}.
\newblock In \emph{{Proceedings of the 2018 Conference on Empirical Methods in Natural Language Processing}}, pp.\  2369--2380, Brussels, Belgium, October-November 2018. {Association for Computational Linguistics}.
\newblock \doi{10.18653/v1/D18-1259}.

\end{thebibliography}
\bibliographystyle{iclr2027_conference}

\clearpage
\appendix

\section{Auxiliary results}\label{sec:proofs}

\subsection{Existence of the fixed point solution}

\begin{lemma}\label{lem:fixed-point-existence}
    Given $\upsilon$ and $X^n$, Eq. \ref{eq:mass-fp-def} has a fixed point.
\end{lemma}
\begin{proof}
For brevity, we will write $\mu(m, \upsilon, X^n)$ as $\mu(m)$.
$\mu(m)$ is the sum of positive polynomials, which are continuous everywhere, so $\mu(m)$ is continuous on $[0, 1]$.
If $m \in [0, 1]$, then $\mu(m) \leq (1-m) \Big( \sum_{u \in X_\neq^n} \hat{p}_u^{ML} \Big) + m = 1$.
Each term of $\mu(m)$ is positive, so $m\in[0,1]$ implies $\mu(m)\in[0,1]$, establishing self-mapping over $[0,1]$.
Therefore, a fixed point exists, by the 1D Brouwer Fixed Point Theorem (i.e., Intermediate Value Theorem).
\end{proof}

\subsection{Results supporting \textsc{SENECA}'s asymptotic consistency}

\begin{lemma}\label{lem:sc-open-interval}
    Let $X^n \sim P$ be a multiset of $n$ IID samples drawn from a multinomial distribution $P$.
    Then $\hat{M}^{SC}(X^n) \in [0, 1)$.
\end{lemma}
\begin{proof}
By Lemma \ref{lem:fixed-point-existence}, $\mu$ is self-mapping over $[0,1]$.
Since $\mu(1) = \Big(1-\frac{1}{\max(\upsilon, 1)}\Big)^n \in [0, 1)$, the fixed point cannot be $1$, and the result follows.
\end{proof}

\begin{lemma}\label{lem:msc-to-zero}
    Let $X^n \sim P$ be a multiset of $n$ IID samples drawn from a multinomial distribution $P$ over finite support.
    If $1 \leq \upsilon \leq f(n)$ for some sub-quadratic function $f(n)=o(n^2)$, then $\hat{M}_{\upsilon}^{SC}(X^n) \rightarrow 0$ as $n \rightarrow \infty$.
\end{lemma}
\begin{proof}
Here, and in proofs to follow, we will use the following notation:
\begin{align}
    \mu(m) &= \textsc{Observed}(m) + \textsc{Unobserved}(m) \text{, where} \nonumber \\
    \textsc{Observed}(m) &\triangleq (1-m) \sum_{u \in X_\neq^n} \hat{p}_u^{ML} \Big(1-(1-m)\hat{p}_u^{ML}\Big)^n \nonumber \\
    \textsc{Unobserved}(m) &\triangleq m\Big(1-\frac{m}{\max(\upsilon, 1)}\Big)^n \nonumber \\
    &= m\Big(1-\frac{m}{\upsilon}\Big)^n \text{ when $\upsilon \geq 1$.} \nonumber
\end{align}
The summand of $\textsc{Observed}(m)$ is maximized by $(1-m)\hat{p}_u^{ML} = \frac{1}{n+1}$ (see Lemma $1$ of \citet{skorski2021mean}), so:
\begin{align}
    \textsc{Observed}(m) &= (1-m) \sum_{u \in X_{\neq}^n} \hat{p}_u^{ML}\Big( 1-(1-m)\hat{p}_u^{ML} \Big)^n \nonumber \\
    &\leq \sum_{u \in X_{\neq}^n} \frac{1}{n+1} \Big(1-\frac{1}{n+1} \Big)^n  \nonumber \\
    &\leq |\mathcal{X}| \frac{1}{n+1} \Big(1-\frac{1}{n+1} \Big)^n \nonumber \\
    &\leq  \frac{|\mathcal{X}|}{n+1}. \label{eq:observed_leq_Onneg}
\end{align}
Next, recall that $e^{-x} \leq \frac{1}{1 + x}$ for all $x \geq 0$ (and equality only at $x=0$), so:
\begin{align}
    \Big(1-\frac{m}{\upsilon} \Big)^n &\leq (e^{-m/\upsilon})^n = e^{-nm/\upsilon} \text{, and} \label{eq:1-mv-exp} \\
    1- \Big(1-\frac{m}{\upsilon} \Big)^n &\geq 1 - e^{-nm/\upsilon} \geq 1 - \frac{1}{1 + nm/\upsilon} = \frac{nm/\upsilon}{1 + nm/\upsilon} = \frac{nm}{\upsilon + nm}. \label{eq:1-mv-exp2}
\end{align}
At any fixed point $m^*$ of $\mu$, we have $\textsc{Observed}(m^*) + \textsc{Unobserved}(m^*) = m^*$.
Applying Ineqs.~\ref{eq:observed_leq_Onneg} and \ref{eq:1-mv-exp2}, we have:
\begin{equation}
    \frac{|\mathcal{X}|}{n+1} \geq \textsc{Observed}(m^*) = m^* - m^*\Big(1-\frac{m^*}{\upsilon} \Big)^n = m^* \Bigg(1 - \Big(1-\frac{m^*}{\upsilon} \Big)^n \Bigg) \geq m^*\frac{nm^*}{\upsilon + nm^*} = \frac{n(m^*)^2}{\upsilon + nm^*}
\end{equation}
Rearranging terms, we see that any fixed point $m^*$ must satisfy:
\begin{align}
    n (m^*)^2 + \Big(\frac{-n|\mathcal{X}|}{n+1}\Big)m^* - \Big(\frac{|\mathcal{X}|}{n+1}\Big) \upsilon &\leq 0, \text{ i.e.,} \nonumber \\
    \Theta(n) (m^*)^2 + O(1) m^* + O\Big(\frac{1}{n}\Big)\upsilon &\leq 0, \nonumber
\end{align}
to abuse asymptotic notation slightly. 
This means an asymptotic bound of $m^*$ is the worse of $\frac{O(1)}{\Theta(n)}$ and $\sqrt{\frac{O(1/n)O(f(n))}{\Theta(n)}} = \sqrt{\frac{O(1/n)o(n^2)}{\Theta(n)}}$, both of which tend to zero.
Then, since every fixed point of $\mu$ is non-negative (Lemma \ref{lem:sc-open-interval}), the result follows from the sandwich theorem.
\end{proof}

\begin{lemma} \label{lem:generalized-vu} (Extension of \citet{vu2007coverage}.)
    Let $X^n \sim P$ be a multiset of $n$ IID samples drawn from a multinomial distribution $P$ with true entropy $\mathbb{H}(P)$ and finite entropy variance.\footnote{i.e., $\sum_{u \in \mathcal{X}} p_u (\log p_u )^2$ \citep{wyner2003lower, vu2007coverage}, which is finite for fixed distributions over finite support.}
    Let $\hat{M}$ be any missing mass estimator satisfying $\hat{M} \in [0, 1)$ and $\hat{M} \xrightarrow{p} 0$ as $n \rightarrow \infty$.
    Let $\hat{H}^{HT}_{\hat{M}}$ denote its corresponding ``Horvitz-Thompson-style'' coverage-adjusted entropy estimator, i.e.:
    \begin{equation}
        \hat{H}^{HT}_{\hat{M}}(X^n) = -\sum_{u \in X_{\neq}^n} \frac{\hat{p}_u^{HT} \log \hat{p}_u^{HT} }{1-(1-\hat{p}_u^{HT})^n}, \nonumber
    \end{equation}
    where $\hat{p}_u^{HT} = ( 1-\hat{M}(X^n) ) \hat{p}_u^{ML} $.
    Then $\hat{H}^{HT}_{\hat{M}} \xrightarrow{p} \mathbb{H}(P)$ as $n \rightarrow \infty$.
\end{lemma}

\begin{proof}
The following inequality, valid for missing mass estimators satisfying $\hat{M} \in [0, 1)$ and $\hat{M} \xrightarrow{p} 0$ over distributions with finite entropy variance, arises from \citet{vu2007coverage} (the proof of the authors' Theorem 1):
\begin{align}
    |\hat{H}^{HT}_{\hat{M}}-\hat{H}^{plugin}| &\leq A_{n} + B_{n} + C_{n} + D_{n} \text{ , where} \nonumber \\
    A_{n} &= \frac{\hat{M}}{1 - (1-\frac{1-\hat{M}}{n})^n} \hat{H}^{plugin} \nonumber \\
    B_{n} &= \frac{(1-\frac{1-\hat{M}}{n})^n}{1 - (1-\frac{1-\hat{M}}{n})^n} \Bigg( \frac{(\log n)^2}{\sqrt{n}} + \log n \sum_{u:p_u \leq \frac{1}{\sqrt{n}}} \hat{p}_u^{ML} \Bigg) \nonumber \\
    C_{n} &= \frac{(1-\frac{\log n}{n})^n}{1-(1-\frac{\log n}{n})^n} \hat{H}^{plugin} \text{ , and} \nonumber \\
    D_{n} &= \frac{|\log (1-\hat{M})|}{1 - (1-\frac{1-\hat{M}}{n})^n}, \nonumber
\end{align}
with $B_{n}$ being $O_P(\frac{1}{\log n})$ and $C_{n}$ being $O_P(\frac{1}{n})$ \citep{vu2007coverage}.
Reviewing the remaining terms:
\begin{align}
    A_{n} &\leq \frac{\hat{M}}{1-\exp(\hat{M}-1)} \hat{H}^{plugin}  \xrightarrow{p} 0 \text{ , and} \nonumber \\
    D_{n} &\leq \frac{|\log (1-\hat{M})|}{1-\exp(\hat{M}-1)} \xrightarrow{p} 0,
\end{align}
Therefore, $|\hat{H}^{HT}_{\hat{M}}-\hat{H}^{plugin}|\xrightarrow{p} 0$ and $\hat{H}^{HT}_{\hat{M}}  \xrightarrow{p} \hat{H}^{plugin}$.
Consequently, $\hat{H}^{HT}_{\hat{M}}  \xrightarrow{p} \hat{H}^{plugin}$ and $\hat{H}^{HT}_{\hat{M}}$ converges in probability to the true entropy.
\end{proof}

\subsection{Results supporting uniqueness of the fixed point solution when $|X_\neq^n| = 1$}

\begin{lemma}\label{lem:fixed-point-uniqueness-all-labels-same}
    If $|X_\neq^n| = 1$, $n \in \mathbb{Z}_{\geq2}$, and $\upsilon \in [1, \frac{n}{2}]$, then the unique solution to Eq. \ref{eq:mass-fp-def} on $[0,1]$ is $m^*=0$.
\end{lemma}
\begin{proof}
Following the shorthand of Lemma \ref{lem:fixed-point-existence}, we will write $\mu(m, \upsilon, X^n)$ as $\mu(m)$.
If $|X_\neq^n| = 1$, then $\mu(m)=(1-m)m^n+m(1-\frac{m}\upsilon)^n$. 
Clearly, $m^*=0$ is one possible fixed point. 
Now, suppose $\mu$ has at least one other fixed point, i.e., there exists some $m^\dagger \in (0, 1]$ such that:
\begin{align}
    \mu(m^\dagger) &= (1-m^\dagger)(m^\dagger)^n+m^\dagger(1-\frac{m^\dagger}\upsilon)^n = m^\dagger \text{, so} \nonumber \\
    1 &= (1-m^\dagger) (m^\dagger)^{n-1} + \Big(1-\frac{m^\dagger}{\upsilon} \Big)^n  \label{eq:lhs-equals-1} \\
    &< (1-m^\dagger) (m^\dagger)^{n-1} + e^{-nm^\dagger/\upsilon} \label{eq:single-class-uniqueness-bound-1} \\
    &\leq (1-m^\dagger) (m^\dagger)^{n-1} + e^{-2m^\dagger} , \label{eq:single-class-uniqueness-bound-2}
\end{align}
since $m^\dagger > 0$, recalling Eq.~\ref{eq:1-mv-exp} from Lemma~\ref{lem:msc-to-zero}.
Note also that $e^{-x} \leq 1 - x + \frac{1}{2}x^2$ for $x>0$.\footnote{This upper bound arises from the Maclaurin expansion of $e^{-x}$.} This allows us to establish a more accessible upper bound for Eq. \ref{eq:single-class-uniqueness-bound-2}:
\begin{align}
(1-m^\dagger) (m^\dagger)^{n-1} + e^{-2m^\dagger}
    &\leq (1-m^\dagger) (m^\dagger)^{n-1} + 1-2m^\dagger +2(m^\dagger)^2  \nonumber \\
    &= (1-m^\dagger) (m^\dagger)^{n-1} + 2m^\dagger (m^\dagger - 1) + 1  \nonumber \\
    &= (1-m^\dagger) \Big((m^\dagger)^{n-1}-2m^\dagger \Big) + 1  \nonumber \\
    &\leq 1, \nonumber
\end{align}
since $(m^\dagger)^{n-1}-2m^\dagger < 0$, contradicting Eq. \ref{eq:lhs-equals-1}.
So, $m^*=0$ is the unique fixed point on $[0, 1]$.
\end{proof}

\begin{lemma}\label{lem:fixed-point-uniqueness-all-labels-same-2}
    Let $|X_\neq^n| = 1$ and $n \in \mathbb{Z}_{>2}$.
    If $\upsilon \in [1, kn]$ for $k \leq ( \log \frac{n-1}{n-2} )^{-1}$, then the unique solution to Eq. \ref{eq:mass-fp-def} on $[0,1]$ is $m^*=0$.
\end{lemma}
\begin{proof}
The proof is identical to that of Lemma \ref{lem:fixed-point-uniqueness-all-labels-same} up to Eq. \ref{eq:single-class-uniqueness-bound-1}, at which point we apply a different upper bound.
For convenience, let $\psi(m, n) = (1-m) (m)^{n-2}$, where:
\begin{equation}
\frac{\partial}{\partial m} \psi(m, n) = m^{n-3} \Big(- (n-1)m+ (n-2) \Big), \nonumber
\end{equation}
which has critical point $m=\frac{n-2}{n-1}$.
$\psi(0, n)= \psi(1, n) = 0$, so $\psi(\frac{n-2}{n-1}, n) = \frac{1}{n-1} \Big(\frac{n-2}{n-1}\Big)^{n-2} < \frac{1}{n-1}$ is the maximum of $\psi(m, n)$ on $m \in [0, 1]$, $n > 2$.
Next, note that $e^{-m/k}$ is strictly convex for $x \in \mathbb{R}$, $k \neq 0$, so, on $m \in [0, 1]$, it is bounded from above by the secant line connecting $(0, 1)$ and $(1, e^{-1/k})$.
In other words, $e^{-m/k} \leq (e^{-1/k}) - 1)m + 1$ on $m \in [0, 1]$, and the inequality is strict on $m \in (0, 1)$.
So, returning to the right-hand side of Eq. \ref{eq:single-class-uniqueness-bound-1}, we have:
\begin{align}
(1-m^\dagger) (m^\dagger)^{n-1} + e^{-nm^\dagger/\upsilon}
    &=  m^\dagger \psi(m^\dagger, n) + e^{-nm^\dagger/\upsilon} \nonumber \\
    &< m^\dagger \frac{1}{n-1} + (e^{-1/k} - 1)m + 1 \nonumber \\
    &= m^\dagger \Big( \frac{1}{n-1} - (1- e^{-1/k}) \Big) + 1. \label{eq:single-class-uniqueness-bound-3}
\end{align}
Finally, observe that Eq. \ref{eq:single-class-uniqueness-bound-3} is less than $1$ when $\frac{1}{n-1} - (1- e^{-1/k}) \leq 0$, which is equivalent to $k \leq - ( \log \frac{n-2}{n-1} )^{-1} = ( \log \frac{n-1}{n-2} )^{-1}$.
\end{proof}

\begin{corollary}\label{cor:fixed-point-uniqueness-all-labels-same}
    Let $|X_\neq^n| = 1$ and $n \in \mathbb{Z}_{\geq 2}$.
    If $\upsilon \in [1, kn]$, for $k$ satisfying:
    \begin{equation}
        k = \begin{cases}
            \frac{1}{2} & n = 2 \\
            ( \log \frac{n-1}{n-2} )^{-1} & \text{otherwise,} \nonumber
        \end{cases}
    \end{equation}
    then the unique solution to Eq. \ref{eq:mass-fp-def} on $[0,1]$ is $m^*=0$.
\end{corollary}
\begin{proof}
This simply combines the conditions from Lemmas \ref{lem:fixed-point-uniqueness-all-labels-same} and \ref{lem:fixed-point-uniqueness-all-labels-same-2}, observing that $( \log \frac{n-1}{n-2} )^{-1} > \frac{1}{2}$ when $n \geq 3$.
\end{proof}

\subsection{Results supporting uniqueness of the fixed point solution when $|X_\neq^n| > 1$ and $1 \leq \upsilon \leq \frac{n}{2}$}

\begin{lemma}\label{lem:fixed-point-uniqueness-right-half}
    If $n \in \mathbb{Z}_{\geq2}$ and $\upsilon \geq 1$, then Eq. \ref{eq:mass-fp-def} has at most one solution on $m\in[\frac{\upsilon}{n+1},1]$.
\end{lemma}
\begin{proof}
To begin, we refer to Lemma \ref{lem:sc-open-interval}, which excludes $m=1$ as a fixed point; the remainder of the proof assumes $m < 1$.
Recalling the notation of $\textsc{Observed}(m)$ and $\textsc{Unobserved}(m)$ from Lemma~\ref{lem:msc-to-zero}:
\begin{align}
\frac{d}{dm} \textsc{Observed}(m) &= \sum_{u \in X_\neq^n} - \hat{p}_u^{ML} \Big(1-(1-m)\hat{p}_u^{ML}\Big)^{n-1}\Big(1-(1-m)(n+1)\hat{p}_u^{ML}\Big) \nonumber \\
    &= \sum_{u \in X_\neq^n} \hat{p}_u^{ML} \Big(1-(1-m)\hat{p}_u^{ML}\Big)^{n-1}\Big((n+1)(1-m)\hat{p}_u^{ML} - 1\Big) \nonumber \\
    &\leq \max_{u \in X_\neq^n} \Big(1-(1-m)\hat{p}_u^{ML}\Big)^{n-1}\Big((n+1)(1-m)\hat{p}_u^{ML} - 1\Big) \nonumber \\
    &\leq \max_{q \in [0, 1)} (1-q)^{n-1}\Big((n+1) q - 1\Big), \nonumber
\end{align}
where the second-to-last step uses the fact that $\hat{p}_u^{ML} > 0$ for all $u \in X_\neq^n$ and $\sum_{u \in X_\neq^n} \hat{p}_u^{ML} = 1$, and the last step applies the shorthand $q_u = (1-m) \hat{p}_u^{ML}$. 
The maximization interval in the last step is half-open because $q = 1$ is excluded by the condition $m \geq \frac{\upsilon}{n+1} > 0$.
Taking the derivative of the max operand, we have:
\begin{align}
\frac{d}{dq} \Bigg( (1-q)^{n-1} \Big((n+1)q - 1\Big) \Bigg) &= - n (1-q)^{n-2} \Big((n+1)q - 2 \Big) \nonumber,
\end{align}
which is positive, zero, and negative for $q < \frac{2}{n+1}$, $q = \frac{2}{n+1}$, and $q > \frac{2}{n+1}$, respectively.
Hence:
\begin{equation}
\frac{d}{dm} \textsc{Observed}(m)
    \leq \Big(1-\frac{2}{n+1}\Big)^{n-1}\Big((n+1)\frac{2}{n+1}-1\Big) = \Big(\frac{n-1}{n+1}\Big)^{n-1} < 1, \nonumber
\end{equation}
for $n \geq 2$. 
Meanwhile:
\begin{equation}
\frac{d}{dm} \textsc{Unobserved}(m) = \frac{\Big(1-\frac{m}{\upsilon}\Big)^n\Big(m(n+1)-\upsilon\Big)}{m-\upsilon}, \nonumber
\end{equation}
which is non-positive if $m \in [\frac{\upsilon}{n+1}, 1)$ and $\upsilon \geq 1$.
Then:
\begin{align}
    \frac{d}{dm} \Big(\mu(m)-m\Big) &= \frac{d}{dm} \textsc{Observed}(m) + \frac{d}{dm} \textsc{Unobserved}(m) - 1 \nonumber \\
    &\leq \frac{d}{dm} \textsc{Observed}(m) - 1 \nonumber \\
    &< 0. \nonumber
\end{align}
Recall that a fixed point occurs when $\mu(m)-m = 0$.
Since $\mu(m)-m$ is strictly decreasing on $[\frac{\upsilon}{n+1}, 1)$, this can occur at most once.
\end{proof}

\begin{lemma}\label{lem:fixed-point-uniqueness-concavity}
    If $n \in \mathbb{Z}_{\geq2}$, $|X_\neq^n| > 1$, and $1\leq \upsilon \leq \frac{n}{2}$, then $\mu(m)-m$ is concave on $m\in[0, \frac{\upsilon}{n+1}]$.
\end{lemma}
\begin{proof}
We will treat the cases of $n=2$ and $n>2$ separately.
In so doing, we use the following facts:
\begin{align}
&\frac{d^2}{dm^2} \textsc{Observed}(m) = n \sum_{u \in X_\neq^n} (\hat{p}_u^{ML})^2 \Big(1-(1-m)\hat{p}_u^{ML} \Big)^{n-2} \Big( (n+1)(1-m)\hat{p}_u^{ML} - 2 \Big) \text{,} \nonumber \\
&\frac{d^2}{dm^2} \textsc{Unobserved}(m) = -\frac{n}{\upsilon} \Big(1-\frac{m}{\upsilon} \Big)^{n-2} \Big(2 - (n+1)\frac{m}{\upsilon} \Big) \text{, and}  \nonumber \\
&\frac{d^2}{dm^2} \Big(\mu(m)-m\Big) = \frac{d^2}{dm^2} \textsc{Observed}(m) + \frac{d^2}{dm^2} \textsc{Unobserved}(m). \nonumber
\end{align}

\textbf{Case 1:} $n=2$.
Since $|X_\neq^n|>1$, we have $\hat{p}_u^{ML}=\frac{1}{2}$ for all $u \in X_\neq^n$.
So:
\begin{align}
&\frac{d^2}{dm^2} \textsc{Observed}(m) = 4 \Big(\frac{1}{2}\Big)^2 (1) \Big(\frac{3(1-m)}{2}-2\Big) = - \frac{3m+1}{2} \text{, and} \nonumber \\
&\frac{d^2}{dm^2} \textsc{Unobserved}(m) = - \frac{2}{\upsilon} \Big(2- \frac{3m}{\upsilon} \Big) = - 2 (2-3m),  \nonumber
\end{align}
since the provided bounds on $\upsilon$ are only satisfied by $\upsilon = 1$ when $n=2$. It follows that:
\begin{equation}
\frac{d^2}{dm^2} \Big(\mu(m)-m\Big) = - \frac{3m+1}{2} - 2 (2-3m) = \frac{9}{2} (m-1), \nonumber
\end{equation}
which is negative for $m<1$.

\textbf{Case 2:} $n>2$.
Since $m \leq \frac{\upsilon}{n+1}$ implies $\frac{m}{\upsilon} \leq \frac{1}{n+1}$:
\begin{align}
\frac{d^2}{dm^2} \textsc{Unobserved}(m) &= -\frac{n}{\upsilon} \Big(1-\frac{m}{\upsilon} \Big)^{n-2} \Big(2-(n+1)\frac{m}{\upsilon} \Big) 
\leq -2 \Big(\frac{n}{n+1} \Big)^{n-2}. \nonumber
\end{align}
As in Lemma \ref{lem:fixed-point-uniqueness-right-half}, we will employ the shorthand $q_u = (1-m) \hat{p}_u^{ML}$. Since $m<1$, we have:
\begin{align}
\frac{d^2}{dm^2} \textsc{Observed}(m) &= \frac{n}{1-m} \sum_{u \in X_{\neq}^n} \hat{p}_u^{ML} q_u (1-q_u)^{n-2} \Big( (n+1) q_u - 2 \Big) \nonumber \\
&\leq \frac{n}{1-m} \max_{u \in X_{\neq}^n} 
\Bigg( q_u (1-q_u)^{n-2} \Big( (n+1) q_u - 2 \Big) \Bigg), \label{eq:d2-observed-max} 
\end{align}
where we again use the fact that $\hat{p}_u^{ML} >0$ for all $u \in X_\neq^n$ and $\sum_{u \in X_\neq^n} \hat{p}_u^{ML} = 1$.
Note that the max operand in the right-hand side (RHS) of Ineq. \ref{eq:d2-observed-max}  is negative for $q_u \in (0, \frac{2}{n+1})$.\footnote{Note that $q_u \neq 0$, because $m < 1$.}
Then, if some $q_u \in (0, \frac{2}{n+1})$ maximizes the RHS of Ineq. \ref{eq:d2-observed-max}, $\frac{d^2}{dm^2} \textsc{Observed}(m) < 0$ and:
\begin{equation}
    \frac{d^2}{dm^2} \Big(\mu(m)-m\Big) < \frac{d^2}{dm^2} \textsc{Unobserved}(m) \leq -2 \Big(\frac{n}{n+1} \Big)^{n-2} < 0. \nonumber
\end{equation}

Now suppose that some $q_u \in [\frac{2}{n+1}, 1)$ maximizes the RHS of Ineq. \ref{eq:d2-observed-max}.\footnote{The half-open interval of maximization reflects that $q_u \neq 1$ (since $|X_\neq^n|>1$ and therefore $\hat{p}_u^{ML} < 1$).}
We will show that $\frac{d^2}{dm^2} (\mu(m)-m) < 0$ in this case, as well.
Revisiting Ineq. \ref{eq:d2-observed-max}:
\begin{align}
\frac{n}{1-m} \max_{u \in X_{\neq}^n} 
\Bigg( q_u (1-q_u)^{n-2} \Big( (n+1) q_u - 2 \Big) \Bigg) &= \frac{n}{1-m} \max_{u \in X_{\neq}^n} \Bigg( (1-q_u)^{n-2} \Big( (n+1) q_u^2 - 2q_u \Big) \Bigg) \nonumber \\
&\leq \frac{n}{1-m} \max_{q \in [\frac{2}{n+1}, 1)} \Bigg( (1-q)^{n-2} \Big( (n+1) q^2 - 2q \Big) \Bigg) \nonumber \\
&\hspace{-1.2cm}\leq \frac{n}{1-m} \max_{q \in [\frac{2}{n+1}, 1)} 
\Bigg( (1-q)^{n-2} \Big( (n+1) q^2 - 2 q + \frac{1}{n+1}\Big) \Bigg) \nonumber \\
&\hspace{-1.2cm}= \frac{n(n+1)}{(1-m)} \max_{q \in [\frac{2}{n+1}, 1)}
\Bigg((1-q)^{n-2} \Big(q - \frac{1}{n+1} \Big)^2 \Bigg). \label{step:case2-step1}
\end{align}
Taking the derivative of the max's operand:
\begin{align}
\frac{d}{dq} \Bigg((1-q)^{n-2} \Big(q - \frac{1}{n+1} \Big)^2 \Bigg) 
&= 2 (1-q)^{n-2}\Big(q-\frac{1}{n+1}\Big) - (n-2)\Big(q-\frac{1}{n+1}\Big)^2 (1-q)^{n-3} \nonumber \\
&= (1-q)^{n-3} \Big(q - \frac{1}{n+1} \Big) \Big(\frac{3n}{n+1} - nq \Big) \nonumber \\
&= n(1-q)^{n-3} \Big(q - \frac{1}{n+1} \Big) \Big(\frac{3}{n+1} - q \Big), \label{eq:operand-deriv}
\end{align}
whose only zero in $[\frac{2}{n+1}, 1)$ is $q=\frac{3}{n+1}$.
Eq. \ref{eq:operand-deriv} is positive on $(\frac{2}{n+1}, \frac{3}{n+1})$ and negative on $(\frac{3}{n+1}, 1)$, so $q=\frac{3}{n+1}$ must maximize Eq. \ref{step:case2-step1}'s max operand on $[\frac{2}{n+1}, 1)$.
Evaluating Eq. \ref{step:case2-step1} at $q=\frac{3}{n+1}$ gives:
\begin{align}
\frac{d^2}{dm^2} \textsc{Observed}(m) &\leq \frac{n(n+1)}{1-m}
\Big(\frac{n-2}{n+1} \Big)^{n-2} \Big(\frac{2}{n+1} \Big)^2  = \frac{n}{1-m} \Big(\frac{4 (n-2)^{n-2}}{(n+1)^{n-1}}\Big) \leq \frac{8n}{n+2} \Big(\frac{n-2}{n+1}\Big)^{n-2}, \nonumber
\end{align}
since $m \leq \frac{\upsilon}{n+1} \leq \frac{n}{2(n+1)}$ implies $\frac{1}{1-m} \leq 2 \frac{n+1}{n+2}$.
Therefore:
\begin{align}
\frac{d^2}{dm^2} \Big(\mu(m)-m\Big) &\leq \Big(\frac{1}{n+1} \Big)^{n-2} \Big(\frac{8n}{n+2} (n-2)^{n-2} - 2n^{n-2} \Big) \label{step:case2-step2} \\
&\leq \Big(\frac{1}{n+1} \Big)^{n-2} \Big(8 (n-2)^{n-2} - 2n^{n-2} \Big). \label{step:case2-step3}
\end{align}
The RHS of Ineq. \ref{step:case2-step3} is negative when $4 < ( \frac{n}{n-2} )^{n-2}$, the right-hand side of which is an increasing function that evaluates to $\frac{125}{27}>4$ at $n=5$, so Ineq. \ref{step:case2-step3} is negative for $n \geq 5$.
For $n=3$, we evaluate Ineq. \ref{step:case2-step2} directly: 
\begin{equation}
\Big(\frac{1}{3+1} \Big)^{3-2} \Big(\frac{8 \cdot 3}{3+2} (3-2)^{3-2} - 2\cdot 3^{3-2} \Big) = \frac{1}{4}\Big(\frac{24}{5}-6 \Big) = -\frac{3}{10} < 0. \nonumber
\end{equation}
Similarly, for $n=4$:
\begin{equation}
\Big(\frac{1}{4+1} \Big)^{4-2} \Big(\frac{8 \cdot 4}{4+2} (4-2)^{4-2} - 2\cdot 4^{4-2} \Big) = -\frac{32}{75} < 0. \nonumber
\end{equation}
To conclude, we have $\frac{d^2}{dm^2} \Big(\mu(m)-m\Big) < 0$ for $m \in [0, \frac{\upsilon}{n+1}]$, and the result follows.
\end{proof}

\begin{lemma}\label{lem:fixed-point-uniqueness-left-half}
    Suppose $n \in \mathbb{Z}_{\geq2}$, $|X_\neq^n| > 1$, and $1\leq \upsilon \leq \frac{n}{2}$.
    If Eq. \ref{eq:mass-fp-def} has a solution on $[0, \frac{\upsilon}{n+1}]$, then it is the unique solution on the entire $[0, 1]$ interval.
\end{lemma}

\begin{proof}
First, we will demonstrate that Eq. \ref{eq:mass-fp-def} has at most one solution on $[0, \frac{\upsilon}{n+1}]$.
Recall again that a fixed point occurs when $\mu(m)-m = 0$.
At $m=0$, we observe:
\begin{equation}
    \mu(0)-0 = \sum_{u \in X_\neq^n} \hat{p}_u^{ML} \Big(1-\hat{p}_u^{ML}\Big)^n > 0, \nonumber
\end{equation}
because $\hat{p}_u^{ML} \in (0, 1)$ when $|X_\neq^n| > 1$.
Then, since $\mu(m)-m$ is concave on $[0, \frac{\upsilon}{n+1}]$ (by Lemma \ref{lem:fixed-point-uniqueness-concavity}), $\mu(m)-m = 0$ at most once on $[0, \frac{\upsilon}{n+1}]$.
Next, suppose such a fixed point $m^* \in [0, \frac{\upsilon}{n+1}]$ does exist.
We know that $\mu(m) - m$ is: (i) positive at $m=0$, (ii) concave on $[0, \frac{\upsilon}{n+1}]$ (Lemma \ref{lem:fixed-point-uniqueness-concavity}), (iii) continuous (Lemma \ref{lem:fixed-point-existence}), and (iv) strictly decreasing in m on $[\frac{\upsilon}{n+1}, 1]$ (proof of Lemma \ref{lem:fixed-point-uniqueness-right-half}).
Then there cannot be any $m^\dagger \in (m^*, 1]$ such that $\mu(m^\dagger) - m^\dagger = 0$, and the result follows.
\end{proof}

\subsection{Results supporting uniqueness of the fixed point solution when $|X_\neq^n| > 1$ and $\upsilon > \frac{n}{2}$}

\begin{proposition} \label{prop:multiple-labels-no-zero}
    If $|X_\neq^n| > 1$, then $m = 0$ is not a solution to Eq. \ref{eq:mass-fp-def}.
\end{proposition}
\begin{proof}
If $|X_\neq^n| > 1$, then $\hat{p}_u^{ML} \in (0, 1)$ for all $u \in X_\neq^n$, so $\textsc{Unobserved}(0) = 0$ and $\textsc{Observed}(0) = \sum_{u \in X_\neq^n} \hat{p}_u^{ML} (1-\hat{p}_u^{ML})^n > 0$.
Then $\mu(0) \neq 0$, and the result follows.
\end{proof}

\begin{lemma} \label{lem:v-upper-bound-lemma}
Define $B(n) \triangleq \min\Bigg(\frac{n(n-2)^2 + n - 2}{2n}, \Big((\frac{n+1}{n-1})^{n-1} - e + 1\Big)n \Bigg)$.
If $n \in \mathbb{Z}_{\geq 4}$ and $\frac{n}{2} \leq \upsilon \leq B(n)$, then
$\frac{d}{dm} \Big(\frac{\mu(m)}{m} \Big) < 0$ on $m \in (0, 1)$.
\end{lemma}
\begin{proof}
To begin, let:
\begin{align}
    F(m) &\triangleq m\frac{d}{dm}\textsc{Observed}(m) - \textsc{Observed}(m) \nonumber \\
    &= \sum_{u \in X_\neq^n} \hat{p}^{ML}_u \Big(1-(1-m)\hat{p}^{ML}_u\Big)^{n-1} \Big((nm+1)(1-m)\hat{p}^{ML}_u - 1 \Big) \nonumber \\
    &\leq \max_{u \in X_\neq^n} \Big(1-(1-m)\hat{p}^{ML}_u\Big)^{n-1} \Big((nm+1)(1-m)\hat{p}^{ML}_u - 1 \Big) \label{eq:max-exchange} \\
    &\leq \max_{q \in [0, 1]} \Big(1-q\Big)^{n-1} \Big((nm+1)q - 1 \Big) \text{, and} \label{eq:max-exchange-q} \\
    G(m) &\triangleq m\frac{d}{dm}\textsc{Unobserved}(m) - \textsc{Unobserved}(m) = -\frac{nm^2}{\upsilon} \Big(1 - \frac{m}{\upsilon}\Big)^{n-1}. \nonumber
\end{align}
As in prior lemmas, Ineq. \ref{eq:max-exchange} applies the fact that $\hat{p}_u^{ML} >0$ for all $u \in X_\neq^n$ and $\sum_{u \in X_\neq^n} \hat{p}_u^{ML} = 1$, and Ineq. \ref{eq:max-exchange-q} applies the shorthand from Lemma \ref{lem:fixed-point-uniqueness-right-half}.
Differentiating the max's operand in Ineq. \ref{eq:max-exchange-q}:
\begin{align}
    \frac{d}{dq} \Bigg(  \Big(1-q \Big)^{n-1} \Big((nm+1)q - 1 \Big) \Bigg) &= -n (1-q)^{n-2} \Big(m(nq-1) + q - 1 \Big) \nonumber \\
    &= n (1-q)^{n-2} \Big(m - mnq - q + 1 \Big) \nonumber \\
    &= n (1-q)^{n-2} \Big(-q(mn+1) + (m+1) \Big), \label{eq:deriv-max-op}
\end{align}
with critical point $q=\frac{m+1}{mn+1}$ maximizing the operand of Ineq. \ref{eq:max-exchange-q}.
Evaluating Ineq.~\ref{eq:max-exchange-q} at $q=\frac{m+1}{mn+1}$:
\begin{equation}
    F(m) \leq \Big(1-\frac{m+1}{mn+1}\Big)^{n-1} \Big((nm+1)\frac{m+1}{mn+1} - 1 \Big) = \Big(\frac{m(n-1)}{mn+1}\Big)^{n-1} m. \nonumber
\end{equation}
Since:
\begin{align}
    \frac{d}{dm} \Big(\frac{\mu(m)}{m} \Big) &= \frac{m\mu^\prime(m)-\mu(m)}{m^2} = \frac{1}{m^2}\Bigg(m \Big(\frac{d}{dm}(\textsc{Observed}(m) + \textsc{Unobserved}(m)) \Big) - \mu(m) \Bigg), \nonumber
\end{align}
the condition $\frac{d}{dm} \Big(\frac{\mu(m)}{m} \Big) < 0$ is equivalent to $F(m) < - G(m)$, which is satisfied when:
\begin{align}
    \Big(\frac{m(n-1)}{mn+1}\Big)^{n-1} m < \frac{nm^2}{\upsilon} \Big(1 - \frac{m}{\upsilon}\Big)^{n-1} &\text{, equivalently} \label{eq:j-step1} \\
     \Big(\frac{m(n-1)}{mn+1}\Big)^{n-1} \Big(1 - \frac{m}{\upsilon}\Big)^{1-n} < \frac{nm}{\upsilon}  &\text{, equivalently} \label{eq:j-step2} \\
    (n-1) \log \Big(\frac{m(n-1)}{mn+1}\Big) + (1-n) \log \Big(1 - \frac{m}{\upsilon}\Big) < \log \Big(\frac{nm}{\upsilon} \Big)   &\text{, equivalently}\nonumber \\
    0 > (n-1) \log \Big(\frac{m(n-1)}{mn+1}\Big) + (1-n) \log \Big(1 - \frac{m}{\upsilon}\Big) - \log \Big(\frac{nm}{\upsilon} \Big) &\triangleq J(m, n, \upsilon). \label{eq:j-def}
\end{align}
Note $\frac{\partial}{\partial m} J(m, n, \upsilon) = \frac{(mn)^2 + \upsilon(n-2 - mn) + m}{m (mn+1) (\upsilon - m)} > 0$ for $m~\in~(0, \frac{n-2}{n}]$.
On $m \in [\frac{n-2}{n}, 1)$, we have $\frac{\partial}{\partial m} J(m, n, \upsilon) > \frac{(mn)^2 - 2 \upsilon + m}{m (mn+1) (\upsilon - m)} \geq \frac{(n-2)^2 - 2 \upsilon + \frac{n-2}{n}}{m (mn+1) (\upsilon - m)}$, which is non-negative when $\upsilon \leq \frac{n(n-2)^2 + n - 2}{2n}$.
So $J$ is increasing on $m \in (0, 1)$, from which we discern that Ineq.~\ref{eq:j-def} is true if $J(1, n, \upsilon) < 0$.
It suffices to show that Ineq.~\ref{eq:j-step2} is true when evaluated at $m=1$, i.e., $(\frac{n-1}{n+1} )^{n-1} (1 - \frac{1}{\upsilon} )^{1-n} < \frac{n}{\upsilon}$ or:
\begin{equation}
    \Big( \frac{n-1}{n+1} \Big)^{n-1} \Big(1 - \frac{1}{\upsilon} \Big)^{1-n} \frac{\upsilon}{n} < 1. \label{eq:j-step3}
\end{equation}

Taking the derivative of the left-hand side:
\begin{equation}
    \frac{\partial}{\partial \upsilon} \Bigg( \Big(\frac{n-1}{n+1} \Big)^{n-1} \Big(1 - \frac{1}{\upsilon} \Big)^{1-n} \frac{\upsilon}{n} \Bigg) = \frac{(\upsilon - n)(\frac{n-1}{n+1})^{n-1}(\frac{\upsilon}{\upsilon-1})^n}{n\upsilon}, \nonumber
\end{equation}
which is negative on $\upsilon \in (1, n)$ and positive for $\upsilon > n$.
So, provided an interval $[a, b]$ such that $b > a > 1$, if Ineq.~\ref{eq:j-step3} is satisfied at the endpoints $\upsilon = a$ and $\upsilon = b$, then Ineq.~\ref{eq:j-step3} is satisfied over the entire interval $[a, b]$.
Since $\upsilon \geq \frac{n}{2} > 1$ for $n \in \mathbb{Z}_{\geq 4}$, the left-hand side of Ineq.~\ref{eq:j-step3} at $\upsilon = \frac{n}{2}$ is:
\begin{equation}
    \Big( \frac{n-1}{n+1} \Big)^{n-1} \Big(1 - \frac{2}{n} \Big)^{1-n} \frac{1}{2} = \frac{1}{2}\Big(1 + \frac{2}{(n+1)(n-2)} \Big)^{n-1}, \nonumber
\end{equation}
which is less than $1$ for $n \geq 4$.
Next, we will identify a sufficient condition on $k \geq 1$ such that Ineq.~\ref{eq:j-step3} is satisfied when $k n \geq \upsilon \geq \frac{n}{2}$, observing that:
\begin{align}
\Big( \frac{n-1}{n+1} \Big)^{n-1} \Big(1 - \frac{1}{kn} \Big)^{1-n} \frac{kn}{n} &= k \Big( \frac{n-1}{n+1} \Big)^{n-1} \Big(1 + \frac{1}{kn-1}\Big)^{n-1} \nonumber \\
&< k \Big( \frac{n-1}{n+1} \Big)^{n-1} e^{(n-1)/(kn-1)}  \nonumber \\
&\leq k \Big( \frac{n-1}{n+1} \Big)^{n-1} e^{n/(kn)}  \nonumber \\
&= k \Big( \frac{n-1}{n+1} \Big)^{n-1} e^{1/k}.  \label{eq:almost-linear-step}
\end{align}
Finally, we employ the same secant line technique of Lemma~\ref{lem:fixed-point-uniqueness-all-labels-same-2}, noting that $e^{x}$ is convex and therefore bounded from above on the interval $x \in [0, 1]$ by the line connecting $(0, 1)$ and $(1, e)$.
So $e^{x} \leq (e-1)x + 1$ when $x \in [0, 1]$, and $e^{1/k} \leq (e-1)\frac{1}{k} + 1$ when $\frac{1}{k} \in [0, 1]$, hence $k e^{1/k} \leq e - 1 + k$ for $k \geq 1$.
So the right-hand side of Eq.~\ref{eq:almost-linear-step} is at most $1$ when $e - 1 + k \leq (\frac{n+1}{n-1} )^{n-1}$.

In summary, the condition $\frac{n}{2} \leq \upsilon \leq B(n)$ guarantees that both (i) $\upsilon \leq \frac{n(n-2)^2 + n - 2}{2n}$ and (ii) $\upsilon \leq ((\frac{n+1}{n-1} )^{n-1} - e + 1)n$ are true, so Ineq.~\ref{eq:j-step3} is satisfied, $J(1, n, \upsilon) < 0$, Ineq.~\ref{eq:j-step2} is satisfied, $F(m) < -G(m)$, and (finally) $\frac{d}{dm} \Big(\frac{\mu(m)}{m} \Big) < 0$, proving the result.
\end{proof}

\subsection{On the behavior of $B(n)$}

\subsubsection{Visualizing the acceptable $\upsilon$-region} \label{sec:v-clipping}
\begin{figure}[t]
    \centering
    \includegraphics[width=\textwidth]{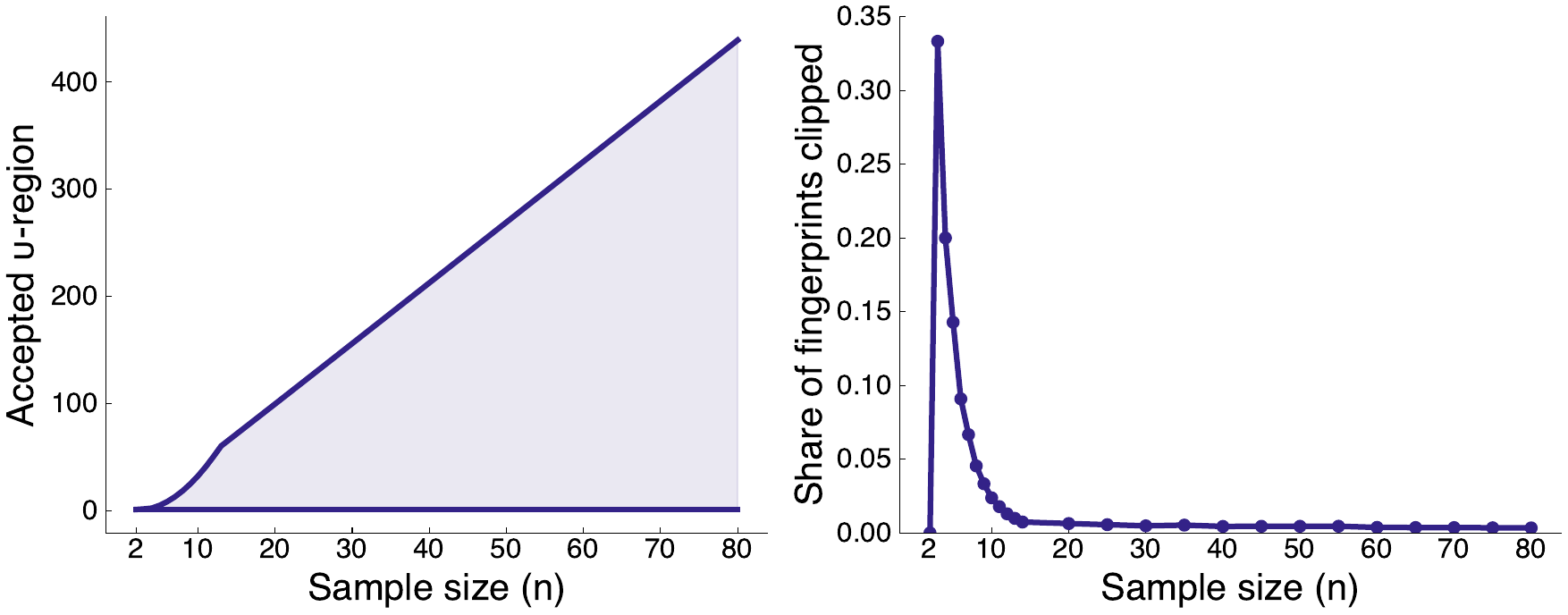}
    \caption{
        \textbf{Impacts of clipping the unobserved support size estimate.}
        \textbf{Left:} Visualizing the accepted region for $\upsilon$ (i.e., the region specified by Eq.~\ref{eq:def-v-clipping}), at various sample sizes.
        \textbf{Right:} At each sample size, we enumerate all fingerprints and report the share of excess support size estimates clipped by our selected upper bound.
        In both panels, $n$ iterates over all integers in $[2, 15]$, followed by every five integers in $[20, 80]$ (i.e., $20$, $25$, \dots, $80$).
    }
    \label{fig:fingerprints-clipped}
\end{figure}

Figure \ref{fig:fingerprints-clipped} provides intuition regarding the nature of our restrictions on $\upsilon$ (i.e., Eq.~\ref{eq:def-v-clipping}).
The left panel illustrates the associated region of permitted values of $\upsilon$.
In the right panel, we iterate over every possible fingerprint at various sample sizes, estimate the excess support size (i.e., unclipped $\upsilon$), and report the share of fingerprints for which our upper bound applies.
At $n=10$, only one fingerprint is clipped ($\Phi_1(X^n)=10$, i.e., all samples are singletons).
We highlight that there are nearly $15.8$ million distinct fingerprints at $n=80$, after which this analysis becomes somewhat intractable computationally.
For the same reason, the right panel visualization employs the bias-corrected Chao1 excess support estimator (i.e., the second term of Eq. \ref{eq:chao1-bias-corrected}, see Appendix~\ref{sec:support-size-details}); it has a closed form, so its substantially faster execution time facilitates the present analysis.

\subsubsection{$B(n)$ satisfies Theorem~\ref{thm:weak-consistency}} \label{sec:b-subquadratic}

For brevity, we omit a formal proof, in favor of a brief intuitive explanation.
Both arguments of $B(n)$ are increasing, and the former is $O(n^2)$.
As for the latter argument, $\lim_{n\rightarrow \infty}(\frac{n+1}{n-1})^{n-1} = e^2$, so the argument is $O(n)$.
Because we take the minimum of the two arguments, $B(n)$ is asymptotically linear (also visualized in Figure~\ref{fig:fingerprints-clipped}) and, therefore, satisfies the sub-quadratic condition of Theorem~\ref{thm:weak-consistency}.

\section{Additional implementation details}  \label{sec:implementation-details}

\subsection{Entropy estimation} \label{sec:entropy-details}

\begin{algorithm}[t!]
\caption{\textsc{SENECA}}\label{alg:sc}
\begin{algorithmic}[1]
\Require Set of label-occurrences $X^n$, initial support size estimate $\widehat{|\mathcal{X}|}$.
\Ensure Entropy estimate $\hat{\mathbb{H}}^{SC}$, missing mass estimate $\hat{M}^{SC}$

\State $P^{ML} \gets \{\frac{N_{u}(X^n)}{n} | u \in X^n_\ne\}$ \Comment{Empirical prevalences $\hat{p}_u^{ML}$ for observed labels}

\If{$|X_\neq^n|$ = 1}
    \State $\hat{M}^{SC} \gets 0$ \Comment{via Corollary~\ref{cor:fixed-point-uniqueness-all-labels-same}}
\Else
    \State Set $\upsilon$ according to Eq.~\ref{eq:def-v-clipping}.\Comment{Clipped estimate of excess support size}
    \State $m_0 \gets \frac{1}{2}$ \Comment{Initial guess}
    \State Perform fixed-point iteration on $\mu$ (Eq. \ref{eq:mu-fp}) to obtain $\hat{M}^{SC}$
    \If{$\hat{M}^{SC} \in (-5 \times 10^{-10}, 0)$}
        \State $\hat{M}^{SC} \gets 0$  \Comment{Floating-point correction (Appendix~\ref{sec:entropy-details})}
    \EndIf
\EndIf
\State Estimate entropy via Eq.~\ref{eq:entropy-sc} \Comment{Horvitz–Thompson estimator using $\hat{M}^{SC}$}
\end{algorithmic}
\end{algorithm}

Algorithm \ref{alg:sc} specifies the procedure for calculating \textsc{SENECA}.
Our method involves fixed-point iteration, for which we employ Steffensen's Method \citep{burden2011numerical}, via the SciPy library \citep{2020SciPy-NMeth}.
We begin the iteration with an initial value of $0.5$.
The true fixed point is guaranteed to lie in $[0, 1)$ (Lemma~\ref{lem:sc-open-interval}), but we observe occasional instances in which the optimizer terminates at an exceedingly small negative value in $(-5 \times 10^{-10}, 0)$, which we attribute to floating-point error; our implementation clips such values to $0$.

Most other entropy estimators - i.e., Grassberger, James-Stein, Bonachela, Chao-Shen, Chao-Wang-Jost, and NSB (see Appendix~\ref{sec:nsb}) -  are implemented via the \textit{infomeasure} library \citep{buth2025infomeasure}.
We note that the library's implementation of the Chao-Shen estimator applies an adjusted Good-Turing estimator $\hat{M}^{GT} = \frac{\Phi(X^n)-1}{n}$ when $\Phi(X^n)=n$, to avoid summands of $\frac{0 \log 0}{0}$.
The Valiant-Valiant entropy estimator is a linear programming-based method, implemented by translating the authors' MATLAB code to Python.

\subsection{Support size estimation} \label{sec:support-size-details}

\textsc{SENECA} uses an estimate of the support size to calculate the ``unobserved'' component of Eq. \ref{eq:mu-fp}.
Except where stated otherwise, we employ the regularized weighted Chebyshev (RWC) estimator of \citet{chien2019regularized, rana2022small}, which is intended for small sample sizes.
In particular, we employ their RWC-S variant, which exhibits the best worst-case performance in the source material \citep{chien2019regularized}.
In this section, we will briefly describe this and three other options and illustrate that our results are generally not sensitive to our choice of support size estimator.

The straightforward-to-calculate Chao1 support estimator employs two FoFs \citep{chao1984nonparametric}:
\begin{equation}\label{eq:chao1-og}
\widehat{|\mathcal{X}|}_{Chao1,1} = |\mathcal{X}_\neq^n| + \frac{\Phi_1(X^n)^2}{2\Phi_2(X^n)}.
\end{equation}
In population ecology, where support size is often known as ``species richness,'' Chao1 tends to underestimate its quantity of interest, though it performs well for power law distributions \citep{elena2025comprehensive}.
A variation that avoids division by zero in the absence of ``doubletons'' is given by:
\begin{equation} \label{eq:chao1-bias-corrected}
\widehat{|\mathcal{X}|}_{Chao1,2} = |\mathcal{X}_\neq^n| + \frac{\Phi_1(X^n)\Big(\Phi_1(X^n)-1\Big)}{2\Big(\Phi_2(X^n)+1\Big)},
\end{equation}
a ``bias-corrected'' version \citep{chao2005speciesrichness} that is "now widely preferred" to Eq. \ref{eq:chao1-og} \citep{deng2024nonparametric}.
Several more recent methods, such as the aforementioned RWC \citep{chien2019regularized, rana2022small}, that of \citet{wu2019chebyshev} (Wu-Yang), and that of \citet{valiant2017estimating} (Valiant-Valiant), rely on more involved techniques at the expense of computational complexity.
We discuss pertinent hyperparameters for the latter three below, but the remaining implementation details are beyond the scope of this work, and we refer interested readers to their corresponding publications.

The Wu-Yang, RWC, and Valiant-Valiant support size estimators require a regularization parameter $p_{min}$, corresponding to the assumption that all labels have true probability mass at least $p_{min} = \min \{p_u \forall u \in \mathcal{X} \}$.
Since $p_{min}$ is generally unknown, we apply a heuristic inspired by that of \citet{valiant2017estimating} and Thm. 1 of \citet{wu2019chebyshev}:
\begin{equation*} \label{eq:min_prob_heuristic}
\hat{p}_{min} = \begin{cases}
        \Big(\frac{\argmin_{i : \Phi_i(X^n) > 0}  \Phi_i(X^n)}{n}\Big)^2, & \text{if }\Phi_1(X^n)=0\\
        \frac{1}{\argmax_{\gamma \in \mathbb{Z}} (n > \frac{\gamma}{\log \gamma})}, & \text{otherwise.}
     \end{cases}
\end{equation*}
The heuristic for the zero-singletons cases is drawn directly from the source code of \citet{valiant2017estimating}.
The other heuristic arises from the result of \citet{wu2019chebyshev} that consistent estimation of $|\mathcal{X}|$ is only possible with at least $\frac{|\mathcal{X}|}{\log |\mathcal{X}|}$ samples; since a uniform distribution over $|\mathcal{X}|$ maximizes the minimum non-zero mass, we select the largest integer value of $\gamma$ such that $n > \frac{\gamma}{\log \gamma}$.
We employ the above heuristic for the RWC and Wu-Yang estimators, which do not have their own heuristic for the regularization parameter.
The Valiant-Valiant estimator is implemented with its own original heuristic.
The Wu-Yang method requires two additional hyperparameters: $\gamma, c_0, \text{ and } c_1$, for which we employ the default settings from the source material's experiments (i.e., $c_0=0.45$, $c_1=0.5$) \citep{wu2019chebyshev}.
The RWC support estimator admits additional hyperparameters $c_0$, dictating the polynomial degree, and $nC$, dictating the number of grid points.
We apply the default of $c_0=0.558$ from the paper and $nC=1000$, as recommended for $n \leq 1000$ in the corresponding source code \citep{chien2019regularized, rana2022small}.
Finally, we clip the lower bound of all support size estimates used in \textsc{SENECA} to ensure that the estimated number of unobserved labels is at least one.

\begin{figure*}[t!]
    \centering
    \includegraphics[width=1.0\textwidth]{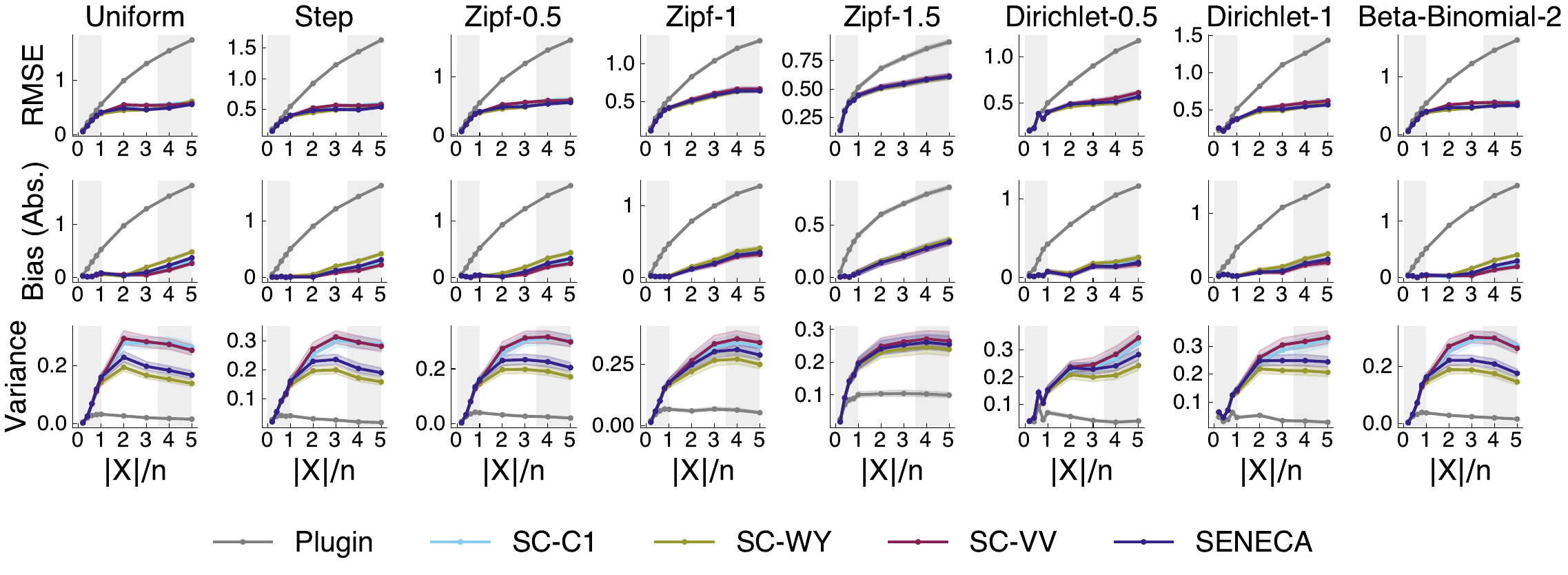}
    \caption{
    \textbf{Support size estimator ablation for \textsc{SENECA} implementations.} 
    The entropy estimators shown are the plugin estimator (Plugin), \textsc{SENECA} with bias-corrected Chao1 support estimation (SC-C1), \textsc{SENECA} with Wu-Yang support estimation (SC-WY), \textsc{SENECA} with Valiant-Valiant support estimation (SC-VV), and \textsc{SENECA} with RWC support estimation, as used in the remainder of this work (\textsc{SENECA}).
    Each item's capture probability is determined uniformly over its label's prevalence probability mass.
    For each of $8$ families of label-prevalence distribution, we generate $100$ items per label.
    The number of labels $|\mathcal{X}|$ varies from $2$ to $50$ across trials.
    For each trial, $n=10$ item-capture samples are drawn, and the missing mass of the label-prevalence distribution is estimated according to each method.
    Results are averaged over $1000$ trials per label count.
    Colored regions about each line plot depict $95\%$ bias-corrected and accelerated confidence intervals, based on $1000$ bootstrap replications.
    In each plot, leftmost grey region corresponds to the well-sampled regime (i.e., $|\mathcal{X}|= \frac{n}{5}, \frac{2n}{5}, \frac{3n}{5}, \frac{4n}{5}, n$), where items to the right correspond to the under-sampled regime (i.e., $|\mathcal{X}|= 2n, 3n, 4n, 5n$).
    The rightmost grey region corresponds to the support-risky regime, where no consistent estimator for support size exists.
    }
    \label{fig:support-estimator-ablation-n10}
\end{figure*}

In Figure \ref{fig:support-estimator-ablation-n10}, we consider the bias-corrected Chao1, Wu-Yang, RWC, and Valiant-Valiant methods as drop-in support size estimators in \textsc{SENECA}, alongside plugin entropy for reference.
We find that alternative choices of support estimator lead to roughly interchangeable results to those presented in the main body of this work.
To gain intuition on this observation, we iterate over all possible fingerprints at $n=10$ and estimate the support size with each of the four aforementioned methods.
Averaged over these fingerprints, the mean pairwise difference between distinct methods' estimates is approximately $0.384 \cdot |X_\neq^n|$.
Such relatively small differences appear to only minutely impact \textsc{SENECA}'s accuracy, since the other factors (e.g., observed label prevalences) are held fixed.
Minor differences are apparent in the bias-variance decomposition. 
For instance, the variants using Chao1 and Valiant-Valiant tend to exhibit slightly lower bias and higher variance in the under-sampled regime, and vice-versa for the Wu-Yang and RWC variants.

\subsection{Text generation details} \label{sec:text-generation-details}

For each model-dataset pair, we sample $n=10$ responses of up to $100$ tokens each, at a temperature of $\tau=1.0$ with nucleus sampling $p=0.9$, following \citet{Farquhar2024}.
Following \citet{kuhn2023semantic}, log-probabilities for these high-temperature sequences are length-normalized (i.e., by number of tokens).
To support the G-NLL method, we also sample a response with temperature $\tau=0$ and nucleus sampling $p=1.0$ \citep{aichberger2026rethinking}; log-probabilities for this greedy sequence are \textit{not} length-normalized, which the authors contend works best.
LLM sampling was performed vLLM \citep{kwon2023efficient} serverless endpoints hosted on Runpod, with KV cache reuse where possible.
All LLMs were hosted on NVIDIA RTX 4090 GPUs (24 GB VRAM), with the exception of Phi-4, which was hosted on NVIDIA A40 instances (48 GB VRAM), and Granite-4.0-H-Small, which ran on RTX PRO 6000s (96 GB VRAM).
Following \citet{Farquhar2024, nikitin2024kernel}, we pre-pend all queries with the following to induce sentence-length responses:

\begin{promptbox}{Template for Pre-pending Prompts}{
Answer the following question in a single brief but complete sentence: 
}
\end{promptbox}

Two LLMs required custom prompt templates to induce well-formed responses.
For Phi-4, we used the following, based on recommendation in its corresponding HuggingFace listing:\footnote{\url{https://huggingface.co/microsoft/phi-4}}
\begin{promptbox}{Prompt Template for Phi-4}
    <|im\_start|>user\\
    \{prompt\}<|im\_end|>\\
    <|im\_start|>assistant
\end{promptbox}
We used the following for Granite-4.0-H-Small, based on recommendation in its HuggingFace listing:\footnote{\url{https://huggingface.co/ibm-granite/granite-4.0-h-small}}
\begin{promptbox}{Prompt Template for Granite-4.0-H-Small}
    <|start\_of\_role|>user<|end\_of\_role|>\\
    \{prompt\}<|end\_of\_text|>\\
    <|start\_of\_role|>assistant<|end\_of\_role|>
\end{promptbox}

\subsection{Semantic clustering} \label{sec:semantic-clustering-details}

To assign ``semantic equivalence class`` labels to LLM responses, we employ the bidirectional entailment clustering (BEC) procedure of \citet{kuhn2023semantic} with ``strict entailment``, where responses are greedily assigned to the same class if they are deemed semantically equivalent, and instantiate new classes otherwise.
To assess semantic equivalence, we employ a DeBERTa-based NLI model, with the prompt pre-pended to each response; in particular, we use a fine-tuned version of \textit{DeBERTaV3} \citep{he2021debertav3}.\footnote{\url{https://huggingface.co/cross-encoder/nli-deberta-v3-base}}
Before prompting the NLI model, we remove the pre-pended portion of the query (i.e., ``Answer the following question in a single brief but complete sentence:'') to limit extraneous context.

\subsection{LLM-as-judge} \label{sec:llm-judge-details}

To assess the correctness of LLM responses, we compare reference answers to a single ``best guess'' response sampled at $\tau=0.1$, following prior work \citep{Farquhar2024, nikitin2024kernel, nguyen-etal-2025-beyond}.
\citet{mccabe2026estimating} uses the same LLM-as-judge prompt as \citet{lin2023generating}, but adds an in-context example and XML-style tags to improve rating parsing.
We add an additional instruction to limit verbose responses:
\begin{promptbox}{Prompt for Assessing Correctness of LLM Generations}
    Rate the level of consistency between the answer to the question and the reference answer, from 0 to 100. Output the float rating inside $<$rating$></$rating$>$ tags. Only provide the rating, without any additional commentary.
    \\
    Here is an example output:
    
    Question: What is the capital of France?
    Reference: Paris is the capital of France.
    Answer: The capital of France is Paris.
    $<$rating$>$100$</$rating$>$
    \\
    Now rate the following:
    \\
    Question: \{question\}
    
    Reference: \{groundtruth\}
    
    Answer: \{pred\}
\end{promptbox}

If multiple reference examples are available, we retain the highest such rating among them.
Like \citet{lin2023generating}, we treat a response as correct if its ultimate rating exceeds $70/100$.
Our judge model is the $32$B-parameter Granite-4.0-H-Small, generating at most $100$ tokens using a temperature of $\tau=0.1$.
Like in the semantic clustering, we remove the pre-pended portion of the query before prompting the judge.
In Appendix \ref{sec:more-llm-results}, we consider an alternative LLM judge also found in related work, with qualitatively interchangeable results.

\subsection{LLM uncertainty quantification methods} \label{sec:llm-uq-details}

In our notation, variability-based uncertainty quantification techniques obtain a set $A^n$ of $n$ LLM responses to a query $q$.
The semantic clustering algorithm (e.g., BEC, Appendix \ref{sec:semantic-clustering-details}) serves as the classification function $h$ that produces the set of label occurrences $X^n$ and corresponding set of unique observed labels $X_\neq^n$.
Discrete semantic entropy (DSE) simply reports the plugin entropy estimate over $X_\neq^n$ \citep{Farquhar2024}.
The ``white-box'' semantic entropy (SE) replaces the empirical label frequencies of plugin entropy with probability estimates based on sequence log-probabilities \citep{kuhn2023semantic}:\footnote{We have detailed DSE and SE in reverse-chronological order of publication for descriptive convenience.}
\begin{align*}
SE(q) &= - \sum_{x \in X_\neq^n} \hat{p}(x | q) \log \hat{p}(x | q), \text{ where} \\
\hat{p}(x | q) &= \frac{\sum_{a} \vmathbb{1}_{h(a)=x} p(a | q)}{\sum_{a} p(a | q)}.
\end{align*}

Kernel language entropy (KLE) does not explicitly use semantic clusters.
Instead, it constructs an NLI-based semantic graph of responses, applies a kernel to the graph's Laplacian, and reports the von Neumann entropy of the resultant matrix.
Letting $f_{ij}=1 \text{, } 0 \text{, or } 0.5$ if the NLI model gauges entailment, contradiction, or neutral, respectively, for  response pair $a_i, a_j$, the semantic graph has $n \times n$ weighted adjacency matrix whose entries are given by $w_{ij}=f_{ij}+f_{ji}$.
Our implementation uses the heat kernel $K_t = exp(-tL)$, which the authors report favorably, with ``reasonable'' default hyperparameter $t=0.3$ \citep{nikitin2024kernel}.

Like KLE, $U_{EigV}$ also relies on an NLI-based semantic graph, though its particular construction differs.
Using the NLI model, the bidirectional entailment probability $e_{ij}$ is calculated for each response pair $a_i, a_j$.
The matrix of weights $w_{ij} = \frac{e_{ij}+e_{ji}}{2}$ serves as the weighted adjacency matrix of a semantic graph.
The metric is then calculated based on the graph Laplacian's spectrum $\Lambda(L)$:
\begin{equation*}
U_{EigV} = \sum_{\lambda \in \Lambda(L)} \max(0, 1-\lambda),
\end{equation*}
which is interpreted as an estimate of the support size $|\mathcal{X}|$ \citep{lin2023generating}.
\citet{mccabe2026estimating} constructs a ``hybrid'' support size estimate based on $U_{EigV}$ and the Good-Turing missing mass estimator (i.e., $\hat{M}^{GT}=\frac{\Phi_1(X^n)}{n}$):
\begin{equation}
    \hat{S}^{Hybrid} = \begin{cases}
            U_{EigV} & \Phi_1(X^n)=n \\
            \max \Big(\frac{|X_\neq^n| n}{n - \Phi_1 (X^n)}, U_{EigV}\Big) & \text{otherwise.}
        \end{cases}
\end{equation}
HE (i.e., the hybrid entropy method of \citet{mccabe2026estimating}) is then a Chao-Shen/Horvitz-Thompson-style entropy estimator using $\frac{|X_\neq^n|}{\hat{S}^{Hybrid}}$ for coverage adjustment.
For consistency, we use the same NLI model for all relevant methods (i.e., those relying on semantic clustering or semantic graphs).

Predictive entropy (PE) and surprise are straightforward functions of the log-probabilities of sampled sequences.
We calculate the former by the average of negative length-normalized sequence log-probabilities \citep{kadavath2022language} and the latter by average exponentiated (base $2$) negative length-normalized sequence log-probabilities \citep{ismayilzada-etal-2025-creative}.
G-NLL calls for one response to be drawn using greedy decoding, reporting the corresponding raw (i.e., not length-normalized) negative sequence log-probability \citep{aichberger2026rethinking}.

\section{Additional entropy estimation results} \label{sec:more-entropy-results}

\begin{figure*}[t!]
    \centering
    \includegraphics[width=1.0\textwidth]{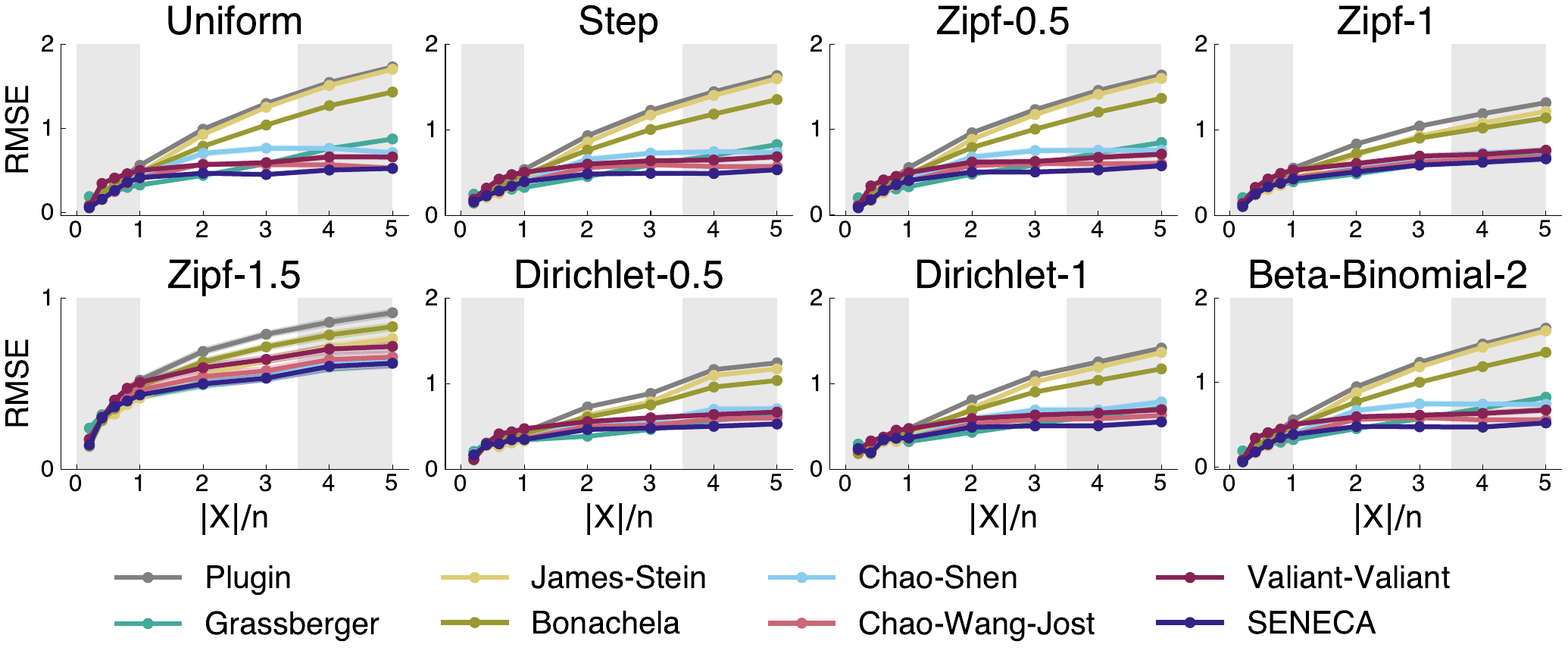}
    \caption{
    \textbf{Entropy estimation in small-sample item capture (all estimators).} 
    For each of $8$ families of label-prevalence distribution, each item's capture probability is determined uniformly over its label's prevalence probability mass.
    The number of labels $|\mathcal{X}|$ varies from $2$ to $50$ across trials.
    For each trial, $n=10$ items are captured, and the entropy of the label-prevalence distribution is estimated according to each of seven methods. 
    Results are averaged over $1000$ trials per label count.
    Colored regions about each line plot depict $95\%$ bias-corrected and accelerated confidence intervals, based on $1000$ bootstrap replications.
    In each plot, the leftmost grey region corresponds to the well-sampled regime (i.e., $|\mathcal{X}|= \frac{n}{5}, \frac{2n}{5}, \frac{3n}{5}, \frac{4n}{5}, n$), where items to the right correspond to the under-sampled regime (i.e., $|\mathcal{X}|= 2n, 3n, 4n, 5n$).
    The rightmost grey region corresponds to the support-risky regime, where no consistent estimator for support size exists.
    Empirical estimator variances are shown in the corresponding Figure \ref{fig:entropy-bias-variance-n10}.
    }
    \label{fig:entropy-rmse-n10-full}
\end{figure*}

We expand Figure \ref{fig:entropy-rmse-n10} to display all considered entropy estimators in Figure \ref{fig:entropy-rmse-n10-full}, where \textsc{SENECA} continues to achieve low RMSE in the under-sampled regime.
In Table \ref{tab:entropy-rmse-n20}, we repeat the analysis of Table \ref{tab:entropy-rmse-n10}, using a twice-large sample size ($n=20$). 
Results are qualitatively similar to the $n=10$ setting, especially in the under-sampled regime, where \textsc{SENECA} achieves the lowest or second-lowest averaged error in all cases but one.
When \textsc{SENECA} is not the lowest-error estimator (e.g., in the well-sampled regime), its error is within a few percentage points of the lowest.
With this larger sample size, the James-Stein estimator is no longer top-performing in the well-sampled regime, and the Chao-Shen estimator improves relatively.

\begin{figure*}[t!]
    \centering
    \includegraphics[width=1.0\textwidth]{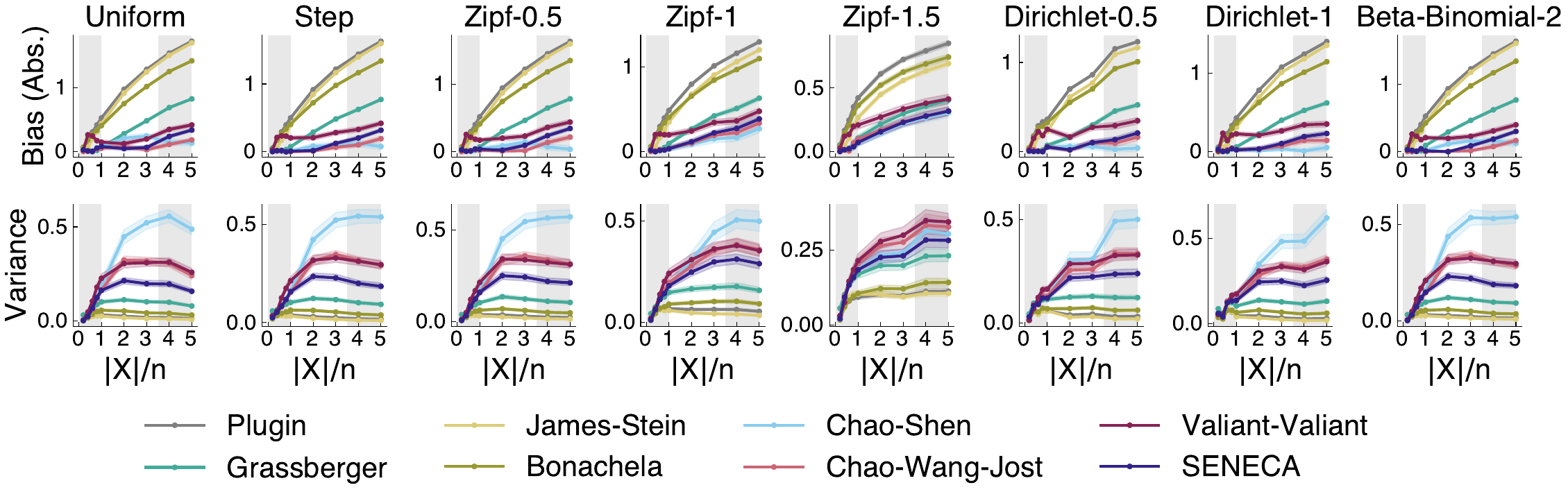}
    \caption{
    \textbf{Empirical bias-variance decomposition of entropy estimation in small-sample item capture.} 
    For each of $8$ families of label-prevalence distribution, we generate $100$ items per label.
    Each item's capture probability is determined uniformly over its label's prevalence probability mass.
    The number of labels $|\mathcal{X}|$ varies from $2$ to $50$ across trials.
    For each trial, $n=10$ item-capture samples are drawn, and the entropy of the label-prevalence distribution is estimated according to each of seven methods. 
    Results are averaged over $1000$ trials per label count.
    Colored regions about each line plot depict $95\%$ bias-corrected and accelerated confidence intervals, based on $1000$ bootstrap replications.
    In each plot, leftmost grey region corresponds to the well-sampled regime (i.e., $|\mathcal{X}|= \frac{n}{5}, \frac{2n}{5}, \frac{3n}{5}, \frac{4n}{5}, n$), where items to the right correspond to the under-sampled regime (i.e., $|\mathcal{X}|= 2n, 3n, 4n, 5n$).
    The rightmost grey region corresponds to the support-risky regime, where no consistent estimator for support size exists.
    }
    \label{fig:entropy-bias-variance-n10}
\end{figure*}

\begin{table*}[t!]
    \caption{
        \textbf{Regime-averaged error for entropy estimation in small-sample item capture.}
        We repeat the analysis of Table \ref{tab:entropy-rmse-n10} using $n=20$ instead of $n=10$. 
        We calculate the root-mean-squared error (RMSE $\downarrow$) for each estimator and average results within the well-sampled (Well, i.e., $|\mathcal{X}| \leq n$) and under-sampled (Under, i.e., $|\mathcal{X}| > n$) regimes (Reg.).
        The estimators are Plugin (ML), Grassberger (G), James-Stein (JS), Bonachela (B), Chao-Shen (CS), Chao-Wang-Jost (CWJ), Valiant-Valiant (VV), and \textsc{SENECA} (highlighted in blue).
        The distributions are Uniform, Step, Zipf-0.5, Zipf-1, Zipf-1.5, Dirichlet-0.5 (Diri-0.5), Dirichlet-1 (Diri-1), and Beta-Binomial-2 (BB-2).
        The lowest and second-lowest values in each row, prior to rounding, are bold and underlined, respectively.
        Bootstrapped confidence intervals (grey text) are slightly more conservative than $95\%$, as described in Section \ref{sec:entropy-experiments-desc}.
        In parentheses, we show the relative difference between \textsc{SENECA}'s averaged RMSE and the lowest value in each row.
    }
    \label{tab:entropy-rmse-n20}
    \centering
    \small
    \setlength{\tabcolsep}{4.25pt}
    \begin{tabular}{l|c|ccccccc>{\columncolor{lightindigocolor!10}}c}
\\ \toprule
Distr. & Reg. & ML & G & JS & B & CS & CWJ & VV & SENECA \\
\midrule
\multirow{2}{*}{Uniform} & Well & 0.31 \tiny{\textcolor{gray}{$\pm$.00}} & 0.17 \tiny{\textcolor{gray}{$\pm$.00}} & 0.22 \tiny{\textcolor{gray}{$\pm$.00}} & \textbf{0.16} \tiny{\textcolor{gray}{$\pm$.00}} & \underline{0.17} \tiny{\textcolor{gray}{$\pm$.00}} & 0.17 \tiny{\textcolor{gray}{$\pm$.00}} & 0.29 \tiny{\textcolor{gray}{$\pm$.00}} & 0.18 \tiny{\textcolor{gray}{$\pm$.00}} (7.2\%) \\
 & Under & 1.40 \tiny{\textcolor{gray}{$\pm$.00}} & 0.65 \tiny{\textcolor{gray}{$\pm$.01}} & 1.35 \tiny{\textcolor{gray}{$\pm$.00}} & 0.65 \tiny{\textcolor{gray}{$\pm$.01}} & 0.65 \tiny{\textcolor{gray}{$\pm$.02}} & 0.58 \tiny{\textcolor{gray}{$\pm$.01}} & \underline{0.56} \tiny{\textcolor{gray}{$\pm$.01}} & \textbf{0.43} \tiny{\textcolor{gray}{$\pm$.01}} (0.0\%) \\
\midrule
\multirow{2}{*}{Step} & Well & 0.32 \tiny{\textcolor{gray}{$\pm$.00}} & 0.20 \tiny{\textcolor{gray}{$\pm$.00}} & 0.23 \tiny{\textcolor{gray}{$\pm$.00}} & \underline{0.20} \tiny{\textcolor{gray}{$\pm$.00}} & \textbf{0.20} \tiny{\textcolor{gray}{$\pm$.00}} & 0.20 \tiny{\textcolor{gray}{$\pm$.00}} & 0.29 \tiny{\textcolor{gray}{$\pm$.00}} & 0.20 \tiny{\textcolor{gray}{$\pm$.00}} (2.4\%) \\
 & Under & 1.32 \tiny{\textcolor{gray}{$\pm$.00}} & 0.61 \tiny{\textcolor{gray}{$\pm$.01}} & 1.26 \tiny{\textcolor{gray}{$\pm$.00}} & 0.62 \tiny{\textcolor{gray}{$\pm$.01}} & 0.59 \tiny{\textcolor{gray}{$\pm$.02}} & \underline{0.54} \tiny{\textcolor{gray}{$\pm$.01}} & 0.55 \tiny{\textcolor{gray}{$\pm$.01}} & \textbf{0.43} \tiny{\textcolor{gray}{$\pm$.01}} (0.0\%) \\
\midrule
\multirow{2}{*}{Zipf-0.5} & Well & 0.31 \tiny{\textcolor{gray}{$\pm$.00}} & 0.19 \tiny{\textcolor{gray}{$\pm$.00}} & 0.21 \tiny{\textcolor{gray}{$\pm$.00}} & \underline{0.19} \tiny{\textcolor{gray}{$\pm$.00}} & \textbf{0.18} \tiny{\textcolor{gray}{$\pm$.00}} & 0.19 \tiny{\textcolor{gray}{$\pm$.00}} & 0.29 \tiny{\textcolor{gray}{$\pm$.00}} & 0.19 \tiny{\textcolor{gray}{$\pm$.00}} (3.0\%) \\
 & Under & 1.31 \tiny{\textcolor{gray}{$\pm$.00}} & 0.62 \tiny{\textcolor{gray}{$\pm$.01}} & 1.25 \tiny{\textcolor{gray}{$\pm$.00}} & 0.63 \tiny{\textcolor{gray}{$\pm$.01}} & 0.59 \tiny{\textcolor{gray}{$\pm$.02}} & \underline{0.55} \tiny{\textcolor{gray}{$\pm$.01}} & 0.55 \tiny{\textcolor{gray}{$\pm$.01}} & \textbf{0.46} \tiny{\textcolor{gray}{$\pm$.01}} (0.0\%) \\
\midrule
\multirow{2}{*}{Zipf-1} & Well & 0.33 \tiny{\textcolor{gray}{$\pm$.00}} & 0.24 \tiny{\textcolor{gray}{$\pm$.00}} & 0.23 \tiny{\textcolor{gray}{$\pm$.00}} & 0.24 \tiny{\textcolor{gray}{$\pm$.00}} & \textbf{0.22} \tiny{\textcolor{gray}{$\pm$.00}} & 0.25 \tiny{\textcolor{gray}{$\pm$.00}} & 0.30 \tiny{\textcolor{gray}{$\pm$.00}} & \underline{0.22} \tiny{\textcolor{gray}{$\pm$.00}} (0.1\%) \\
 & Under & 1.05 \tiny{\textcolor{gray}{$\pm$.01}} & 0.57 \tiny{\textcolor{gray}{$\pm$.01}} & 0.88 \tiny{\textcolor{gray}{$\pm$.01}} & 0.61 \tiny{\textcolor{gray}{$\pm$.01}} & \underline{0.51} \tiny{\textcolor{gray}{$\pm$.01}} & \textbf{0.50} \tiny{\textcolor{gray}{$\pm$.01}} & 0.54 \tiny{\textcolor{gray}{$\pm$.01}} & 0.52 \tiny{\textcolor{gray}{$\pm$.01}} (2.8\%) \\
\midrule
\multirow{2}{*}{Zipf-1.5} & Well & 0.33 \tiny{\textcolor{gray}{$\pm$.00}} & 0.28 \tiny{\textcolor{gray}{$\pm$.00}} & 0.27 \tiny{\textcolor{gray}{$\pm$.00}} & 0.27 \tiny{\textcolor{gray}{$\pm$.00}} & \underline{0.26} \tiny{\textcolor{gray}{$\pm$.00}} & 0.29 \tiny{\textcolor{gray}{$\pm$.00}} & 0.30 \tiny{\textcolor{gray}{$\pm$.00}} & \textbf{0.26} \tiny{\textcolor{gray}{$\pm$.00}} (0.0\%) \\
 & Under & 0.70 \tiny{\textcolor{gray}{$\pm$.01}} & 0.47 \tiny{\textcolor{gray}{$\pm$.01}} & 0.55 \tiny{\textcolor{gray}{$\pm$.01}} & 0.50 \tiny{\textcolor{gray}{$\pm$.01}} & \textbf{0.43} \tiny{\textcolor{gray}{$\pm$.01}} & 0.47 \tiny{\textcolor{gray}{$\pm$.01}} & 0.51 \tiny{\textcolor{gray}{$\pm$.01}} & \underline{0.43} \tiny{\textcolor{gray}{$\pm$.01}} (0.7\%) \\
\midrule
\multirow{2}{*}{Diri-0.5} & Well & 0.28 \tiny{\textcolor{gray}{$\pm$.00}} & 0.22 \tiny{\textcolor{gray}{$\pm$.00}} & \textbf{0.22} \tiny{\textcolor{gray}{$\pm$.00}} & 0.23 \tiny{\textcolor{gray}{$\pm$.00}} & \underline{0.22} \tiny{\textcolor{gray}{$\pm$.00}} & 0.23 \tiny{\textcolor{gray}{$\pm$.00}} & 0.28 \tiny{\textcolor{gray}{$\pm$.00}} & 0.22 \tiny{\textcolor{gray}{$\pm$.00}} (3.4\%) \\
 & Under & 0.98 \tiny{\textcolor{gray}{$\pm$.01}} & 0.47 \tiny{\textcolor{gray}{$\pm$.01}} & 0.86 \tiny{\textcolor{gray}{$\pm$.00}} & 0.49 \tiny{\textcolor{gray}{$\pm$.01}} & \underline{0.44} \tiny{\textcolor{gray}{$\pm$.01}} & 0.46 \tiny{\textcolor{gray}{$\pm$.01}} & 0.48 \tiny{\textcolor{gray}{$\pm$.01}} & \textbf{0.43} \tiny{\textcolor{gray}{$\pm$.01}} (0.0\%) \\
\midrule
\multirow{2}{*}{Diri-1} & Well & 0.29 \tiny{\textcolor{gray}{$\pm$.00}} & 0.21 \tiny{\textcolor{gray}{$\pm$.00}} & \underline{0.20} \tiny{\textcolor{gray}{$\pm$.00}} & 0.21 \tiny{\textcolor{gray}{$\pm$.00}} & \textbf{0.20} \tiny{\textcolor{gray}{$\pm$.00}} & 0.21 \tiny{\textcolor{gray}{$\pm$.00}} & 0.28 \tiny{\textcolor{gray}{$\pm$.00}} & 0.20 \tiny{\textcolor{gray}{$\pm$.00}} (1.1\%) \\
 & Under & 1.13 \tiny{\textcolor{gray}{$\pm$.00}} & 0.53 \tiny{\textcolor{gray}{$\pm$.01}} & 1.05 \tiny{\textcolor{gray}{$\pm$.00}} & 0.53 \tiny{\textcolor{gray}{$\pm$.01}} & 0.51 \tiny{\textcolor{gray}{$\pm$.01}} & \underline{0.50} \tiny{\textcolor{gray}{$\pm$.01}} & 0.52 \tiny{\textcolor{gray}{$\pm$.01}} & \textbf{0.44} \tiny{\textcolor{gray}{$\pm$.01}} (0.0\%) \\
\midrule
\multirow{2}{*}{BB-2} & Well & 0.31 \tiny{\textcolor{gray}{$\pm$.00}} & 0.18 \tiny{\textcolor{gray}{$\pm$.00}} & 0.21 \tiny{\textcolor{gray}{$\pm$.00}} & 0.18 \tiny{\textcolor{gray}{$\pm$.00}} & \textbf{0.17} \tiny{\textcolor{gray}{$\pm$.00}} & \underline{0.18} \tiny{\textcolor{gray}{$\pm$.00}} & 0.29 \tiny{\textcolor{gray}{$\pm$.00}} & 0.18 \tiny{\textcolor{gray}{$\pm$.00}} (3.5\%) \\
 & Under & 1.32 \tiny{\textcolor{gray}{$\pm$.00}} & 0.61 \tiny{\textcolor{gray}{$\pm$.01}} & 1.27 \tiny{\textcolor{gray}{$\pm$.00}} & 0.61 \tiny{\textcolor{gray}{$\pm$.01}} & 0.61 \tiny{\textcolor{gray}{$\pm$.02}} & 0.56 \tiny{\textcolor{gray}{$\pm$.01}} & \underline{0.55} \tiny{\textcolor{gray}{$\pm$.01}} & \textbf{0.44} \tiny{\textcolor{gray}{$\pm$.01}} (0.0\%) \\
\bottomrule
\end{tabular}
\end{table*}

\section{On the ``natural'' assumption} \label{sec:natural}

In our problem statement, one does not know a priori the shape of the unobserved portion of the data-generating distribution.
For this reason, we employ a ``clipped'' estimate of the unobserved support size and assign all elements of the estimated support equal prevalences.
In practice, these support size and prevalence estimates may diverge from ground truth, but the results of Section~\ref{sec:numerical} do not indicate that this is particularly problematic.
First, our analysis of the ``support-risky'' regime indicates that \textsc{SENECA} performs favorably even where no consistent estimator for support size exists.
Additionally, the ``equal probabilities'' assumption fails by construction for several distributions therein - e.g., Zipf-$1.5$, where \textsc{SENECA} nonetheless has the lowest regime-averaged error in the under-sampled setting (Table~\ref{tab:entropy-rmse-n10}).

\section{Comparison to the NSB estimator} \label{sec:nsb}

\begin{figure*}[t!]
    \centering
    \includegraphics[width=1.0\textwidth]{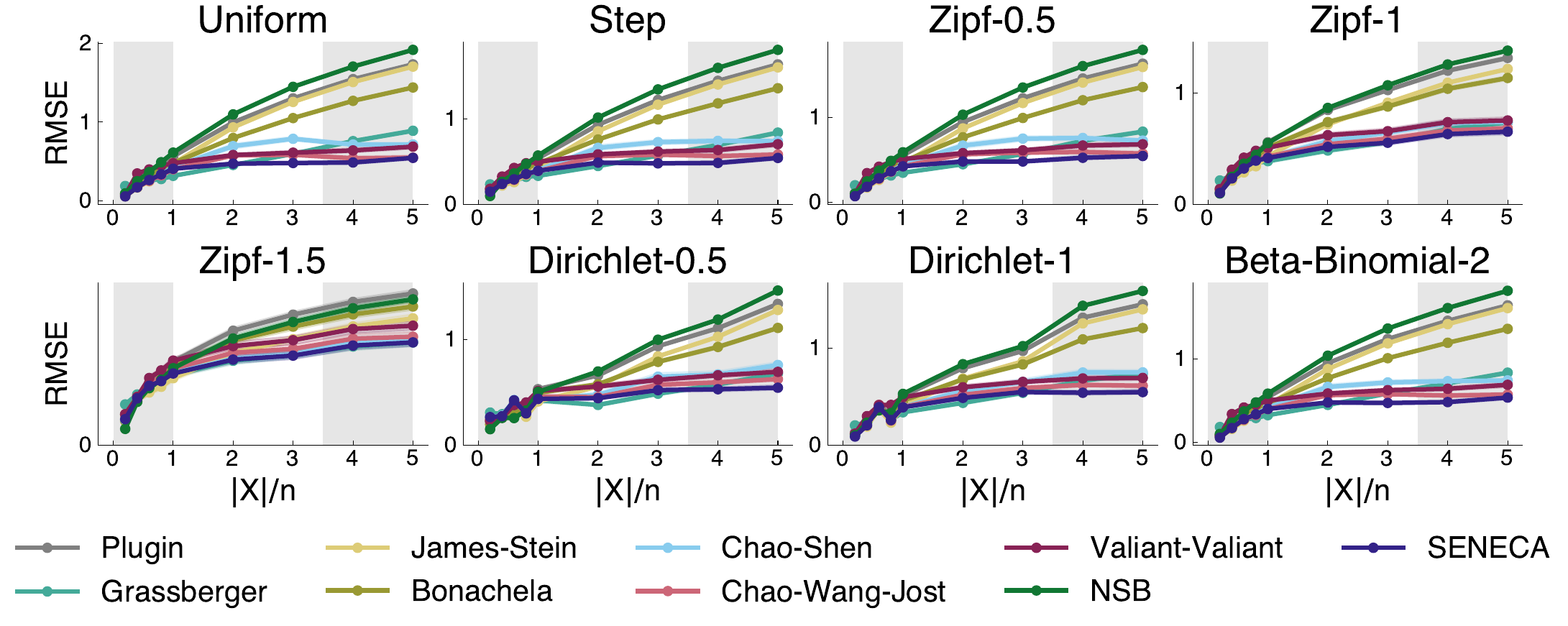}
    \caption{
    \textbf{Entropy estimation in small-sample item capture (all estimators, plus the NSB estimator).} 
    For each of $8$ families of label-prevalence distribution, each item's capture probability is determined uniformly over its label's prevalence probability mass.
    The number of labels $|\mathcal{X}|$ varies from $2$ to $50$ across trials.
    For each trial, $n=10$ items are captured, and the entropy of the label-prevalence distribution is estimated according to each of seven methods. 
    Results are averaged over $1000$ trials per label count.
    Colored regions about each line plot depict $95\%$ bias-corrected and accelerated confidence intervals, based on $1000$ bootstrap replications.
    Instances for which the NSB estimator returned null values are ignored.
    In each plot, the leftmost grey region corresponds to the well-sampled regime (i.e., $|\mathcal{X}|= \frac{n}{5}, \frac{2n}{5}, \frac{3n}{5}, \frac{4n}{5}, n$), where items to the right correspond to the under-sampled regime (i.e., $|\mathcal{X}|= 2n, 3n, 4n, 5n$).
    The rightmost grey region corresponds to the support-risky regime, where no consistent estimator for support size exists.
    }
    \label{fig:entropy-rmse-n10-full-with-nsb}
\end{figure*}

In this section, we briefly consider the Nemenman-Shafee-Bialek (NSB) estimator, a Bayesian approach introduced in \citet{nemenman2001entropy}.
We benchmark against this estimator here, and not in prior sections, because NSB periodically returns null results (i.e., when $\Phi_1(X^n)=n$).
Therefore, we discard said null instances (\textasciitilde$6\%$ of trials in the simulation study), but this makes the confidence intervals for NSB not directly comparable to the others'.
Figure \ref{fig:entropy-rmse-n10-full-with-nsb} adds the NSB estimator to Figure \ref{fig:entropy-rmse-n10-full}.
There are two settings - the well-sampled instances of Zipf-1.5 and Dirichlet-0.5 - for which NSB appears quite strong.
On the other hand, NSB does not perform reliably in the under-sampled settings herein.

\begin{figure*}[t!]
    \centering
    \includegraphics[width=0.95\textwidth]{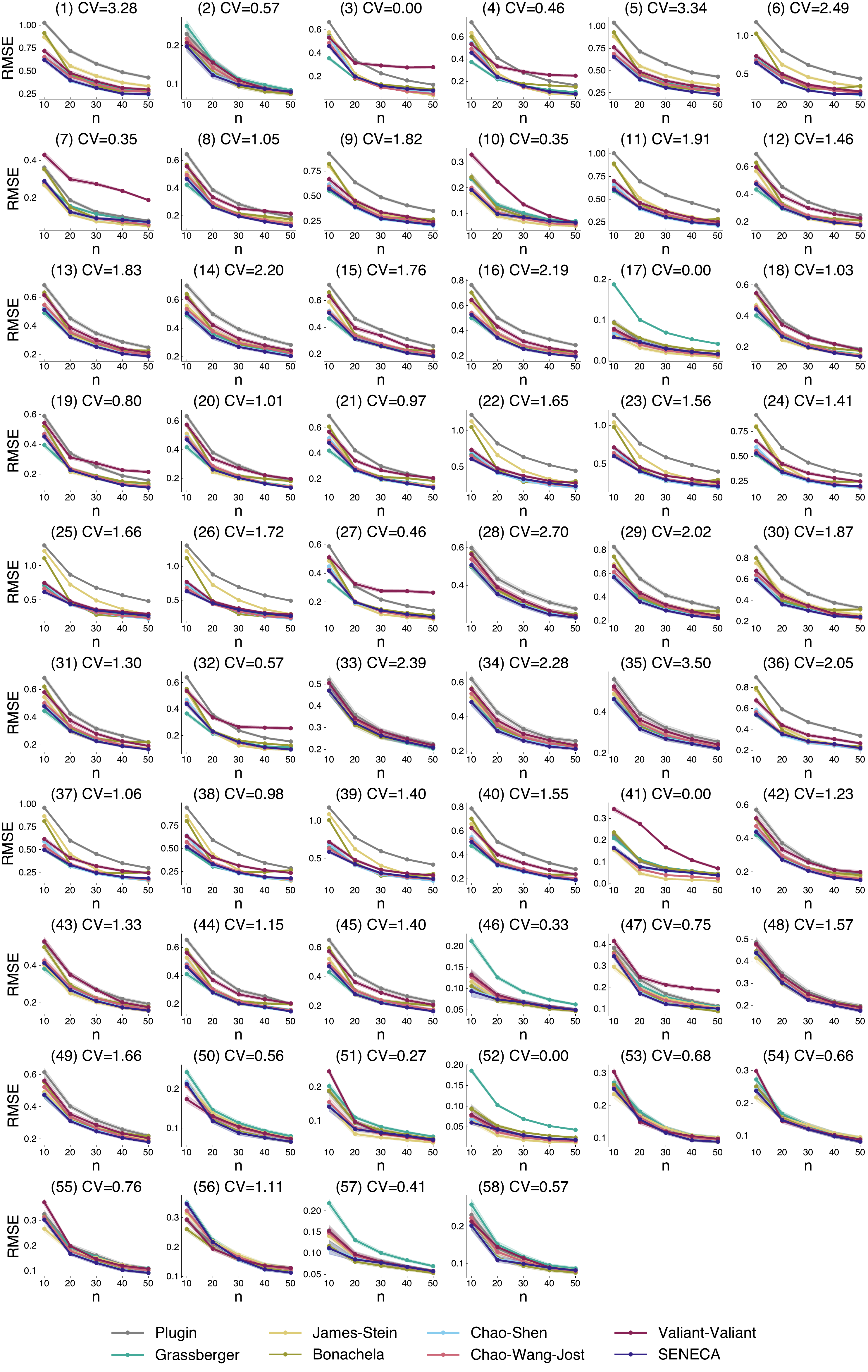}
    \caption{
    \textbf{Entropy estimation in $58$ real species-prevalence distributions.} 
    Collections of true bugs from \citet{janzen1973sweep2} are treated as full populations, similar to \citet{chao2003nonparametric}, for each of which we compare $8$ entropy estimators.
    Values are averaged over $1000$ trials per sample size and population.
    Color regions about each line plot depict $95\%$ bias-corrected and accelerated CIs, based on $1000$ bootstrap replications.
    CI lower bounds are clipped to ensure non-negativity.
    }
    \label{fig:biodiversity-rmse}
\end{figure*}

\clearpage
\begin{figure*}[t!]
    \centering
    \includegraphics[width=1.0\textwidth]{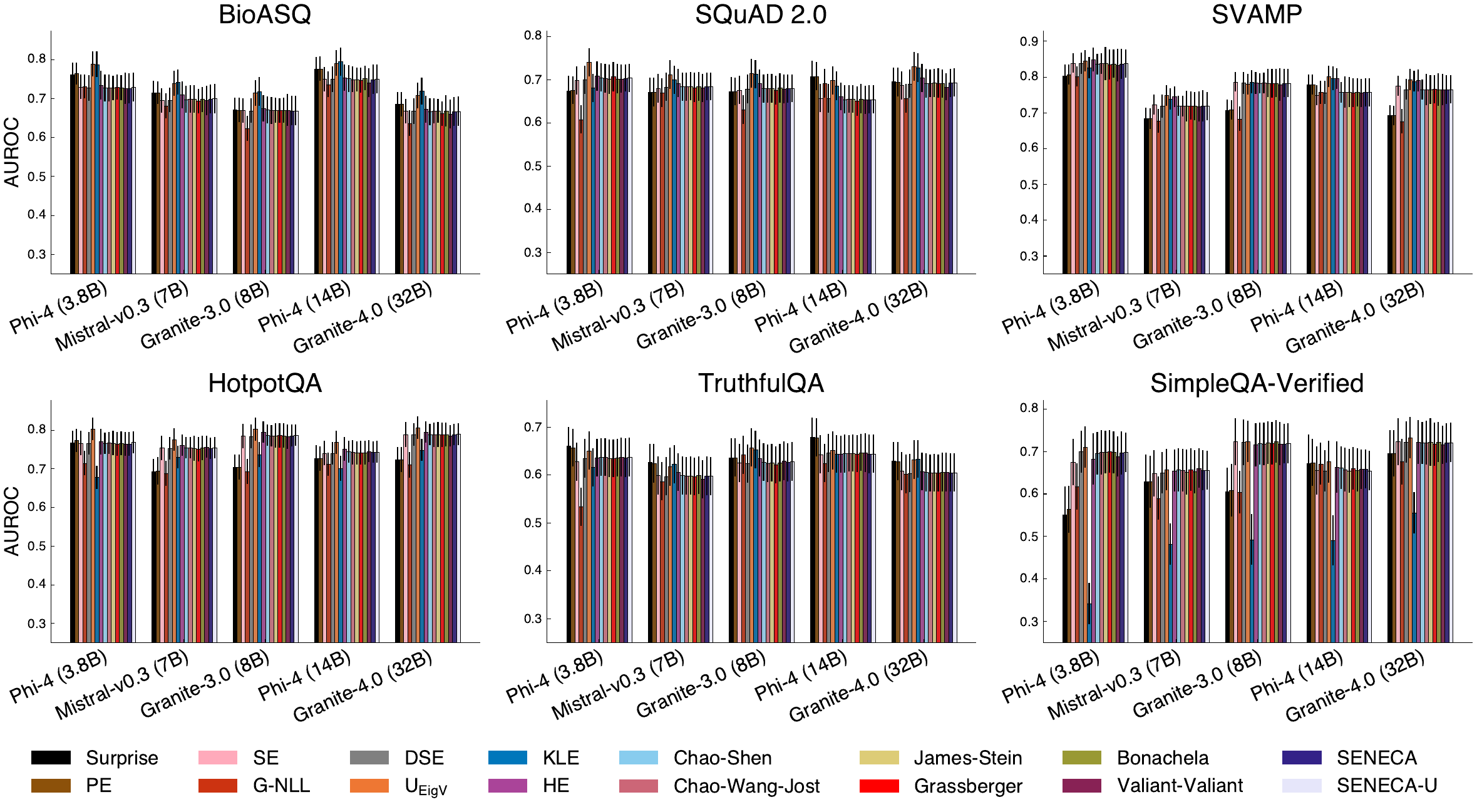}
    \caption{
    \textbf{LLM uncertainty quantification on six question-answering datasets (consistency judge).}
    $8$ methods are employed to detect incorrect LLM responses, as determined by the consistency-based judge described in Section \ref{sec:llm-judge-details}.
    Methods are evaluated by area under the receiver operating characteristic curve (AUROC).
    95\% CIs about AUROC estimates are calculated via the DeLong method \citep{delong1988comparing, sun2014fast}.
    Results are aggregated and summarized in Figure \ref{fig:applications-summary-uq-consistency}.
    }
    \label{fig:auroc}
\end{figure*}

\section{Additional LLM UQ results} \label{sec:more-llm-results}

In the main body of this work and in Figure \ref{fig:auroc}, our LLM UQ experiments gauge the correctness of an LLM response using a consistency-based judge, described in Appendix \ref{sec:llm-judge-details}.
Other related works employ a binary LLM judge that simply rates responses as correct or incorrect.
We consider this approach, as well, using the following prompt:
\begin{promptbox}{Binary LLM-as-Judge}
We are assessing the quality of answers to the follwoing question: \{question\}

The expected answer is: \{groundtruth\}

The proposed answer is: \{pred\}

Within the context of the question, does the proposed answer mean the same \
as the expected answer? Respond only with yes or no inside <rating></rating> \
tags. Only provide the rating, without any additional commentary.
\end{promptbox}

\begin{wrapfigure}[31]{r}{0.49\textwidth}
    \centering
    \includegraphics[width=0.49\textwidth]{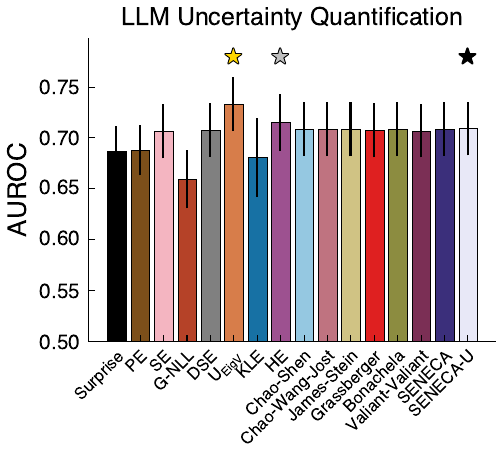}
    \caption{
    \textbf{Summary of LLM UQ results using binary LLM judge of correctness.}
    Using the binary LLM-as-judge prompt described in Appendix \ref{sec:more-llm-results}, ``best guess'' responses are labeled as correct/incorrect and UQ methods are assessed for their ability to detect incorrect responses.
    AUROC values and parameter uncertainties are aggregated across models and datasets, as described in Section \ref{sec:llm-experiments-desc}, to produce the above per-method summary.
    Gold and silver stars indicate the methods with highest and second-highest aggregated parameter estimates, respectively.
    }
    \label{fig:auroc-binary-summary}
\end{wrapfigure}

The above prompt is drawn near-identically from \citet{Farquhar2024}, but we request rating tags for easy parsing, like in the consistency-based prompt.
A summary figure of the results, similar to Figure \ref{fig:applications-summary-uq-consistency}, is shown in Figure \ref{fig:auroc-binary-summary}, with full results in Figure \ref{fig:auroc-binary-summary}.
We observe no qualitative difference arising from the difference in judge strategies.

\section{The missing mass estimator} \label{sec:more-missing-mass-results}

Although our primary goal is estimating entropy from small sample sizes, our method for doing so involves a new ``self-consistent'' missing mass estimator. In this section, we evaluate $\hat{M}^{SC}$ in isolation.
For baseline comparison, we consider the following alternatives, illustrating a range from high bias/low variance to low bias/high variance estimators: the classic Good-Turing method \citep{good1953population}; the so-called ``minimal bias'' estimator of \citet{leeHowMuchUnseen2025} (Lee-B{\"o}hme); and the closed-form, two-FoF estimator of \citet{painsky2023generalized}, which is arises from minimax risk minimization (Painsky).
We also include a missing mass estimator based on the support size estimator of \citet{wu2019chebyshev} (Wu-Yang), which is also described in Section \ref{sec:support-size-details}.
\citet{wu2019chebyshev} convert the Good-Turing missing mass into a support size estimator via $\widehat{|\mathcal{X}|}^{GT} = \frac{|X_\neq^n|}{1-\hat{M}^{GT}}$. 
By similar reasoning, \citet{leeHowMuchUnseen2025} construct a missing mass estimator from the Wu-Yang support size via $\hat{M}^{WY} = 1 - \frac{|X_\neq^n|}{\widehat{|\mathcal{X}|}^{WY}}$; we use the same, bounding values to the $[0, 1]$ interval.

\begin{figure*}[t!]
    \centering
    \includegraphics[width=1.0\textwidth]{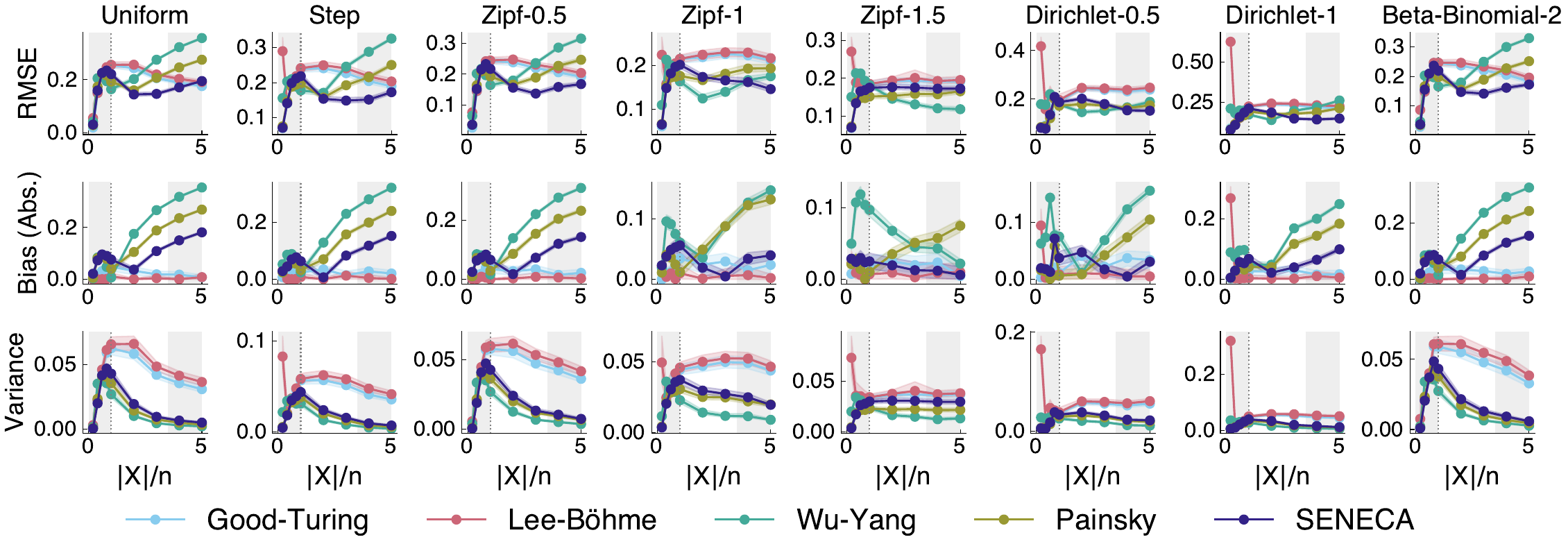}
    \caption{
    \textbf{Empirical RMSE, bias, and variance of missing mass estimation in small-sample item capture.} 
    For each of $8$ families of label-prevalence distribution, we generate $100$ items per label.
    Each item's capture probability is determined uniformly over its label's prevalence probability mass.
    The number of labels $|\mathcal{X}|$ varies from $2$ to $50$ across trials.
    For each trial, $n=10$ item-capture samples are drawn, and the missing mass of the label-prevalence distribution is estimated according to each of four methods. 
    Results are averaged over $1000$ trials per label count.
    Colored regions about each line plot depict $95\%$ bias-corrected and accelerated confidence intervals, based on $1000$ bootstrap replications.
    In each plot, leftmost grey region corresponds to the well-sampled regime (i.e., $|\mathcal{X}|= \frac{n}{5}, \frac{2n}{5}, \frac{3n}{5}, \frac{4n}{5}, n$), where items to the right correspond to the under-sampled regime (i.e., $|\mathcal{X}|= 2n, 3n, 4n, 5n$).
    The rightmost grey region corresponds to the support-risky regime, where no consistent estimator for support size exists.
    }
    \label{fig:mass-mse-bias-variance-n10}
\end{figure*}

In Figure \ref{fig:mass-mse-bias-variance-n10}, we conduct a similar analysis to that of Figures \ref{fig:entropy-rmse-n10} and \ref{fig:entropy-bias-variance-n10}, displaying the RMSE, bias, and variance of missing mass estimators instead of those of entropy estimators.
In general, our method exhibits comparable (albeit higher) variance to that of Painsky's two-FoF method, but does so with lower bias in the under-sampled regime.
True to its name, the Lee-B{\"o}hme ``minimal bias'' estimator exhibits the lowest bias overall.
Similar to Tables \ref{tab:entropy-rmse-n10} and \ref{tab:entropy-rmse-n20}, Table \ref{tab:mass-mse-n10} averages RMSE values from the data for Figure \ref{fig:mass-mse-bias-variance-n10} within the well-sampled and under-sampled regimes.
\textsc{SENECA} especially excels in the latter, with the lowest or second-lowest regime-averaged RMSE in $7/8$ under-sampled settings.
This is particularly apparent among distributions that are fatter tailed (i.e., Uniform, Step, Zipf-0.5, Beta-Binomial-2), where the differences between the smallest and next-smallest errors are large.
On the other hand, \textsc{SENECA} exhibits the second-lowest regime-averaged RMSE in $5/8$ well-sampled settings, where Painsky's method is generally preferable.
In those settings, however, \textsc{SENECA}'s aggregated RMSE is typically within a few percentage points of the lowest, underscoring balanced performance across regimes.

\begin{figure*}[t!]
    \centering
    \includegraphics[width=1.0\textwidth]{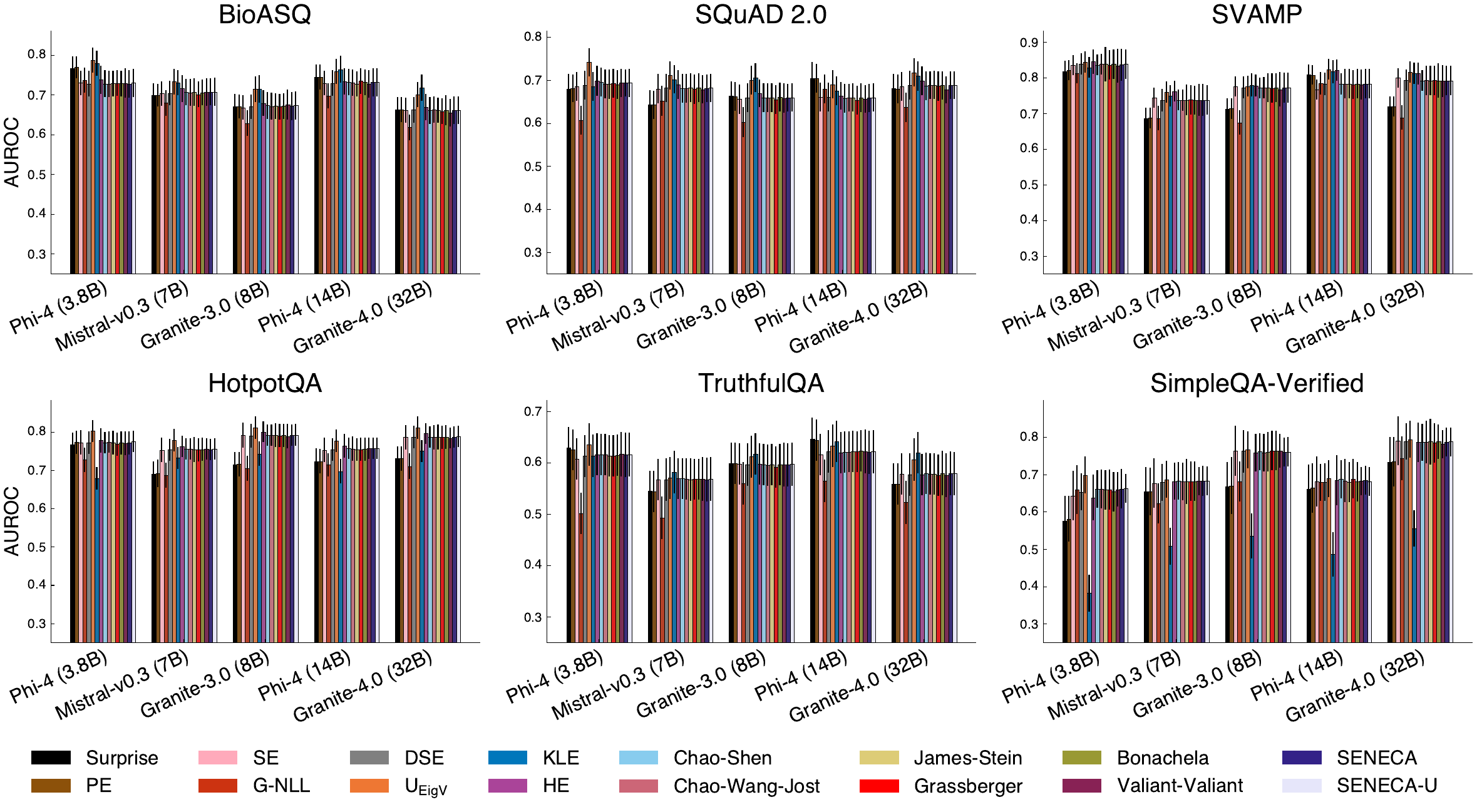}
    \caption{
    \textbf{LLM uncertainty quantification on six question-answering datasets (binary judge).}
    $8$ methods are employed to detect incorrect LLM responses, as determined by the binary judge described in Section \ref{sec:more-llm-results}.
    Methods are evaluated by area under the receiver operating characteristic curve (AUROC).
    95\% CIs about AUROC estimates are calculated via the DeLong method \citep{delong1988comparing, sun2014fast}.
    Results are aggregated and summarized in Figure \ref{fig:auroc-binary-summary}.
    }
    \label{fig:auroc-binary}
\end{figure*}

\begin{table*}[t!]
    \caption{
        \textbf{Regime-averaged error for missing mass estimation in small-sample item capture.}
        Using the data from Figure \ref{fig:mass-mse-bias-variance-n10} ($n=10$), we calculate the root-mean-squared error (RMSE) for each estimator and average results within the well-sampled ($|\mathcal{X}| \leq n$) and under-sampled ($|\mathcal{X}| > n$) regimes.
        The lowest and second-lowest values in each row are bold and underlined, respectively.
        The estimators are Good-Turing (GT), Lee-B{\"o}hme (LB), Wu-Yang (WY), Painsky (P), and the underlying missing mass estimator used in \textsc{SENECA} (highlighted in blue).
        In parentheses, we show the amount by which \textsc{SENECA}'s averaged RMSE is from the lowest value in each row.
    }
    \label{tab:mass-mse-n10}
    \centering
    \small
    \setlength{\tabcolsep}{7.25pt}
    \begin{tabular}{l|c|cccc>{\columncolor{lightindigocolor!10}}c}
    \\ 
    \toprule
Distribution & Regime & GT & LB & WY & P & SENECA \\
\midrule
\multirow{2}{*}{Uniform} & $|S| \leq n$ & 0.177 \tiny{\textcolor{gray}{$\pm$.004}} & 0.185 \tiny{\textcolor{gray}{$\pm$.008}} & \textbf{0.165} \tiny{\textcolor{gray}{$\pm$.004}} & \underline{0.167} \tiny{\textcolor{gray}{$\pm$.003}} & 0.174 \tiny{\textcolor{gray}{$\pm$.003}} (5.4\%) \\
 & $|S| > n$ & \underline{0.204} \tiny{\textcolor{gray}{$\pm$.004}} & 0.218 \tiny{\textcolor{gray}{$\pm$.004}} & 0.289 \tiny{\textcolor{gray}{$\pm$.002}} & 0.222 \tiny{\textcolor{gray}{$\pm$.003}} & \textbf{0.164} \tiny{\textcolor{gray}{$\pm$.003}} (0.0\%) \\
\midrule
\multirow{2}{*}{Step} & $|S| \leq n$ & 0.172 \tiny{\textcolor{gray}{$\pm$.004}} & 0.217 \tiny{\textcolor{gray}{$\pm$.008}} & 0.187 \tiny{\textcolor{gray}{$\pm$.003}} & \textbf{0.160} \tiny{\textcolor{gray}{$\pm$.003}} & \underline{0.168} \tiny{\textcolor{gray}{$\pm$.003}} (4.9\%) \\
 & $|S| > n$ & 0.215 \tiny{\textcolor{gray}{$\pm$.004}} & 0.228 \tiny{\textcolor{gray}{$\pm$.004}} & 0.257 \tiny{\textcolor{gray}{$\pm$.003}} & \underline{0.204} \tiny{\textcolor{gray}{$\pm$.004}} & \textbf{0.157} \tiny{\textcolor{gray}{$\pm$.004}} (0.0\%) \\
\midrule
\multirow{2}{*}{Zipf-0.5} & $|S| \leq n$ & 0.173 \tiny{\textcolor{gray}{$\pm$.004}} & 0.184 \tiny{\textcolor{gray}{$\pm$.009}} & \underline{0.169} \tiny{\textcolor{gray}{$\pm$.004}} & \textbf{0.163} \tiny{\textcolor{gray}{$\pm$.003}} & 0.171 \tiny{\textcolor{gray}{$\pm$.004}} (4.9\%) \\
 & $|S| > n$ & 0.214 \tiny{\textcolor{gray}{$\pm$.004}} & 0.225 \tiny{\textcolor{gray}{$\pm$.005}} & 0.254 \tiny{\textcolor{gray}{$\pm$.003}} & \underline{0.207} \tiny{\textcolor{gray}{$\pm$.003}} & \textbf{0.155} \tiny{\textcolor{gray}{$\pm$.004}} (0.0\%) \\
\midrule
\multirow{2}{*}{Zipf-1} & $|S| \leq n$ & 0.161 \tiny{\textcolor{gray}{$\pm$.004}} & 0.197 \tiny{\textcolor{gray}{$\pm$.010}} & 0.174 \tiny{\textcolor{gray}{$\pm$.003}} & \textbf{0.148} \tiny{\textcolor{gray}{$\pm$.003}} & \underline{0.159} \tiny{\textcolor{gray}{$\pm$.003}} (7.0\%) \\
 & $|S| > n$ & 0.219 \tiny{\textcolor{gray}{$\pm$.005}} & 0.225 \tiny{\textcolor{gray}{$\pm$.005}} & \textbf{0.151} \tiny{\textcolor{gray}{$\pm$.004}} & 0.183 \tiny{\textcolor{gray}{$\pm$.004}} & \underline{0.162} \tiny{\textcolor{gray}{$\pm$.004}} (6.8\%) \\
\midrule
\multirow{2}{*}{Zipf-1.5} & $|S| \leq n$ & 0.148 \tiny{\textcolor{gray}{$\pm$.003}} & 0.203 \tiny{\textcolor{gray}{$\pm$.012}} & 0.190 \tiny{\textcolor{gray}{$\pm$.003}} & \textbf{0.131} \tiny{\textcolor{gray}{$\pm$.003}} & \underline{0.142} \tiny{\textcolor{gray}{$\pm$.003}} (8.7\%) \\
 & $|S| > n$ & 0.192 \tiny{\textcolor{gray}{$\pm$.004}} & 0.196 \tiny{\textcolor{gray}{$\pm$.005}} & \textbf{0.129} \tiny{\textcolor{gray}{$\pm$.003}} & \underline{0.159} \tiny{\textcolor{gray}{$\pm$.004}} & 0.174 \tiny{\textcolor{gray}{$\pm$.003}} (35.4\%) \\
\midrule
\multirow{2}{*}{Dirichlet-0.5} & $|S| \leq n$ & 0.144 \tiny{\textcolor{gray}{$\pm$.003}} & 0.239 \tiny{\textcolor{gray}{$\pm$.013}} & 0.191 \tiny{\textcolor{gray}{$\pm$.003}} & \textbf{0.134} \tiny{\textcolor{gray}{$\pm$.003}} & \underline{0.139} \tiny{\textcolor{gray}{$\pm$.003}} (3.8\%) \\
 & $|S| > n$ & 0.238 \tiny{\textcolor{gray}{$\pm$.005}} & 0.244 \tiny{\textcolor{gray}{$\pm$.005}} & \textbf{0.164} \tiny{\textcolor{gray}{$\pm$.004}} & 0.173 \tiny{\textcolor{gray}{$\pm$.004}} & \underline{0.172} \tiny{\textcolor{gray}{$\pm$.004}} (5.3\%) \\
\midrule
\multirow{2}{*}{Dirichlet-1} & $|S| \leq n$ & 0.153 \tiny{\textcolor{gray}{$\pm$.003}} & 0.261 \tiny{\textcolor{gray}{$\pm$.007}} & 0.194 \tiny{\textcolor{gray}{$\pm$.003}} & \textbf{0.144} \tiny{\textcolor{gray}{$\pm$.003}} & \underline{0.150} \tiny{\textcolor{gray}{$\pm$.003}} (4.3\%) \\
 & $|S| > n$ & 0.226 \tiny{\textcolor{gray}{$\pm$.004}} & 0.236 \tiny{\textcolor{gray}{$\pm$.005}} & 0.205 \tiny{\textcolor{gray}{$\pm$.003}} & \underline{0.190} \tiny{\textcolor{gray}{$\pm$.004}} & \textbf{0.158} \tiny{\textcolor{gray}{$\pm$.004}} (0.0\%) \\
\midrule
\multirow{2}{*}{Beta-Binomial-2} & $|S| \leq n$ & 0.174 \tiny{\textcolor{gray}{$\pm$.005}} & 0.186 \tiny{\textcolor{gray}{$\pm$.011}} & \textbf{0.165} \tiny{\textcolor{gray}{$\pm$.004}} & \underline{0.165} \tiny{\textcolor{gray}{$\pm$.004}} & 0.172 \tiny{\textcolor{gray}{$\pm$.004}} (4.4\%) \\
 & $|S| > n$ & 0.212 \tiny{\textcolor{gray}{$\pm$.005}} & 0.224 \tiny{\textcolor{gray}{$\pm$.005}} & 0.265 \tiny{\textcolor{gray}{$\pm$.003}} & \underline{0.206} \tiny{\textcolor{gray}{$\pm$.003}} & \textbf{0.156} \tiny{\textcolor{gray}{$\pm$.004}} (0.0\%) \\
\bottomrule
\end{tabular}
\end{table*}

\section{AI use statement} \label{sec:gen-ai-usage}

LLMs were employed for coding, statistical question-answering, and LaTeX formatting in the preparation of this work.
The original ideas and written content were human-generated.
Mathematical content was either human-drafted and revised based on LLM feedback or facilitated by iterative collaboration with LLMs.
Any LLM-assisted mathematical content was reviewed, digested, and rewritten by the human authors.
Existing literature were discussed conversationally with LLMs during the preparation process.

\end{document}